\documentclass{ar-1col}

\usepackage{amsmath,array,amssymb,natbib,aas_macros}
\usepackage{url}

\newcommand{\rqq}{\textquotedblright}
\newcommand{\lqq}{\textquotedblleft}

\jname{Annu. Rev. Astron. Astrophys. 58:617-59.}
\jvol{The Annual Review of Astronomy and Astrophysics is online at astro.annualreviews.org}
\jyear{2020}
\doi{10.1146/annurev-astro-032620-021859}

\begin{document}

\markboth{Ouchi et al.}{Ly$\alpha$ Universe}

\title{Observations of the Lyman-$\alpha$ Universe}

\author{Masami Ouchi,$^{1,2,3}$ Yoshiaki Ono,$^2$ and Takatoshi Shibuya$^4$
\affil{$^1$National Astronomical Observatory of Japan, 2-21-1 Osawa, Mitaka, Tokyo 181-8588, Japan; email: ouchims@icrr.u-tokyo.ac.jp}
\affil{$^2$Institute for Cosmic Ray Research, The University of Tokyo, 5-1-5 Kashiwanoha, Kashiwa, Chiba 277-8582, Japan}
\affil{$^3$Kavli Institute for the Physics and Mathematics of the Universe (Kavli IPMU, WPI), The University of Tokyo, 5-1-5 Kashiwanoha, Kashiwa, Chiba, 277-8583, Japan}
\affil{$^4$Kitami Institute of Technology, 165 Koen-cho, Kitami, Hokkaido 090-8507, Japan}}

\begin{abstract}
%
Hydrogen Lyman-$\alpha$ (Ly$\alpha$) emission has been one of the major observational probes for the high redshift universe, since the first discoveries of high-$z$ Ly$\alpha$ emitting galaxies in the late 1990s. Due to the strong Ly$\alpha$ emission originated by resonant scattering and recombination of the most-abundant element, Ly$\alpha$ observations witness not only {\sc Hii} regions of star formation and AGN but also diffuse {\sc Hi} gas in the circum-galactic medium (CGM) and the inter-galactic medium (IGM). Here we review Ly$\alpha$ sources, and present theoretical interpretations reached to date. We conclude that: 1) A typical Ly$\alpha$ emitter (LAE) at $z\gtrsim 2$ with a $L^*$ Ly$\alpha$ luminosity is a high-$z$ 
counterpart of a local dwarf galaxy,
a compact metal-poor star-forming galaxy (SFG) with an approximate stellar (halo) mass and star-formation rate of $10^{8-9} M_\odot$ ($10^{10-11} M_\odot$) and $1-10 M_\odot$ yr$^{-1}$, respectively; 2) High-$z$ 
SFGs 
ubiquitously have a diffuse Ly$\alpha$ emitting halo in the CGM extending to the halo virial radius and beyond; 3) Remaining neutral hydrogen at the epoch of reionization makes a strong dimming of Ly$\alpha$ emission for galaxies at $z>6$ that suggest the late reionization history. The next generation large telescope projects will combine Ly$\alpha$ emission data with {\sc Hi} Ly$\alpha$ absorptions and 21cm radio data 
that map out the majority of
hydrogen ({\sc Hi}$+${\sc Hii}) gas, uncovering the exchanges of i) matter by outflow/inflow and ii) radiation, relevant to cosmic reionization, between galaxies and the CGM/IGM.
%
\end{abstract}

\maketitle

\tableofcontents


\section{Introduction}
\label{sec:introduction}
Hydrogen Ly$\alpha$ emission is a critical probe for understanding the high-redshift universe. Due to the abundant hydrogen and the atomic electron transition between the lowest energy levels,
%
from the $n=2$ state to the ground ($n=1$) state, 
%
Ly$\alpha$ is one of the strongest emission lines produced in the universe. 
%
%
Although Ly$\alpha$ has 
a physical nature of resonance line
\footnote{
Because a chance of electron's staying at the ground state is very high, due to a short-time scale required for ground-state transitions, Ly$\alpha$ photons are resonantly scattered by hydrogen.
}
whose photons have more chances of dust attenuation in gaseous nebulae in the process of resonant scattering,
%
%
the short wavelength ($1216$\AA) of Ly$\alpha$ emission allows us to pinpoint faint high-$z$ objects by optical and near-infrared (NIR) observations 
very efficiently.
Ly$\alpha$ is used as an excellent probe of high-$z$ objects near the observational redshift frontier.

Historically, \citet{partridge1967} first predict an importance of a high-$z$ galaxy that emits a strong Ly$\alpha$ line. \citet{partridge1967} argue that an early galaxy produces strong Ly$\alpha$ emission by the recombination process in the inter-stellar medium (ISM) heated by young massive stars, and that up to 6-7\% of the total galaxy luminosity can be converted to a Ly$\alpha$ luminosity that is as bright as $\sim 2\times 10^{45}$ erg s$^{-1}$ at $z\sim 10-30$. Moreover,
\citet{partridge1967} claim a possibility of strong free-electron scattering in ionized gas of the inter-galactic medium (IGM) that smears radiation from young galaxies. Although recent studies do not agree with the extremely bright Ly$\alpha$ luminosity and the notable smearing of the radiation in the ionized IGM, 
%
%
\citet{partridge1967} is an excellent imaginative theory paper (published more than a half century ago) suggesting that strong Ly$\alpha$ emission of high-$z$ galaxies is physically related to early galaxies and the ionization state of the IGM that are two major topics discussed today with strong Ly$\alpha$ emission, namely galaxy formation and cosmic reionization.

After the theoretical predictions of \citet{partridge1967}, a number of programs searched for high-$z$ galaxies with strong Ly$\alpha$ emission by observations. However, no such objects were found until the mid-1990s, due to the limited sensitivities of the available telescopes. Finally, by the operation starts of large (8-10m class) ground-based telescopes and \textit{Hubble Space Telescope} (HST), a few high-$z$ galaxies with strong Ly$\alpha$ emission were successfully identified by narrowband (NB) imaging
\footnote{
Ly$\alpha$ is redshifted to the 
NB transmission, showing a 
NB photometry excess.
}
and spectroscopy on the sky around QSO BR2237-0607 at $z=4.55$ \citep{hu1996} and a radio galaxy 53W002 at $z=2.39$ \citep{pascarelle1996} as well as in the blank field  \citep{cowie1998}. Subsequently, a number of deep observation programs have been conducted for Ly$\alpha$ emitting objects at $z\gtrsim 2$, including the Hawaii Survey \citep{cowie1998}, the Large Area Lyman Alpha Survey \citep{rhoads2000}, the Subaru surveys \citep{ouchi2003}, and the Multiwavelength Survey by Yale-Chile \citep{gawiser2007}, 
the Hobby-Eberly Telescope Dark Energy Experiment (HETDEX) 
Pilot Survey \citep{adams2011}, 
and 
Very Large Telescope (VLT) 
/Multi Unit Spectroscopic Explorer (MUSE) survey \citep{bacon2017}. Moreover, Ly$\alpha$ emitting objects in the local and low-$z$ universe are investigated by space-based UV observations such by the HST program of Lyman-Alpha Reference Sample \citep{ostlin2014} and the \textit{Galaxy Evolution Explorer} (GALEX) grism programs \citep{deharveng2008,cowie2011}.
Ly$\alpha$ emitting galaxies with no AGN, thus found, have a Ly$\alpha$ luminosity of $10^{41}-10^{44}$ erg s$^{-1}$, more than an order of magnitude fainter than the one predicted by \citet{partridge1967}.

\begin{figure}[h]
\includegraphics[width=4.8in]{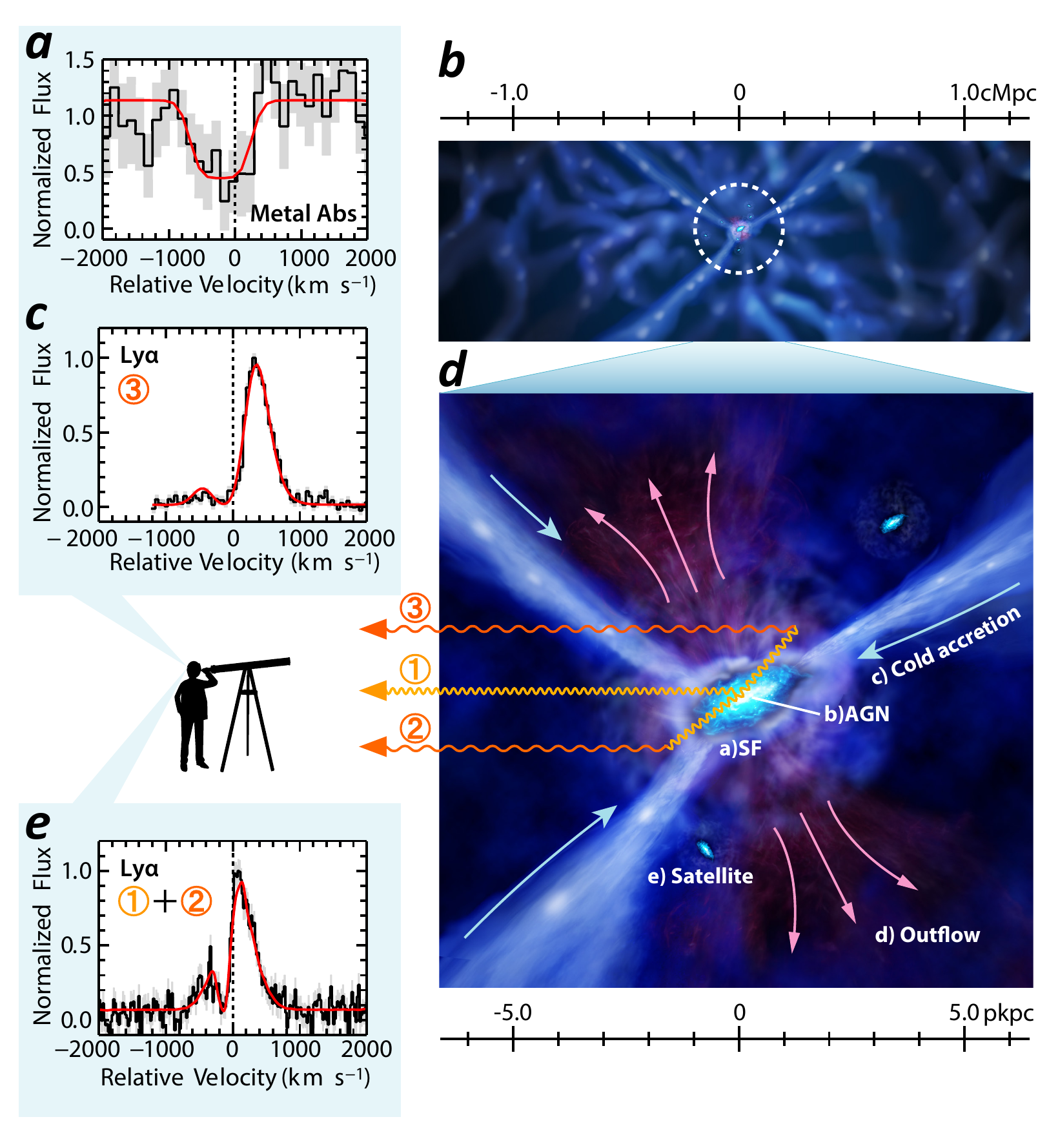}
\caption{
Conceptual figure of an LAE with a moderately high mass. In panel $b$, the LAE is located at the central node of the filamentary structure surrounded with satellite galaxies. The white dashed circle represents the virial radius of the host dark matter halo of the LAE. Panel $d$ is a zoom-in of the LAE. Possible origins of Ly$\alpha$ emission are labeled: a) star formation (SF), b) AGN, c) cold accretion, d) outflow, and e) satellite galaxies. In the LAE with outflowing gas, the wavy lines ($\textcircled{\scriptsize1}-\textcircled{\scriptsize3}$) indicate three light paths for the observed Ly$\alpha$ lines. Panels $e$ and $c$ present spectra that are dominated by the contributions of the light paths of $\textcircled{\scriptsize1}+\textcircled{\scriptsize2}$ and $\textcircled{\scriptsize3}$, respectively, with the observed spectrum (black line) and error (gray region) and the best-fit models (red line). The light paths of $\textcircled{\scriptsize2}$ and $\textcircled{\scriptsize3}$ indicate that Ly$\alpha$ photons are resonantly scattered in the outflowing gas. In panel $a$, the continuum spectrum of the LAE with a metal absorption line is shown. The spectral data are adapted with permission from \cite{hashimoto2015} and \cite{sugahara2019}. 
%
%
}
\label{fig:LAE_Concept}
\end{figure}

Ly$\alpha$ emitting objects are called Ly$\alpha$ emitters (LAEs). Conventionally, LAEs are defined as objects with a rest-frame Ly$\alpha$ equivalent width 
(EW) of 
\begin{equation}
EW_0\gtrsim 20{\rm \AA}
\end{equation}
This 
Ly$\alpha$ $EW_0$
limit corresponds to that of the samples made by classical observations for LAEs with a 
NB whose wavelength transmission width is a $\sim 1$\% of the central wavelength.\footnote{Note that some of the recent studies refer to objects with weak Ly$\alpha$ of $EW_0>0 {\rm \AA}$ as LAEs.
}
It is known that LAEs are young 
star-forming galaxies (SFGs) 
or AGNs (Section \ref{sec:Lya_emitter_obs}).
%
Figure \ref{fig:LAE_Concept}
%
\footnote{
We use abbreviations of cMpc, pMpc, and pkpc for comoving megaparsec, physical megaparsec, and physical kiloparsec, respectively.
}
is an illustration of 
a conceptual 
LAE with spectroscopic properties that are detailed in Section \ref{sec:physical_picture}.


\begin{marginnote}[]
\entry{cMpc}{comoving megaparsec}
\entry{pMpc}{physical megaparsec}
\entry{pkpc}{physical kiloparsec}
\end{marginnote}

Because Ly$\alpha$ photons are resonantly scattered by neutral hydrogen {\sc Hi}, Ly$\alpha$ is unfortunately thought to be a poor probe of galaxy dynamics or the ionizing photon budget. However, the distribution and dynamics of {\sc Hi} gas are encoded to Ly$\alpha$ emission via the resonant scattering. The resonantly scattered Ly$\alpha$ photons allow us to investigate {\sc Hi} gas in and around high-$z$ objects that is generally hard to be probed by observations. 
The circum-galactic medium (CGM) of high-$z$ 
SFGs 
are characterized by diffuse Ly$\alpha$ emission extending over 
$>10$ physical kpc (pkpc) even to the large scale reaching the IGM (\citealt{kakuma2019}; Section \ref{sec:CGM}).
%
%
Interestingly, such diffuse Ly$\alpha$ emission of high-$z$ galaxies 
cover the entire sky
in any observational lines of sight \citep{wisotzki2018}.
%
%
%
%
Figure 1$b,d$ 
illustrate Ly$\alpha$ emission traveling in the CGM and the IGM. Theoretical models predict that some fractions of Ly$\alpha$ photons are produced by cold accretion, outflow, and unresolved faint satellites (Section \ref{sec:physical_picture}) especially in the CGM and the IGM.
%
%
At the epoch of reionization (EoR), 
the 
ionized fraction and distribution of {\sc Hi} in the IGM 
are imprinted in the observed Ly$\alpha$ emission (Section \ref{sec:reionization}). Ly$\alpha$ is an important probe for gas of the CGM and the IGM for characterizing galaxies and cosmic reionization.

\section{Physical Picture of LAEs}
\label{sec:physical_picture}

\subsection{Origins of Ly$\alpha$ emission}
\label{sec:Lya_origins}

%


Ly$\alpha$ emission from galaxies have five major origins. 
In the ISM 
near the central region of a galaxy, 
Ly$\alpha$ emission can be originated from recombinations of hydrogen atoms 
that are ionized mainly by two sources:  
1) young massive stars (star formation)
and 
2) an AGN  
if any. 
In the CGM 
and the outer region, 
Ly$\alpha$ can be emitted from three origins: 
3) outflow that can collisionally excite {\sc Hi} atoms (shock heating),\footnote{In the case 
of collisional excitation,  
the Ly$\alpha$ to H$\alpha$ flux ratio can be higher than 
that of the case B recombination 
(Figure 7 of \citealt{dijkstra2017}).}
4) infalling gas (cold accretion), 
which is predicted to release a significant amount of its gravitational energy 
in Ly$\alpha$ through collisional excitation (gravitational cooling), 
and 
5) fluorescence from hydrogen 
in the CGM and IGM 
photoionized by UV background radiation powered by energetic sources such as QSOs. 
In addition, 
star formation in unresolved faint satellite galaxies can also contribute. 
%
%
Note that no Ly$\alpha$ sources dominated by 
gravitational cooling
have been definitively identified so far 
(Sections \ref{sec:ISM} and \ref{sec:CGM}). 
%
%

Another important physical process for Ly$\alpha$ emisssion is resonant scattering, 
%
since 
{\sc Hi} gas in typical galaxies is optically thick to Ly$\alpha$. 
The cross section of Ly$\alpha$ for a collection of moving atoms 
can be obtained by convolving the single atom cross section 
with their velocity distribution. 
Assuming a Maxwellian velocity distribution 
and introducing the dimensionless frequency variable $x \equiv (\nu - \nu_0)/\Delta \nu_{\rm D}$, 
where $\nu_0$ is the line center frequency and $\Delta \nu_{\rm D}$ is the Doppler width, 
the average cross section is\footnote{A 
detailed derivation of this equation is given in Section 6.4 of \cite{dijkstra2017}.}  
\begin{equation} 
\sigma_x (\nu, T)  
	= \frac{3 \lambda_0^2 a_v}{2 \sqrt{\pi}} H(a_v, x) 
	\simeq 5.9 \times 10^{-14} \left( \frac{T}{10^4 \, {\rm K}} \right)^{-1/2} H(a_v, x) \,\, {\rm cm}^2, 
\label{eq:cross_section}
\end{equation}
where 
$\lambda_0$ is the Ly$\alpha$ wavelength, 
$a_v$ is the Voigt parameter, 
and $H(a_v,x)$ is the Voigt function, 
\begin{equation}
H(a_v,x) 
	= \frac{a_v}{\pi} \int^\infty_{-\infty} \frac{e^{-y^2} dy}{(y-x)^2 + a_v^2} 
	\approx \begin{cases}
	e^{-x^2} & {\rm central \,\, resonant \,\, core,} \\
	\frac{a_v}{\sqrt{\pi} x^2} & {\rm damping \,\, wing.}
	\end{cases}
\end{equation}
The transition between the core and wing happens at around $e^{-x^2} = a_v / (\sqrt{\pi} x^2)$.
Based on these equations, 
{\sc Hi} gas is optically thick to Ly$\alpha$ at the line center 
%
(hereafter referred to as 
{\lqq}optically thick{\rqq})
%
when the {\sc Hi} column density is higher than 
$N_\mathrm{HI} = 1/\sigma_x (\nu_0, T) \simeq 2 \times 10^{13} (T / 10^4 \, {\rm K})^{1/2}$ cm$^{-2}$, 
which is much lower than those of typical galaxies. 
Although this resonance nature of Ly$\alpha$ makes it difficult to 
pinpoint the original Ly$\alpha$ source position, 
it enables investigations of the distribution and the kinematics of {\sc Hi} gas 
via theoretical modeling. 


\subsection{Modeling Ly$\alpha$ Emission}
\label{sec:Lya_model}

%
%
%
%
%
%
%
%

One of the key issues to modeling Ly$\alpha$ emission from galaxies is 
its complicated radiative transfer due to the resonance nature of Ly$\alpha$
\footnote{
In other words, Ly$\alpha$ photons scatter multiple times in the optically thick line core before diffusing into the wings, where they finally escape.
}.
Analytical solutions have been obtained only in very limited cases.  
%
In the simple case of an optically thick dust-free static slab with a central plane source emitting Ly$\alpha$, 
the emergent Ly$\alpha$ line profile is given by 
\cite{harrington1973}
%
\footnote{See also Equation 3.51 and the footnote on p.29 of \citet{laursen2010}}
as
\begin{equation} 
J (x) 
	= \frac{\sqrt{6}}{24 \sqrt{\pi} a_v \tau_0} \frac{x^2}{\cosh \left[ \sqrt{\frac{\pi^3}{54}} \frac{x^3}{a_v \tau_0}  \right]}, 
\end{equation} 
where $\tau_0$ is the optical depth at the line center from the center to the boundary of the slab. 
The spectral shape is symmetric around $x=0$ 
and double peaked at $x \approx \pm 1.1 (a_v \tau_0)^{1/3}$. 
The peaks are more separated with higher $\tau_0$, 
because Ly$\alpha$ photons need to shift their velocities more into the wings to escape.

For more general cases, 
there are some theoretical models, such as the expanding shell (ES) model (\citealt{ahn2004}; \citealt{verhamme2006}), that adopt the Monte Carlo radiative transfer technique, which
successfully explain the diversity of observed Ly$\alpha$ profiles 
with a relatively small number of physical parameters. 
The ES model assumes a simple geometry 
where a Ly$\alpha$ source is located at the center of a spherically symmetric expanding shell of homogeneous and isothermal {\sc Hi} gas, modeling a galaxy-scale supershell made by multiple supernovae in star-forming regions
\footnote{Such 
galaxy-scale expanding supershells are found in nearby starbursts 
(\citealt{marlowe1995}).}
 (Figure \ref{fig:LAE_Concept}$d$).


To explain the basic idea of the ES model, Figure \ref{fig:LAE_Concept} shows the predictions of observed Ly$\alpha$ emission line profiles for three different light paths ($\textcircled{\scriptsize1}-\textcircled{\scriptsize3}$). Ly$\alpha$ photons along the light path (\textcircled{\scriptsize1}) directely come from the central Ly$\alpha$ source penetrating the shell. Along the light path ($\textcircled{\scriptsize2}$), Ly$\alpha$ photons escape from the shell approaching the observer via scattering, although some of them are absorbed by {\sc Hi} in the shell, resulting in a red component escaping the red wing of Ly$\alpha$ absorption in the blueshifted shell and a small blue component escaping the blue wing of the Ly$\alpha$ absorption. Figure \ref{fig:LAE_Concept}$e$ shows the Ly$\alpha$ spectrum that is dominated by these two components ($\textcircled{\scriptsize1}+\textcircled{\scriptsize2}$). The velocity shift of the red component relative to the systemic redshift is small, because a small number of Ly$\alpha$ photons escape from the far side of the shell blowing away by scattering that is referred to as backscattering. Ly$\alpha$ photons tracking the light path ($\textcircled{\scriptsize3}$) experience the backscattering at the shell. In this case, the Ly$\alpha$ spectrum can have a peak at $\sim 2V_{\rm exp}$, where $V_{\rm exp}$ is the radial expansion velocity or the outflow velocity (see Figure \ref{fig:LAE_Concept}$c$). The $\sim 2V_{\rm exp}$ shift of the peak is given by an effect similar to that of a reflection in a moving mirror; the Ly$\alpha$ photons enter the shell experiencing a Doppler shift by $V_{\rm exp}$ and are then backscattered to the observer being further Doppler shifted by $V_{\rm exp}$. This simple model has successfully reproduced observed Ly$\alpha$ profiles of LAEs as well as of Lyman break galaxies (LBGs) at both low and high redshifts (Section \ref{sec:ISM}).

\begin{figure}[h]
\includegraphics[width=5.3in]{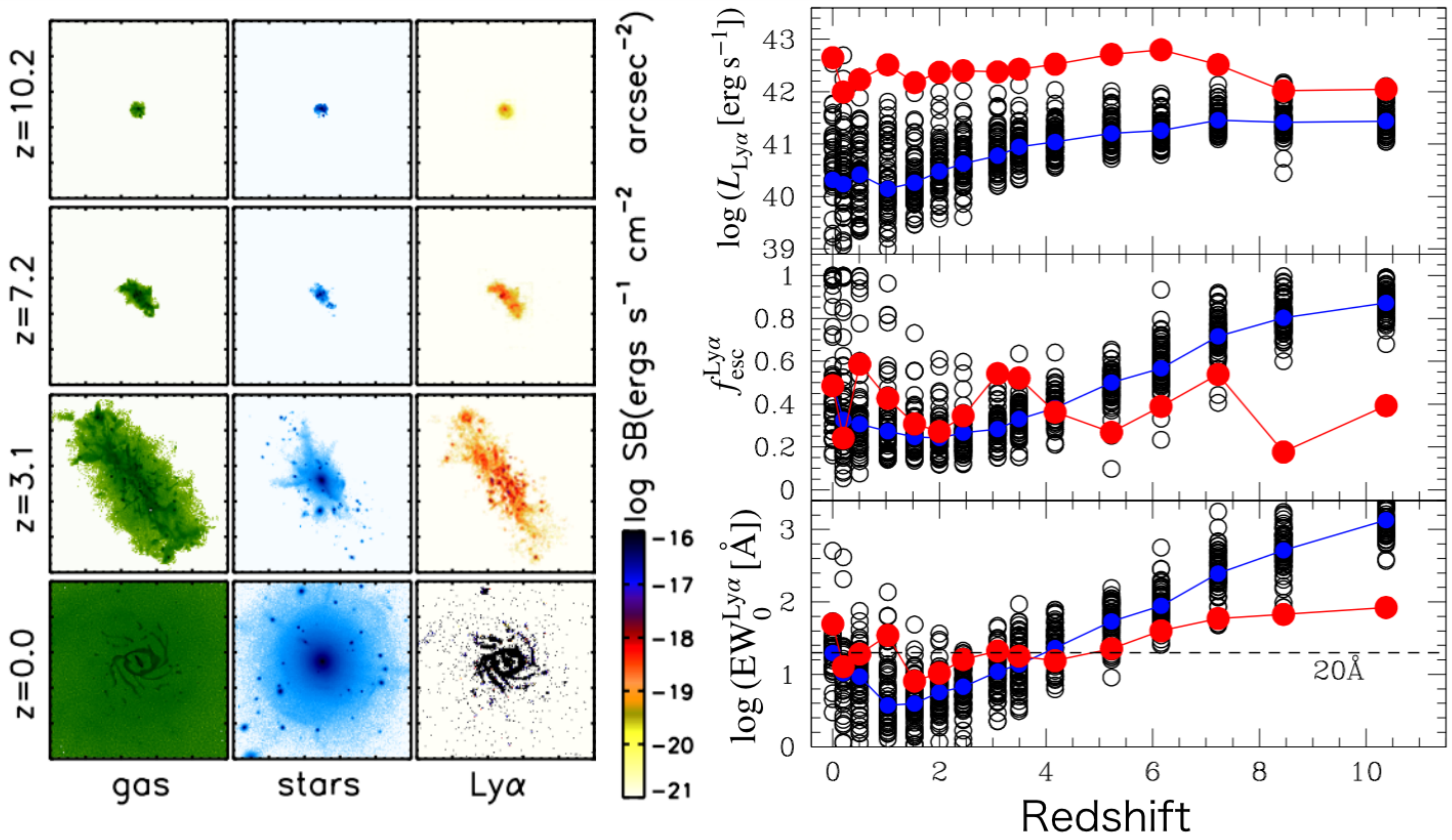}
\caption{
Left: 
Snapshots of a MW-like galaxy at different redshifts 
in a cosmological hydrodynamic simulation. 
From left to right, 
the distributions of gas 
and stars 
as well as 
the SB of Ly$\alpha$ 
are shown. 
The box size is $200$ pkpc. 
%
Right: 
Evolution of the Ly$\alpha$ properties of the 60 most massive progenitors. 
From top to bottom, 
the Ly$\alpha$ luminosity, 
the escape fraction of Ly$\alpha$ photons over whole solid angle, 
and Ly$\alpha$ 
$EW_0$
are presented. 
The red filled circle corresponds to 
the most massive progenitor at each redshift. 
The blue filled circles represent the median value of the $60$ galaxies. 
Note that observational effects of reionization, Ly$\alpha$ dimming in the IGM (Section \ref{sec:reionization}), are not included in the models.
%
Adapted from \cite{yajima2012} with permission. 
}
\label{fig:LAE_Simulation}
\end{figure}

Combining with Monte Carlo radiative transfer 
calculations, 
some theoretical studies have conducted cosmological hydrodynamic simulations
including dust attenuation effects
%
to investigate possible descendants of high-$z$ LAEs as well as their physical properties.  
\cite{yajima2012} have found that 
progenitors 
of a present-day Milky Way (MW) like galaxy with a halo mass $\sim 10^{12} M_\odot$ 
show bright Ly$\alpha$ emission at high redshifts 
comparable to Ly$\alpha$ $L^\ast$ ($L_{{\rm Ly}\alpha}^\ast$) of LAEs at $z \sim 2-6$. 
The left panel of Figure \ref{fig:LAE_Simulation} shows 
the distributions of gas and stars of the most massive progenitor of a MW-like galaxy 
as well as the Ly$\alpha$ surface brightness (SB) at different redshifts. 
%
%
%
The right panel of Figure \ref{fig:LAE_Simulation} represents 
the evolution of Ly$\alpha$ properties of the $60$ most massive progenitors.
As shown in the bottom right panel of Figure \ref{fig:LAE_Simulation}, 
most of the galaxies at high redshifts 
are classified as LAEs, 
and some of them are bright enough to be detected in previous observations (top right panel). 
The escape fractions of Ly$\alpha$ photons
over whole solid angle 
are also presented  
in the middle right panel of Figure \ref{fig:LAE_Simulation}. 
The median value of the Ly$\alpha$ escape fraction
is about $30${\%} at $z\sim3$, 
which is consistent with those obtained in observational studies 
(Section \ref{sec:Lya_emitter_obs}). 
Note that at low redshifts of $z \lesssim 2$, 
the Ly$\alpha$ escape fraction 
has a relatively large dispersion 
due to the diversities in their physical properties. 
%
These results suggest that 
some of the typical observed LAEs at $z\sim 2-6$ 
would evolve into present-day MW-like galaxies, 
which is consistent with the results from the clustering analyses
(\citealt{gawiser2007}; \citealt{ouchi2010}). 
\section{Ly$\alpha$ Emitter Observations}
\label{sec:Lya_emitter_obs}

 %
 %
 %
 
 %
 %

\begin{figure}[h]
\begin{minipage}{3.3in}
\begin{center}
\includegraphics[width=3.3in]{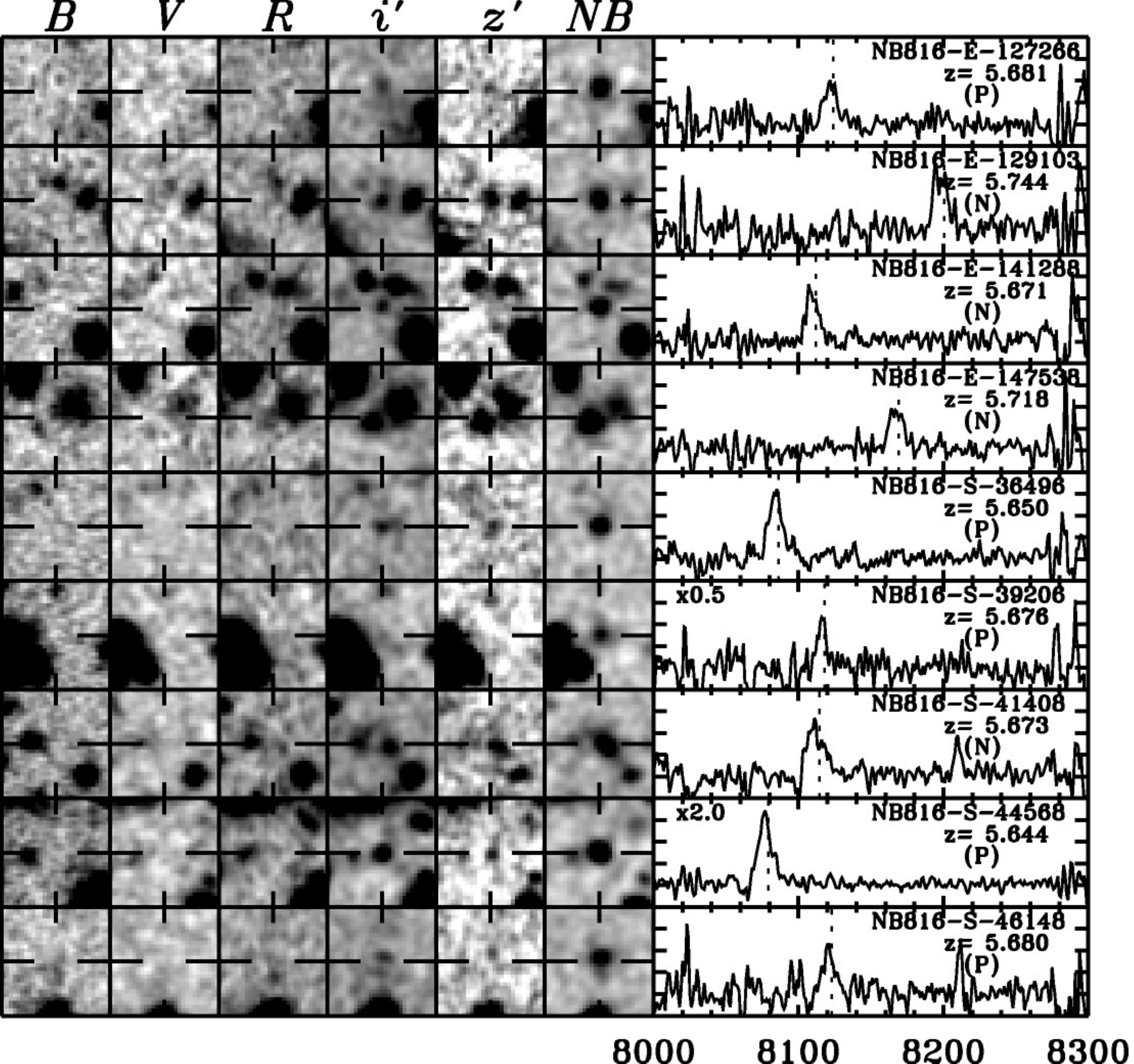}
\end{center}
\end{minipage}
\begin{minipage}{2.2in}
\begin{center}
\includegraphics[width=2.2in]{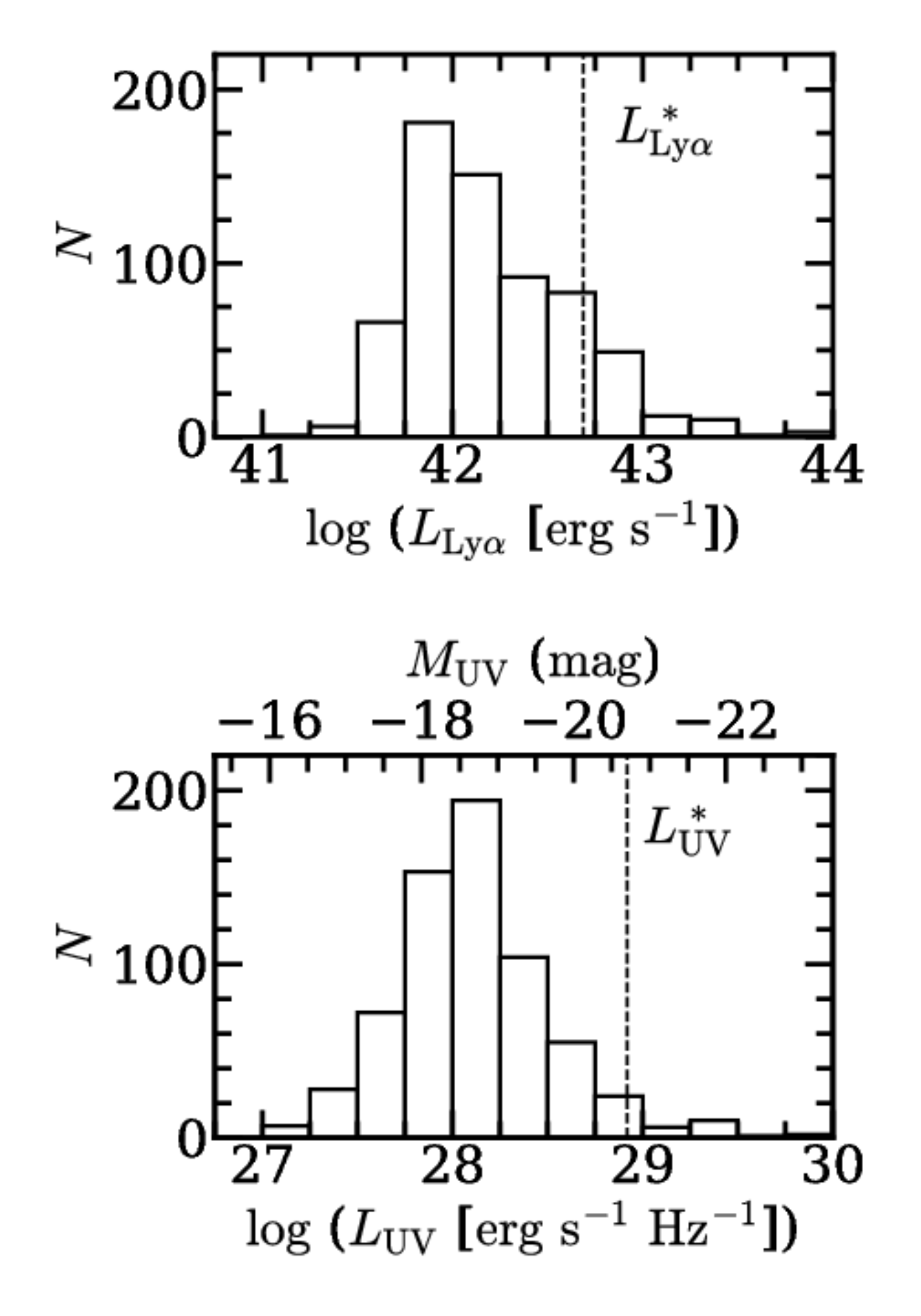}
\end{center}
\end{minipage}
\caption{Left: 
Examples of cutout images and spectra of LAEs residing at $z=5.7$. 
The size of each cutout is $6'' \times 6''$. 
In the spectrum panels, 
the vertical dashed line is the central wavelength of the detected emission line.  
Adapted from \cite{ouchi2008} with permission. 
%
Top right: 
Histogram of the Ly$\alpha$ luminosities of $z\sim2-3$ LAEs, so far observed
\citep{shibuya2019}. 
The characteristic Ly$\alpha$ luminosity $L_{{\rm Ly}\alpha}^\ast$ at $z=2.2$
is presented with a vertical dashed line. 
Bottom right: 
Same as the top right panel, but for UV luminosity. 
The vertical dashed line is the characteristic UV luminosity $L_{\rm UV}^\ast$ at $z\sim 2$.
%
}
\label{fig:Fig_z5p7_LAE}
\end{figure}

As introduced in Section \ref{sec:introduction}, 
LAEs have been identified by many observational studies. 
The successful classical technique to find LAEs is 
NB imaging. 
In this technique, LAEs are selected 
based on their NB excesses compared to broadbands 
(Figure \ref{fig:Fig_z5p7_LAE}). 
Imaging surveys with wide-field cameras such as Subaru Suprime-Cam 
have 
constructed large samples of LAEs 
and their follow-up spectroscopic campaigns have confirmed 
the validity of the NB imaging technique with low fractions of contaminants. 
The recent advent of new wide-field cameras including Hyper Suprime-Cam (HSC) 
now allows LAE surveys over cosmological volumes  ($\sim 0.5$ comoving Gpc$^2$; Ouchi et al. 2018).   

Another successful technique is blind spectroscopy. 
In particular, the integral field spectrograph of VLT/MUSE has recently yielded complementary results 
to those of the NB imaging. 
Thanks to the wider wavelength coverage and higher spectral resolution, 
MUSE is suitable to search for LAEs over a wide redshift range down to a flux limit fainter than 
the NB observations  
(\citealt{drake2017b}),  
although the field-of-view of MUSE is small. 
Another remarkable effort is made by 
a wide-field fiber spectrograph survey, 
HETDEX (\citealt{hill2008}), 
to cover a $\sim400$ deg$^2$ sky 
via a blind search for relatively bright LAEs. 

\begin{figure}[h]
\includegraphics[width=3.5in]{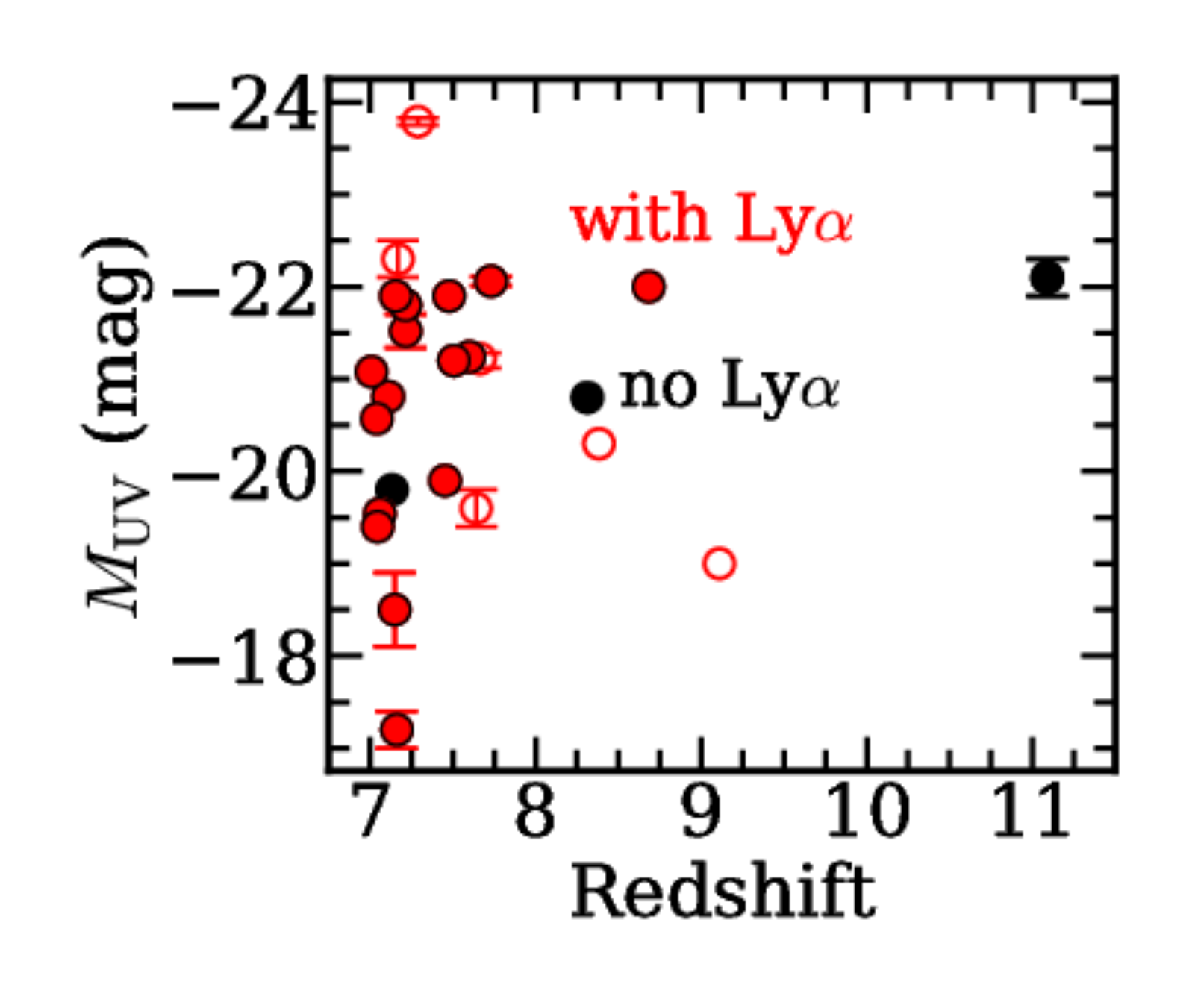}
\caption{
UV absolute magnitude, $M_{\rm UV}$, as a function of redshift 
for spectroscopically confirmed galaxies at $z_{\rm spec} > 7$ 
taken from Table \ref{tab:list_specz} and the literature. 
%
The red and black circles represent  
galaxies with and without Ly$\alpha$ detection, respectively.  
In particular, galaxies with Ly$\alpha$ 
$EW_0 > 20${\AA} 
are 
shown with the red filled circles.
%
}
\label{fig:Redshift_Histogram}
\end{figure}

Because strong Ly$\alpha$ emission from high-$z$ galaxies 
is redshifted in a wavelength range of optical and NIR 
where deep spectroscopic data can be obtained, 
Ly$\alpha$ emission has been detected in many
galaxies by spectroscopy at the high-redshift frontier.
Table \ref{tab:list_specz} presents the list of galaxies  identified, to date,
at a spectroscopic redshift $z_{\rm spec} > 7.2$. 
Some high-$z$ galaxies are found only with the Ly$\alpha$ continuum break \citep{oesch2016} 
or 
[{\sc Oiii}]$88\mu$m emission
(\citealt{tamura2019}).
Previously it was thought that 
Ly$\alpha$ photons cannot escape from galaxies especially at the early 
EoR 
due to the resonant scattering effect in the neutral IGM. 
However, as summarized in Figure \ref{fig:Redshift_Histogram}, 
a majority of galaxies, found to date, show Ly$\alpha$ in emission even at $z \sim 9$, beyond the heart of reionization,  
suggesting a patchy nature of reionization 
where relatively luminous ionizing sources such as the spectroscopically confirmed galaxies 
are located in large ionized bubbles in the IGM, which 
allow Ly$\alpha$ photons to escape from the neutral IGM. 

\begin{table}[h]
\tabcolsep7.5pt
\caption{List of spectroscopically identified galaxies at $z_{\rm spec} > 7.2$}
\label{tab:list_specz}
\begin{center}
\scalebox{0.63}{
\begin{tabular}{@{}l|c|c|c|c|c|c|c|c@{}}
\hline
ID & R.A. & Decl. & $z_{\rm spec}$ & $M_{\rm UV}$ & Ly$\alpha$ $EW_0$ & Probe & Other & Reference \\
& (J2000) & (J2000) & & (mag) & ({\AA}) & & Lines & \\
(1) & (2) & (3) & (4) & (5) & (6) & (7) & (8) & (9) \\
\hline
GN-z11          & 12:36:25.46  & $+$62:14:31.4  & 11.09  & $-22.1 \pm 0.2$          & ---                  & Lyman break          & ---                  & O16 \\
MACS1149-JD     & 11:49:33.59  & $+$22:24:45.80 & 9.1096 & $-19.0$                  & $11.4$               & [{\sc Oiii}]$88\mu$m & Ly$\alpha$           & H18, Ho18 \\
EGSY-2008532660 & 14:20:08.50  & $+$52:53:26.60 & 8.683  & $-22.0$                  & $28  $               & Ly$\alpha$           & ---                  & Z15 \\
A2744-YD4       & 00:14:24.9   & $-$30:22:56.1  & 8.382  & $-20.3$                  & $10.7 \pm 2.7$       & [{\sc Oiii}]$88\mu$m & Ly$\alpha$           & L17 \\
MACS0416-Y1     & 04:16:09.40  & $-$24:05:35.5  & 8.3118 & $-20.8$                  & ---                  & [{\sc Oiii}]$88\mu$m & ---                  & T19 \\
EGS-zs8-1       & 14:20:34.89  & $+$53:00:15.4  & 7.7302 & $-22.06 \pm 0.05$        & $21 \pm 4$           & Ly$\alpha$           & {\sc Ciii}]1908      & O15, S17 \\
z7-GSD-3811     & 03:32:32.03  & $-$27:45:37.1  & 7.6637 & $-21.22^{+0.06}_{-0.10}$ & $15.6^{+5.9}_{-3.6}$ & Ly$\alpha$           & ---                  & S16 \\
MACS1423-z7p64  & 14:23:46.18  & $+$24:04:10.76 & 7.640  & $-19.6 \pm 0.2$          & $9 \pm 2$            & Ly$\alpha$           & ---                  & H17 \\
z7-GND-16863    & 12:37:19.94  & $+$62:15:26.05 & 7.599  & $-21.24$                 & $61.28 \pm 5.85$     & Ly$\alpha$           & ---                  & J19 \\
z8-GND-5296     & 12:36:37.90  & $+$62:18:08.5  & 7.506  & $-21.2$                  & $33.19 \pm 3.20$     & Ly$\alpha$           & {\sc Ciii}]1908      & F13, J19, Hu19 \\
EGS-zs8-2       & 14:20:12.09  & $+$53:00:26.97 & 7.4770 & $-21.9$                  & $20.2$               & Ly$\alpha$           & ---                  & RB16, S17 \\
GS2-1406        & 03:33:09.14  & $-$27:51:55.47 & 7.452  & $-19.9$                  & $140.3 \pm 19.0$     & Ly$\alpha$           & ---                  & L18 \\
SDF-NB1006-2    & 13:24:35.418 & $+$27:27:27.81 & 7.288  & $-23.79 \pm 0.04$        & $1.99 \pm 0.37$      & Ly$\alpha$           & ---                  & S12 \\
SXDF-NB1006-2   & 02:18:56.523 & $-$05:19:58.79 & 7.215  & $-21.52 \pm 0.18$        & $32$                 & Ly$\alpha$           & [{\sc Oiii}]$88\mu$m & S12, I16 \\
GN-108036       & 12:36:22.68  & $+$62:08:07    & 7.213  & $-21.8$                  & $33$                 & Ly$\alpha$           & ---                  & O12 \\
\hline
\end{tabular}
}
\end{center}
\begin{tabnote}
Note: 
(1) Object ID. 
(2) Right ascension. 
(3) Declination. 
(4) Spectroscopic redshift. 
(5) Intrinsic UV absolute magnitude. 
(6) Ly$\alpha$ EW in the rest-frame. 
(7) Emission lines used for spectroscopic redshift determination. 
(8) Other detected emission lines. 
(9) Reference: 
F13 = \cite{finkelstein2013}, 
H18 = \cite{hashimoto2018a}, 
Ho17 = \cite{hoag2017}, 
Ho18 = \cite{hoag2018}, 
Hu19 = \cite{hu2019}, 
I16 = \cite{inoue2016}, 
J19 = \cite{jung2019}, 
L17 = \cite{laporte2017a}, 
L18 = \cite{larson2018}, 
O12 = \cite{ono2012}, 
O15 = \cite{oesch2015}, 
O16 = \cite{oesch2016}, 
RB16 = \cite{roberts-borsani2016}, 
S12 = \cite{shibuya2012}, 
S16 = \cite{song2016}, 
S17 = \cite{stark2017}, 
T19 = \cite{tamura2019}, 
and 
Z15 = \cite{zitrin2015}. 
\end{tabnote}
\end{table}

%
%

\begin{figure}[h]
\includegraphics[width=4.0in]{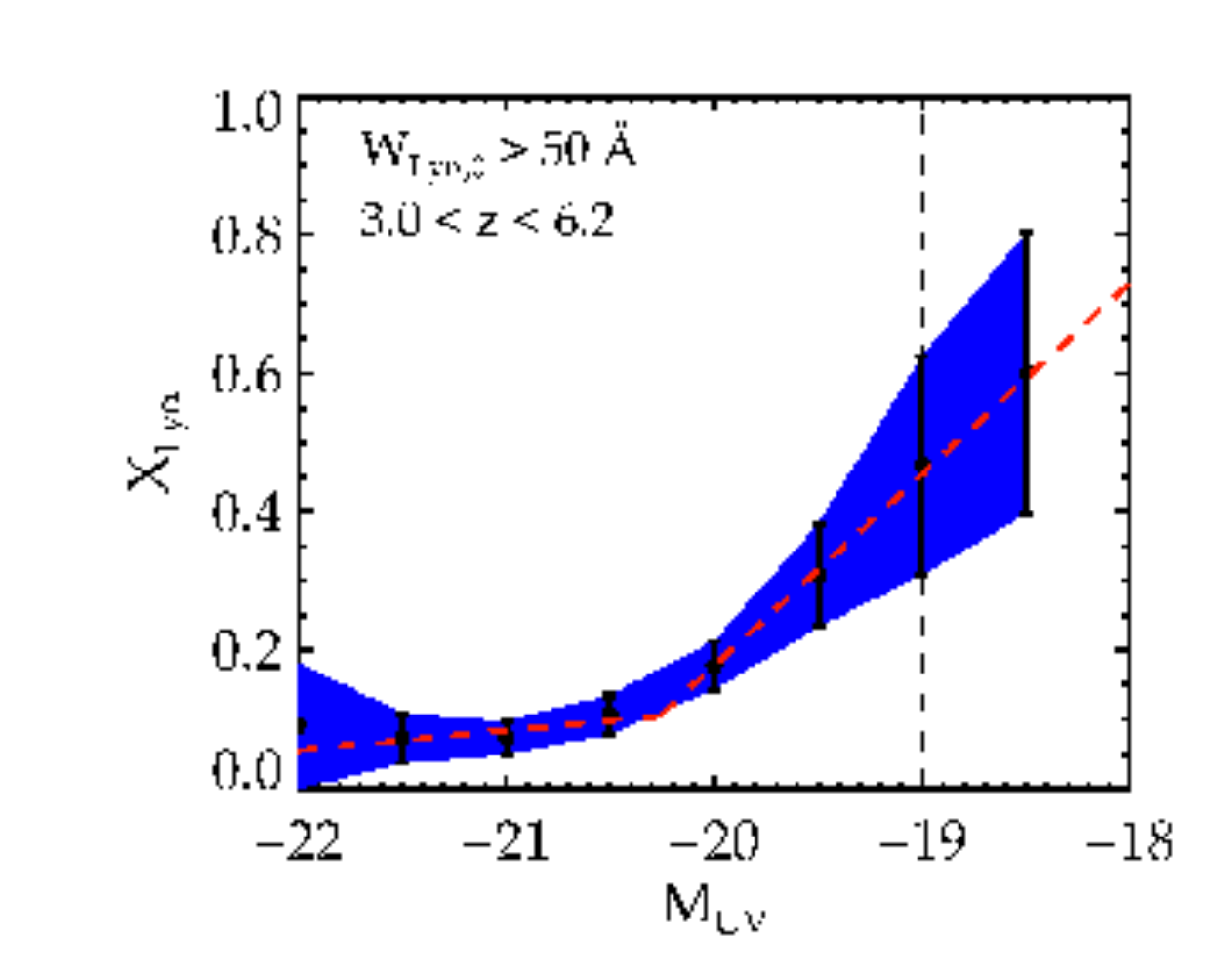}
\caption{
Fraction of spectroscopically confirmed galaxies at $z=3.0$--$6.2$ 
emitting strong Ly$\alpha$ lines 
(Ly$\alpha$ $EW_0 > 50${\AA}) as a function of $M_{\rm UV}$. 
The black vertical dashed line denotes the
$\simeq 90${\%} detection completeness limit. 
The red dashed lines are the best-fit lines of the first-order polynomials in the ranges of $-22.0 < M_{\rm UV} < -20.5$ and $-20.0 < M_{\rm UV} < -18.5$. 
Adapted from \cite{stark2010} with permission. 
}
\label{fig:Stark2010_fig13}
\end{figure}

In total, thanks to the successful selection techniques and intensive spectroscopic follow-up campaigns, 
until now 
$>$1,000 ($>$20,000) LAEs 
have been spectroscopically identified (photometrically selected) 
in the literature 
(e.g., \citealt{drake2017b}; \citealt{sobral2018}; \citealt{shibuya2019}). 
As shown in the bottom right panel of Figure \ref{fig:Fig_z5p7_LAE}, 
LAEs tend to have 
faint sub $L_{\rm UV}^*$ ($\sim 0.1 L_{\rm UV}^*$) luminosities,
%
UV-continuum
\footnote{
Throughout this review, the UV continuum indicates
the continuum emission at the rest-frame $\simeq 1500$\AA. 
No extinction corrections are applied unless otherwise specified.
}
luminosities fainter
than the characteristic luminosity 
$L_{\rm UV}^*$
%
of SFGs \citep{reddy2009} at similar redshifts 
by about an order of magnitude.
The numerous LAEs with faint UV continua can be understood by the observational fact that a UV-continuum faint SFG has a Ly$\alpha$ emitting galaxy fraction higher than the one of a UV-continuum bright SFG (Figure \ref{fig:Stark2010_fig13}), which is known as the Ando effect \citep{ando2006}.

\begin{figure}[h]
\includegraphics[width=6.2in]{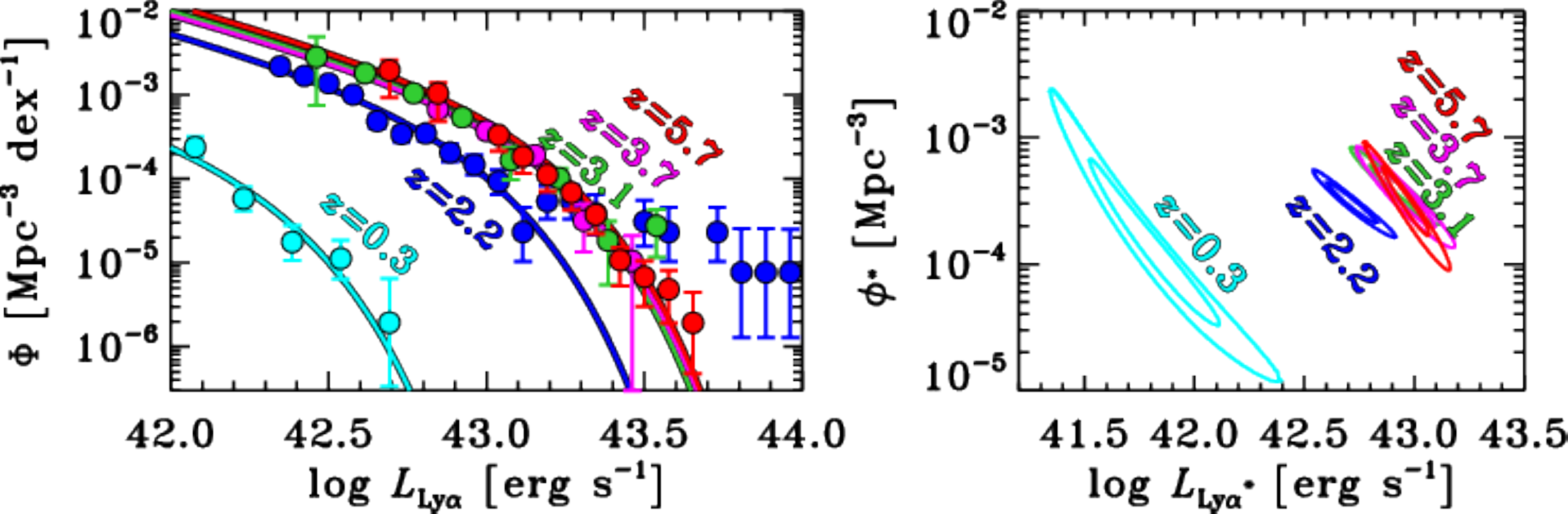}
\includegraphics[width=6.2in]{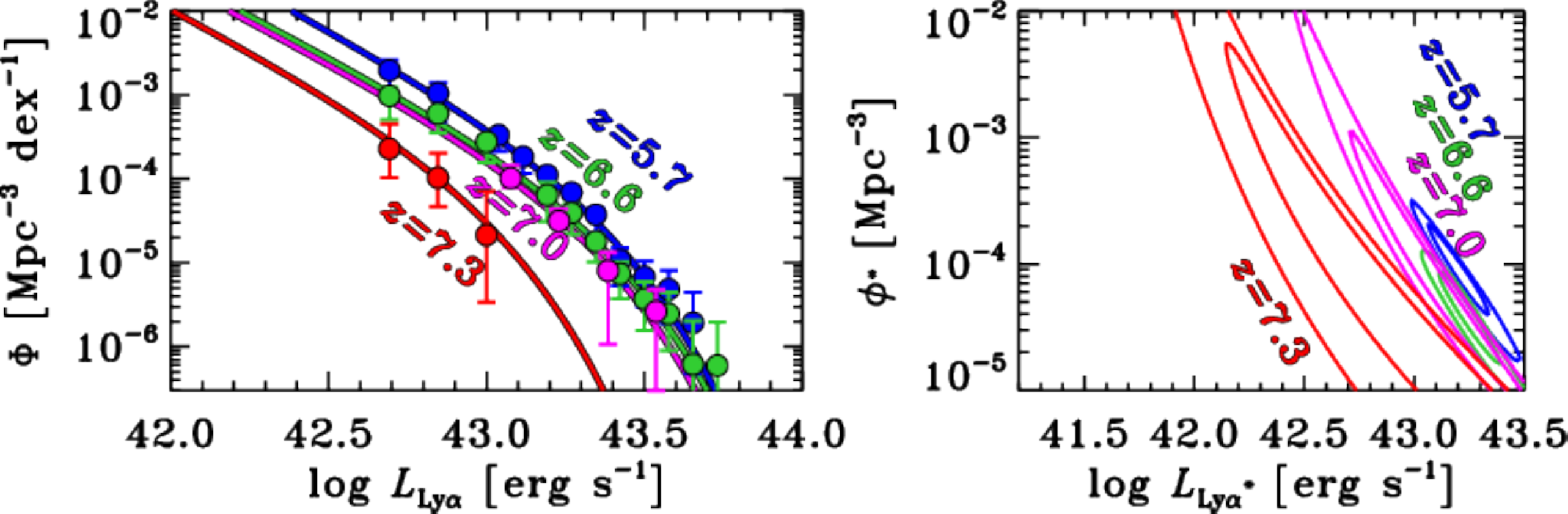}
\caption{
Left: Ly$\alpha$ LFs from $z=0.3$ to $z=5.7$ (top) and from $z=5.7$ to $z=7.3$ (bottom). 
The colored curves are their best-fit Schechter functions. 
Right: 
$1\sigma$- and $2\sigma$-level error contours of their Schechter parameters, $L_{{\rm Ly}\alpha}^\ast$ and $\phi_{{\rm Ly}\alpha}^\ast$. 
The references of the data are summarized in Table \ref{tab:Schechter_parameters_Lya}. 
}
\label{fig:LyaLF}
\end{figure}

One of the most fundamental observational quantities to characterize galaxy properties is 
the luminosity function (LF). 
The left panels of Figure \ref{fig:LyaLF} compile Ly$\alpha$ LF measurements 
for LAEs 
over a wide redshift range of $z \sim 0.3 - 7.3$.
The Ly$\alpha$ LF is often parameterized with a Schechter function \citep{schechter1976}, 
\begin{equation}
\phi(L_{{\rm Ly}\alpha}) dL_{{\rm Ly}\alpha}
	= \phi^\ast \left( \frac{L_{{\rm Ly}\alpha}}{L_{{\rm Ly}\alpha}^\ast} \right)^\alpha 
		\exp \left( -\frac{L_{{\rm Ly}\alpha}}{L_{{\rm Ly}\alpha}^\ast} \right) 
		d \left( \frac{L_{{\rm Ly}\alpha}}{L_{{\rm Ly}\alpha}^\ast} \right), 
\label{eq:Lya_LF}
\end{equation} 
where 
$L_{{\rm Ly}\alpha}$ is the observed Ly$\alpha$ luminosity, 
$L_{{\rm Ly}\alpha}^\ast$ is the characteristic Ly$\alpha$ luminosity, 
$\phi^\ast$ is the normalization, 
and $\alpha$ is the faint-end slope. 
A Schechter function is also expressed with a Ly$\alpha$ luminosity in the logarithmic form,
\begin{equation}
\Phi(\log L_{{\rm Ly}\alpha}) 
	= (\ln 10) \, \phi^\ast 
		10^{(\alpha+1) (\log L_{{\rm Ly}\alpha} - \log L_{{\rm Ly}\alpha}^\ast)}
		\exp \left( - 10^{(\log L_{{\rm Ly}\alpha} - \log L_{{\rm Ly}\alpha}^\ast)} \right). 
\label{eq:Lya_LF_log}
\end{equation} 
The best-fit Schechter functions derived in the literature are also plotted in the left panels of Figure \ref{fig:LyaLF} 
and their best-fit Schechter parameters are summarized in Table \ref{tab:Schechter_parameters_Lya}. 
The right panels of Figure \ref{fig:LyaLF} show 
the $1\sigma$ and $2\sigma$ confidence intervals 
for the combinations of the Schechter parameters of $L_{{\rm Ly}\alpha}^\ast$ and $\phi^\ast$, 
where the $\alpha$ values are fixed at fiducial values of 
$-1.8$ for low redshifts (top) and $-2.5$ for high redshifts (bottom).

\begin{table}[h]
\tabcolsep7.5pt
\caption{Schechter parameters for Ly$\alpha$ 
LFs of LAEs}
\label{tab:Schechter_parameters_Lya}
\begin{center}
\begin{tabular}{@{}l|c|c|c|c@{}}
\hline
redshift & $L^\ast_{{\rm Ly}\alpha}$ & $\phi^\ast$ & $\alpha$ &Reference \\
& ($10^{42}$ erg s$^{-1}$) & ($10^{-4}$ Mpc$^{-3}$) & & \\
\hline
0.3 & $0.71^{+0.32}_{-0.29}$ & $1.12^{+2.45}_{-0.61}$ & $-1.8$ (fixed) & \cite{cowie2010}, \cite{konno2016} \\
2.2 & $4.87^{+0.83}_{-0.68}$ & $3.37^{+0.80}_{-0.66}$ & $-1.8$ (fixed) & \cite{konno2016} \\
3.1 & $8.49^{+1.65}_{-1.46}$ & $3.90^{+1.27}_{-0.90}$ & $-1.8$ (fixed) & \cite{ouchi2008}, \cite{konno2016} \\
3.7 & $9.16^{+2.03}_{-1.67}$ & $3.31^{+1.42}_{-0.98}$ & $-1.8$ (fixed) & \cite{ouchi2008}, \cite{konno2016} \\
5.7 & $9.09^{+3.67}_{-2.70}$ & $4.44^{+4.04}_{-2.05}$ & $-1.8$ (fixed) & \cite{ouchi2008}, \cite{konno2016} \\
5.7 & $16.4^{+21.6}_{-6.2}$ & $0.849^{+1.87}_{-0.771}$ & $-2.56^{+0.53}_{-0.43}$ & \cite{konno2018} \\
6.6 & $16.6^{+3.0}_{-6.9}$ & $0.467^{+1.44}_{-0.442}$ & $-2.49^{+0.50}_{-0.50}$ & \cite{konno2018} \\
7.0 & $15.0^{+4.2}_{-3.1}$ & $0.45^{+0.26}_{-0.18}$ & $-2.5$ (fixed) & \cite{itoh2018} \\
7.3 & $5.5^{+94.5}_{-3.3}$ & $0.94^{+12.03}_{-0.93}$ & $-2.5$ (fixed) & \cite{konno2014}, \cite{itoh2018} \\
\hline
\end{tabular}
\end{center}
\end{table}

\begin{figure}[h]
\includegraphics[width=5.0in]{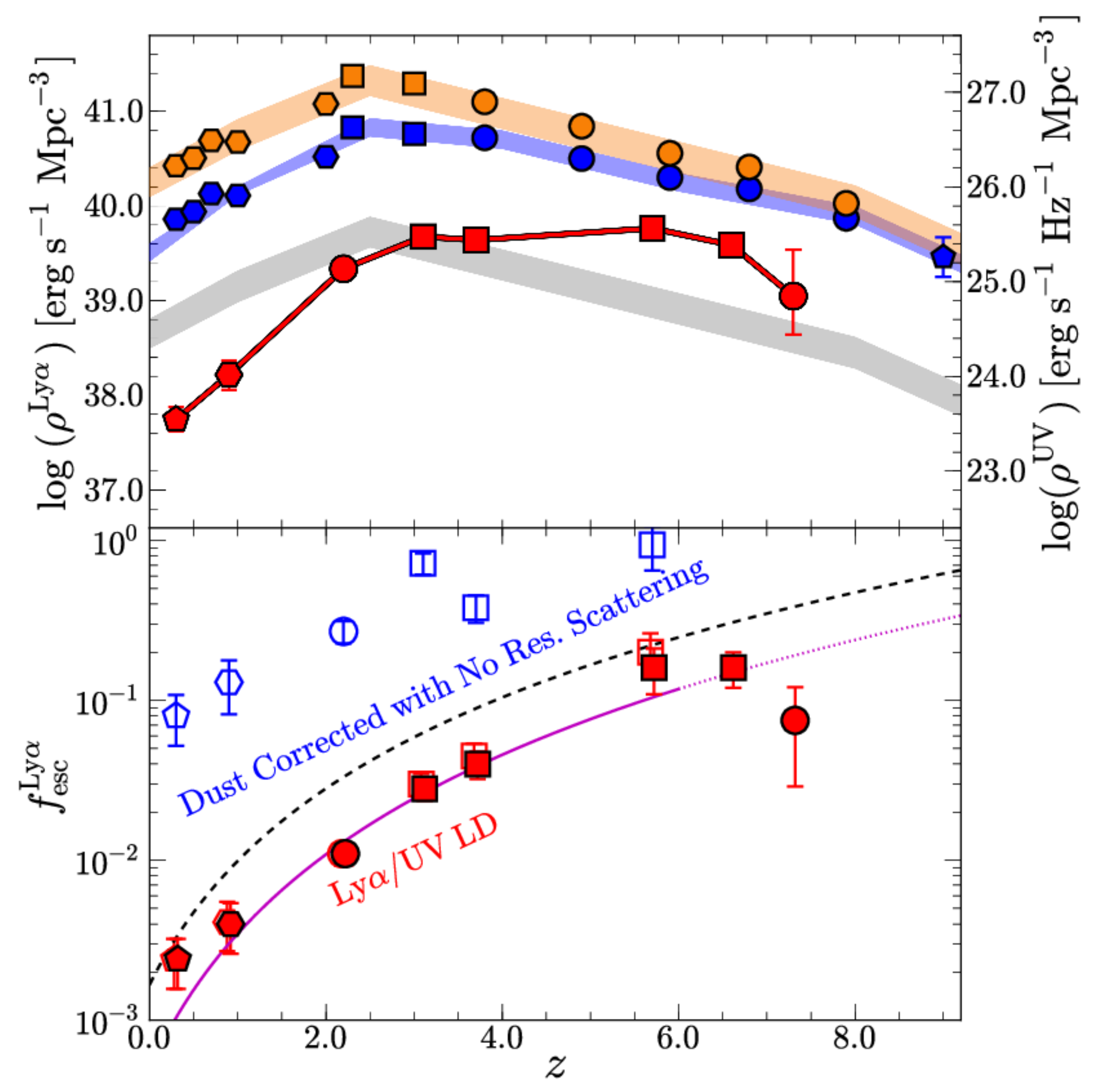}
\caption{
Top: 
Ly$\alpha$ LDs and UV LDs as a function of redshift. 
The red symbols indicate Ly$\alpha$ LDs. 
The orange (blue) symbols and shaded area denote the UV LDs and the errors, respectively, corrected for (no) dust extinction.  
The gray area is the evolutionary trend of the dust-corrected UV LDs scaled to the Ly$\alpha$ LD at $z \sim 3$. 
Bottom: 
Cosmic average
%
Ly$\alpha$ escape fraction, $f^{\mathrm{Ly}\alpha}_{\mathrm{esc}}$, as a function of redshift.
The red filled (open) symbols are 
$f^{\mathrm{Ly}\alpha}_{\mathrm{esc}}$ 
estimated from the observed Ly$\alpha$ LDs and the dust-corrected UV LDs 
without (with) considering the effect of IGM absorption. 
The blue symbols are 
those 
corrected for dust extinction in the case of no Ly$\alpha$ resonant scattering,
which are discussed in the text of Section \ref{sec:ISM} regarding Ly$\alpha$ escape fraction.
%
The magenta solid line represents the best-fit function for the Ly$\alpha$ escape fraction evolution from $z = 0$ to $6$. 
Adapted from \cite{konno2016} with permission. 
}
\label{fig:LyaLD}
\end{figure}

Ly$\alpha$ LFs show three evolutionary trends: 
a monotonic increase from $z \sim 0$ to $z \sim 3$, 
no evolution from $z \sim 3$ to $z \sim 6$, 
and a rapid drop beyond $z \sim 6$. 
First, 
at $z \sim 0-3$, 
the Ly$\alpha$ LFs show a strong increase with increasing redshift 
(\citealt{deharveng2008}; \citealt{cowie2010}). 
The rise of Ly$\alpha$ LFs is much larger than that of UV-continuum LFs based on the comparisons of Ly$\alpha$ luminosity densities (LDs), i.e.,  
\begin{equation} 
\rho_{{\rm Ly}\alpha} 
	= \int^\infty_{L_{{\rm Ly}\alpha}^{\rm lim}} 
		L_{{\rm Ly}\alpha} \phi_{{\rm Ly}\alpha} (L_{{\rm Ly}\alpha}) dL_{{\rm Ly}\alpha}, 
\label{eq:rho_Lya}
\end{equation}
where $L_{{\rm Ly}\alpha}^{\rm lim}$ is the limiting Ly$\alpha$ luminosity, 
and the UV-continuum LDs, 
\begin{equation}
\rho_{\rm UV}
	= \int^\infty_{L_{\rm UV}^{\rm lim}} 
		L_{\rm UV} \phi_{\rm UV} (L_{\rm UV}) dL_{\rm UV}, 
\label{eq:rho_UV}
\end{equation} 
where $\phi_{\rm UV}$, $L_{\rm UV}$, and $L_{\rm UV}^{\rm lim}$ are 
the UV-continuum LF, 
the UV-continuum luminosity, 
and the limiting UV-continuum luminosity, respectively. 
The top panel of Figure \ref{fig:LyaLD} (\citealt{konno2016}) presents the comparison of the Ly$\alpha$ and UV-continuum LD evolution, and indicates that the rise of the Ly$\alpha$ LD is larger than the one of the UV LD from $z\sim 0$ to $3$.
%
%
Second, from $z \sim 3$ to $z \sim 6$, the Ly$\alpha$ LFs show no significant evolution (\citealt{ouchi2008}), while the UV LFs decrease strongly in the volume number density at $M_{\rm UV} = -21$ mag by a factor of about $5$ \citep{bouwens2015}. %
These differences of the evolutionary trends can be explained by the fact that 
%
the escape fraction of Ly$\alpha$ photons and/or the ionizing photon production rate from galaxies 
increase with increasing redshift from $z\sim 0$ to $z \sim 6$
(\citealt{hayes2011}; see Section \ref{sec:ISM} for more details). 
%
The Ly$\alpha$ escape fraction is defined as  
\begin{equation}
f_{\rm esc}^{{\rm Ly}\alpha}
	= \frac{L_{{\rm Ly}\alpha}}{L_{{\rm Ly}\alpha}^{\rm int}},
\label{eq:f_esc}
\end{equation}
where $L_{{\rm Ly}\alpha}^{\rm int}$ is the intrinsic Ly$\alpha$ luminosity 
that can be estimated from a star-formation rate (SFR), 
$L_{{\rm Ly}\alpha}^{\rm int}$ [erg s$^{-1}$] 
$= 1.1 \times 10^{42}\ {\rm SFR}$ [$M_\odot$ yr$^{-1}$], 
under the assumption of the case B recombination (Footnote 11 of \citealt{henry2015})
and the relation between the H$\alpha$ luminosity and SFR 
(i.e., a constant ionizing photon production rate; \citealt{kennicutt1998}). 
The bottom panel of Figure \ref{fig:LyaLD} presents evolution of the cosmic average $f^{\mathrm{Ly}\alpha}_{\mathrm{esc}}$ values that are estimated with the Ly$\alpha$ and UV LDs via eq. (\ref{eq:f_esc}). The cosmic average $f^{\mathrm{Ly}\alpha}_{\mathrm{esc}}$ monotonically increases by two orders of magnitude 
with increasing redshift from $z \sim 0$ to $z \sim 6$. 
%
Lastly, 
the Ly$\alpha$ LD 
%
%
appears to drop 
faster than the UV LD from $z\sim 6$ to a higher redshift
(Figure \ref{fig:LyaLD}). 
Because this evolutional trend at $z\gtrsim 6$ is closely related to cosmic reionoization, 
we discuss this evolutional trend thoroughly in Section \ref{sec:reionization}.

Another notable feature in Ly$\alpha$ LFs is the shape of the bright end.  
At $z\sim2-3$, the significant bright-end LF excesses beyond the Schechter functions are found and explained by AGNs by the multiwavelength analysis
(\citealt{konno2016}) 
and spectroscopy (\citealt{sobral2018}). 
At redshifts higher than $z\sim 2-3$, such bright-end LF excess features are not clearly found, probably because the number densities of AGNs decrease with increasing redshift. 
Interestingly, at the 
EoR $z\sim 7$, 
some studies have reported possible bright-end LF excess detections, 
arguing that 
bright LAEs are presumably surrounded by large ionized bubbles  
(\citealt{matthee2015}; \citealt{zheng2017}). 
However, with the great statistical accuracy, 
the Subaru HSC survey has recently claimed that 
the Ly$\alpha$ LF at $z\sim 7$ can be explained by the Schechter functional form with no significant bright-end LF excess
but with a steep slope of $\alpha \sim -2.5$ (\citealt{konno2018,itoh2018}). 
The shape of the bright-end LF is under debate (\citealt{hu2019}).


%
%
Although the previous studies have 
successfully characterized the overall evolution of Ly$\alpha$ LFs, 
there is still a fundamental but important open question: 
What is the lowest mass of dark matter halos of SFGs that can emit Ly$\alpha$?   
The previous observations are not deep enough to detect very faint LAEs. 
In fact,  as can be seen in Table \ref{tab:Schechter_parameters_Lya}, 
most of the previous studies do not obtain good constraints on the faint end slope $\alpha$ 
and thus have fixed $\alpha$ at a fiducial value in the Schechter function fitting. 
From the theoretical point of view, 
the LF is expected to have a turnover at a faint luminosity, 
because it is difficult for very low-mass halos to host SFGs 
due to inefficiency of gas cooling 
(\citealt{liu2016}).

\section{Morphological Properties}
\label{sec:morphology}



\begin{figure}[h]
\begin{minipage}{2.4in}
\begin{center}
\includegraphics[width=2.4in]{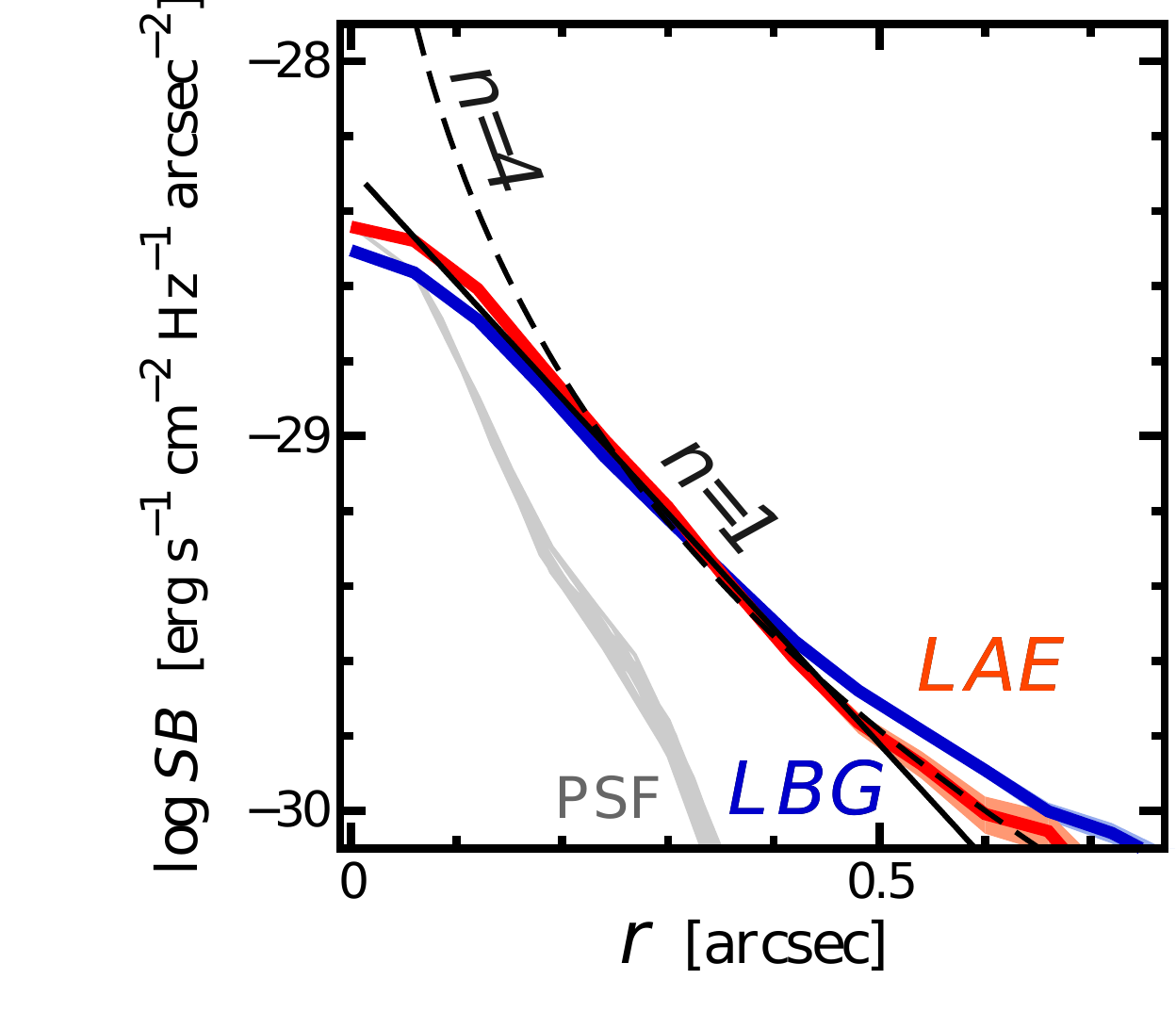} 
\end{center}
\end{minipage}
\begin{minipage}{2.7in}
\begin{center}
\includegraphics[width=2.7in]{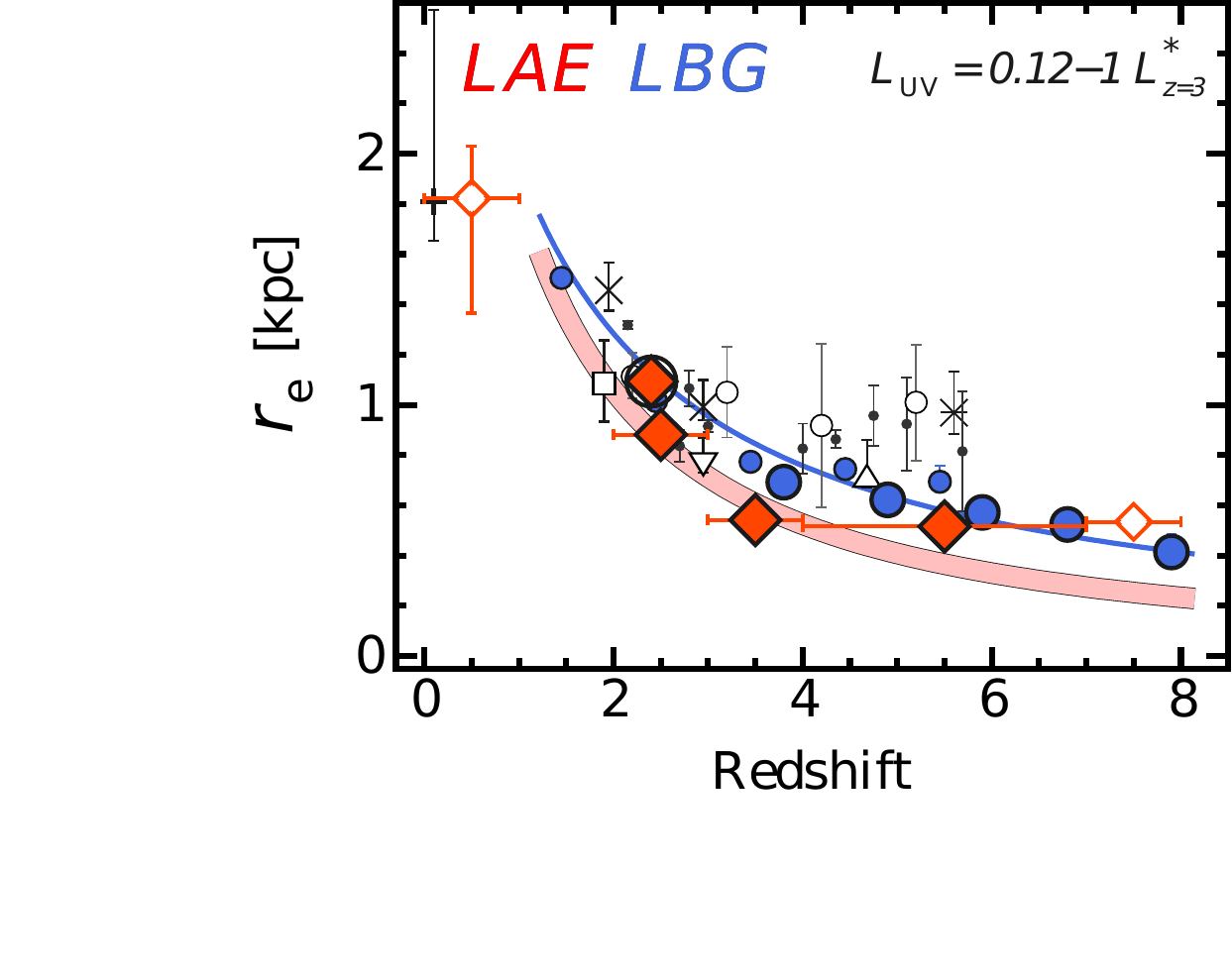}
\end{center}
\end{minipage}
\caption{Left:  Radial SB profiles of continuum in the rest-frame UV wavelength for LAEs (red curve) and LBGs (blue curve) at $z\sim3-4$ \citep{shibuya2019}. The shaded regions associated with the red and blue curves show the 1$\sigma$ uncertainties of the radial SB profiles. The solid and dashed black 
lines
%
depict the best-fit S\'ersic profiles with $n_{\rm s}=1$ and $n_{\rm s}=4$, respectively. The gray line denotes a point spread function (PSF) of the imaging data.
Right:  Redshift evolution of $r_{\rm e}$ for LAEs \citep[red diamonds; ][]{shibuya2019} and SFGs/LBGs (small and large blue filled circles; \citealt{shibuya2015}) in the $L_{\rm UV}$ range of $0.12-1$ $L_{z=3}^*$, where $L_{z=3}^*$ is the characteristic UV luminosity at $z\sim3$ \citep{steidel1999}. The red filled diamonds with and without an open circle represent $r_{\rm e}$ measured 
in the rest-frame optical and UV wavelengths,
%
respectively. 
The red open diamonds are observational estimates of $r_{\rm e}$ for the luminosity range of $L_{\rm UV}=0.12-1 L^*_{\rm z=3}$ (see \citealt{shibuya2019}).
%
The magenta region and blue line present the best-fit $(1+z)^\alpha$ functions for the LAEs and SFGs/LBGs, respectively, where $\alpha$ is the power-law index. The 
black
symbols denote measurements of LAEs summarized in \citet{shibuya2019}. }
\label{fig_rad_prof_z_re}
\end{figure}

A compact nature of LAEs has been known by HST observations since the early discoveries of LAEs \citep{pascarelle1996}. Deep HST extra-galactic legacy data have indicated that LAEs have half light radii of $r_{\rm e}\sim1$ pkpc on average in the rest-frame UV and optical stellar continua \citep{malhotra2012, paulino-afonso2018, shibuya2019}. Although sub-structures are found in some cases, the main stellar components of LAEs basically have a disk-like radial SB profile with a S\'ersic index of $n_{\rm s}\sim1$ \citep{gronwall2011, taniguchi2009} similar to LBGs (left panel of Figure \ref{fig_rad_prof_z_re}). This typical radial SB profile of LAEs is not largely changed over cosmic time similar to other SFG populations \citep{paulino-afonso2018, shibuya2019}. 

Early HST studies have reported that the UV morphology of LAEs shows the nearly 
no
%
$r_{\rm e}$ evolution over cosmic time \citep{malhotra2012, paulino-afonso2018}.
However, it is claimed that the 
no
%
$r_{\rm e}$ evolution results may be obtained by the bias raised by heterogeneous LAE samples in luminosity over the redshift range \citep{shibuya2019}. Note that the UV continuum morphology of LAEs follows the $r_{\rm e}$-luminosity relation similar to the one of LBGs. The $r_{\rm e}$-luminosity relation indicates that faint continuum LAEs have a small $r_{\rm e}$ \citep{leclercq2017, shibuya2019}. If sample selections at different redshifts are not well controlled, observers identify many faint LAEs at low $z$ that has a small $r_{\rm e}$, which diminishes a trend of $r_{\rm e}$ evolution. Comparing bias-controlled and uncontrolled samples, \citet{shibuya2019} find that the 
no
%
$r_{\rm e}$ evolution of LAEs is mistakenly concluded, due to the existence of the $r_{\rm e}$-luminosity relation and the sample bias. With the bias-controlled samples, the median $r_{\rm e}$ values of LAEs monotonically evolve as $r_{\rm e}\sim(1+z)^{-1.37}$ similar to those of SFGs and LBGs at a given UV-continuum luminosity (\citealt{shibuya2019}; right panel of Figure \ref{fig_rad_prof_z_re}).

The spatial offset between Ly$\alpha$ and stellar continuum emission peaks, referred to as the Ly$\alpha$ spatial offset or $\delta_{\rm Ly\alpha}$, is an important clue to physical properties of LAEs \citep{ouchi2013, jiang2013b}. 
Statistical studies suggest that LAEs with a low Ly$\alpha$ $EW_0$ typically have a large $\delta_{\rm Ly\alpha}$ value \citep{shibuya2014a,hoag2019}. 
This Ly$\alpha$ $EW_0 - \delta_{\rm Ly\alpha}$ anti-correlation implies that Ly$\alpha$ photons are 
selectively
%
attenuated by dust, due to the long mean-free path of Ly$\alpha$ photons by resonant scattering in 
abundant H{\sc i} gas of low-Ly$\alpha$ $EW_0$ galaxies
that makes large $\delta_{\rm Ly\alpha}$ values.
%

In summary, the disk-like radial SB profile and the $r_{\rm e}$ values indicate that LAEs have stellar components similar to that of LBGs at a given UV continuum luminosity. This morphological similarity suggests that the Ly$\alpha$ escape is not strongly related to the morphology of stellar components, but instead, is governed by the column density, geometry, kinematics, and/or ionization states of the ISM and CGM. This interpretation is also supported by the $\delta_{\rm Ly\alpha}$ measurements. This physical picture is explained by the theoretical models of the viewing angle effect where Ly$\alpha$ photons easily escape in the direction of disk face-on \citep{zheng2014, verhamme2012}. 
%
Of course, the morphology of LAEs may not be a simple disk, but a disk-like shape with sub-structures on $<1$ pkpc scales that are poorly understood. Clumpy structures are identified in various high-$z$ galaxies including some LAEs \citep{shibuya2016, cornachione2018, ritondale2019}. Further studies are needed to test the viewing angle effect of the Ly$\alpha$ escape.


%
%


%
%

\section{Stellar Population}
\label{sec:stellar_population}


Studying the stellar population of LAEs is important to 
understand their physical nature and to reveal the relationship 
between LAEs and other high-$z$ galaxies at similar redshifts. 
Stellar populations of galaxies 
are characterized with physical properties such as stellar mass and stellar age,    
and can be investigated from comparisons 
between their observed 
spectral energy distributions (SEDs) 
with those of stellar population synthesis models. 
Since LAEs are typically faint in the continuum, 
it is not easy to derive their SEDs on an individual basis, 
and thus many studies have performed stacking analyses
%
of low spectral-resolution broadband photometry
%
to obtain their typical SEDs with a high signal-to-noise ratio.\footnote{There are 
two major stacking methods, i.e., average and median, and each method has its pros and cons. 
In general, average stacking can consider all objects in a sample 
yielding a good representative value, 
unless the sample includes outliers 
such as very bright AGN and/or low-$z$ emission line galaxy contaminants. 
Median stacking is less likely to be affected by such contaminants, 
although it does not take into account the fluxes of all objects in a sample. 
Thus, one should check the consistency of the results with these two methods. 
Note that,
although these two stacking methods mostly give reasonable results for typical values, 
no stacking method can identify the large dispersion of properties in a sample if any \citep{vargas2014}.} 
Early studies have demonstrated that 
the rest-frame UV to optical SEDs of high-$z$ LAEs can be obtained 
from the combination of deep optical data with 
NIR images such taken with the \textit{Spitzer Space Telescope} 
(\citealt{gawiser2007}; \citealt{lai2008}) .

Since 
typical LAEs host star formation, 
their SEDs
of broadband photometry
%
are expected to be characterized with 
not only stellar 
emission 
but also nebular emission from their star-forming regions. 
In fact, deep 
NIR spectroscopy results targeting $z\sim 2-3$ LAEs 
have revealed the presence of strong nebular emission lines 
in the rest-frame optical wavelength
such as H$\alpha$ and [{\sc Oiii}]5007 
(\citealt{guaita2013}; \citealt{nakajima2013}).
A study of local LAEs indicates that a majority of LAEs have H$\alpha$ $EW_0$ of $\sim 50-1000$\AA\ \citep{cowie2010}.
%
To include the influence of nebular emission 
on the determination of stellar population parameters, 
\cite{ono2010b} have fitted stellar population synthesis models 
with and without nebular emission to averaged SEDs of $z \sim 6-7$ LAEs
by adopting the prescription for nebular emission suggested by \citet{schaerer2009}.  
\cite{ono2010b} have shown that 
the best-fit parameters in these two cases are extremely different
for the LAEs,
%
mainly because strong nebular lines mimic a substantial Balmer break feature. 
\footnote{
It is confirmed that flux ratios of {\sc[Oiii]}/H$\alpha$ in best-fit SED+nebular emission models are consistent with those directly obtained by spectroscopy (Figure 9 of \citealt{harikane2018}).
}
For example, 
the best-fit stellar mass (age) in the case of considering nebular emission 
is about an order (two orders) of magnitude smaller than that without nebular emission, which is more serious than the case of the continuum-bright LBGs \citep{schaerer2009}. 
It is critical for including the effect of nebular emission in stellar population analyses of LAEs.

\begin{figure}[h]
\begin{minipage}{2.5in}
\begin{center}
\includegraphics[width=2.5in]{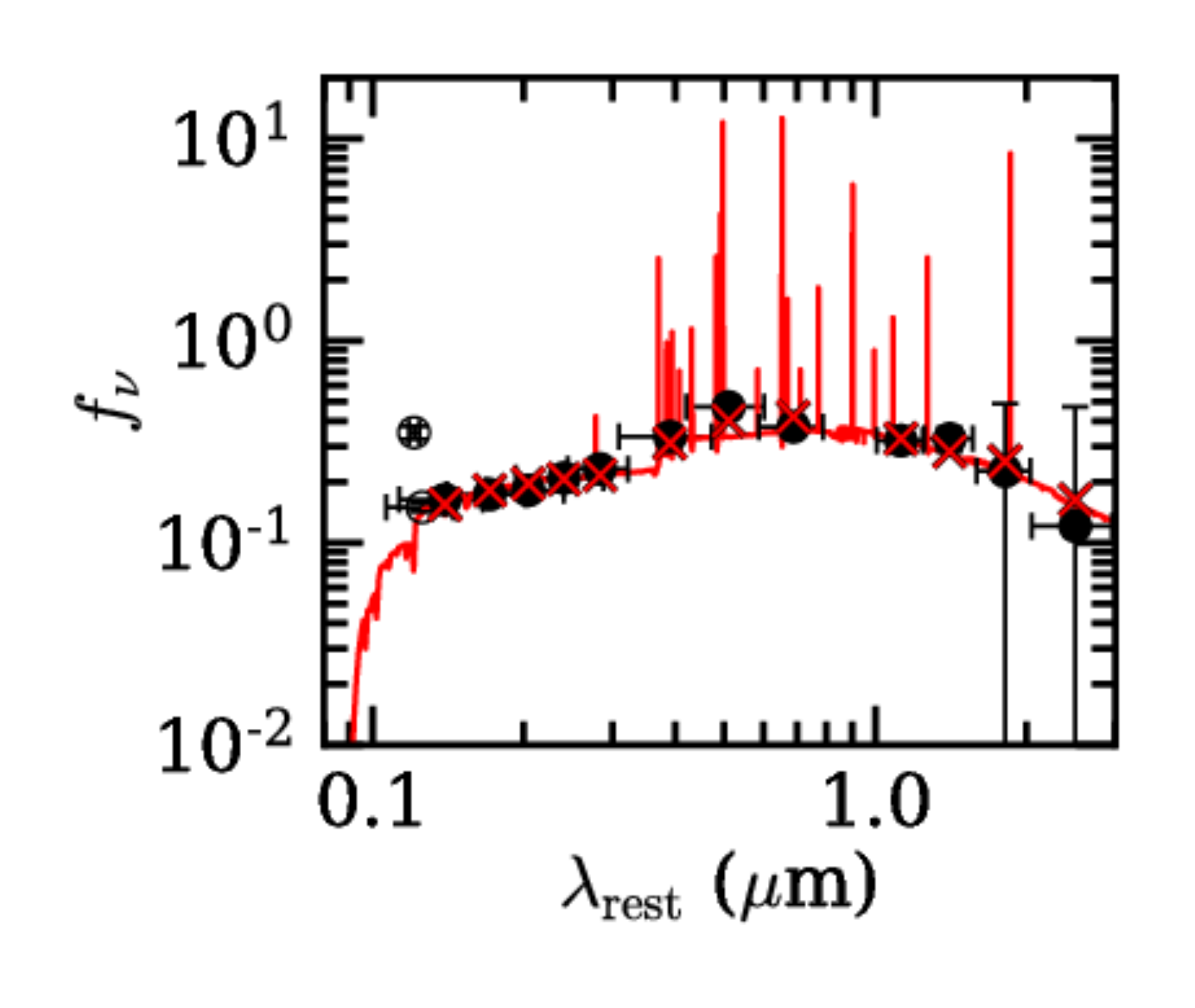}
\end{center}
\end{minipage}
\begin{minipage}{3.5in}
\begin{center}
\includegraphics[width=3.5in]{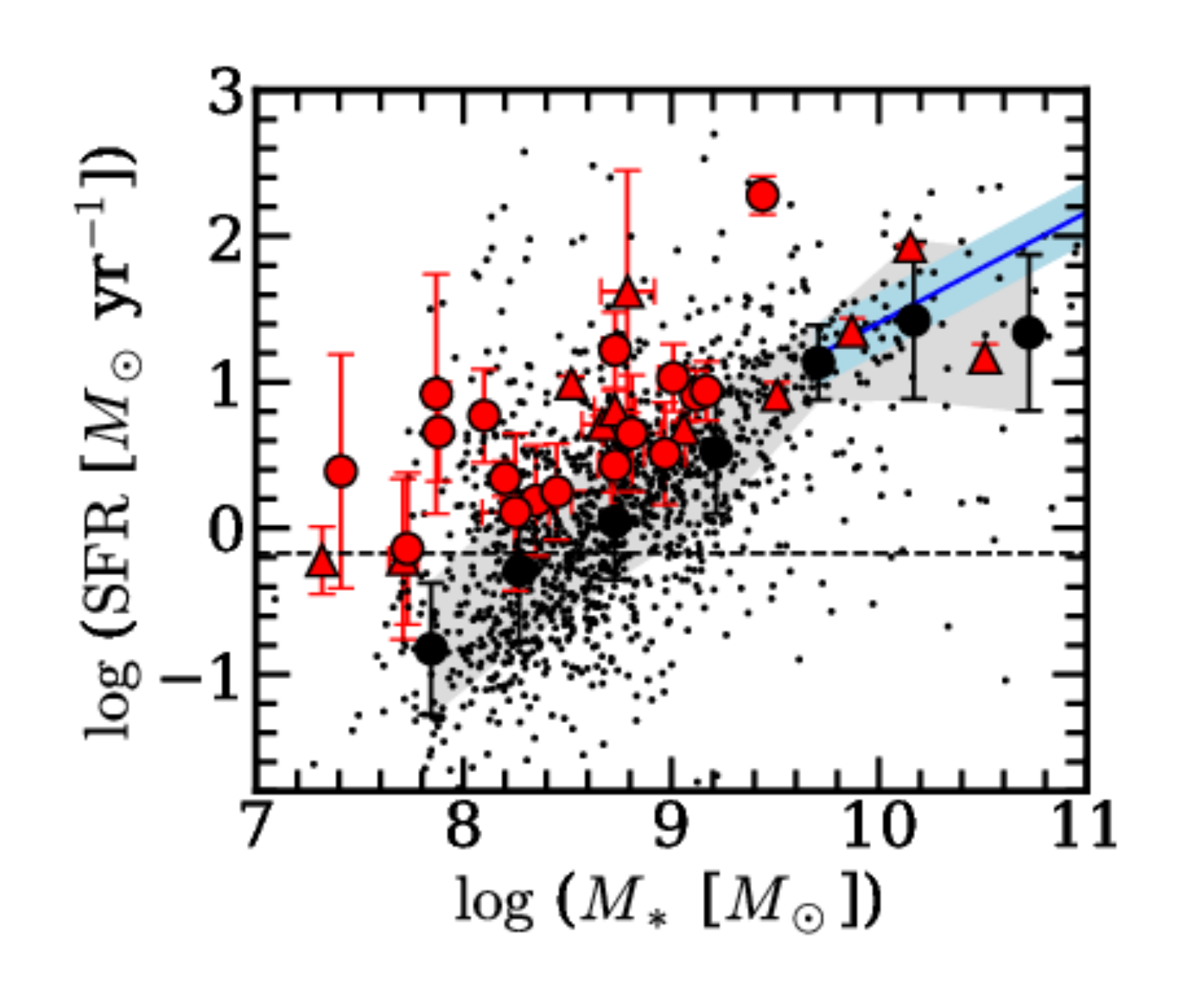}
\end{center}
\end{minipage}
\caption{
Left: 
Example of SED fitting results of LAEs \citep{nakajima2012}. 
The flux density $f_\nu$ in arbitrary units is plotted against wavelength in the rest frame. 
The black filled (open) circles represent the observed flux densities that are (not) used for SED fitting. 
The red curve indicates the best-fit SED 
and the red crosses correspond to synthesized 
broadband flux densities that include nebular-emission contributions.
%
Right: 
SFR as a function of stellar mass. 
The red circles and triangles are LAEs at $z \sim 2$ 
that are identified in the NB survey 
and 
the HETDEX Pilot Survey, 
respectively \citep{hagen2016},
being obtained with model SEDs including nebular emission.
%
The black dashed horizontal line corresponds to the $5\sigma$ detection limit. 
The black dots are photo-$z$ selected SFGs at similar redshifts, 
while the large black circles 
correspond to their average values \citep{santini2017}. 
The blue solid line and shaded area 
show the average values and the errors, respectively, for relatively massive SFGs that are regularly studied by deep observations at similar redshifts
\citep{speagle2014}. 
}
\label{fig:SEDfit}
\end{figure}

Subsequent results including the effect of nebular emission 
have led to a general consensus that 
typical LAEs are low-mass 
(stellar mass $\sim$ $10^{8-9} M_\odot$), 
young 
stellar age
%
($\sim 10$ Myr)
\footnote{
This young stellar age can be found, probably due to intermittent star formation.
}
SFGs 
with a SFR of $\sim 1-10 \, M_\odot$ yr$^{-1}$
(Figure \ref{fig:SEDfit}; \citealt{nakajima2012}; see also \citealt{hagen2014}), 
suggesting that typical LAEs are 
high-$z$ counterparts of local dwarf galaxies, i.e.
high-$z$ analogs of local dwarf galaxies. 
%
%
%
The right panel of 
Figure \ref{fig:SEDfit} shows 
compilations of stellar mass and SFR estimates for LAEs at $z \sim 2$ (\citealt{hagen2016}) 
as well as photo-$z$ selected SFGs at similar redshifts (\citealt{santini2017}). 
%
At fixed stellar masses of 
$M_\ast \sim 10^9 M_\odot$, 
where the completeness for the NB-selected LAEs is high 
enough for fair comparisons of LAEs and SFGs,
LAEs have higher SFRs than the average values of the SFGs, 
indicating that LAEs have higher specific SFRs than typical SFGs 
due to the young stellar ages of LAEs.

\section{Inter-Stellar Medium (ISM)}
\label{sec:ISM}

In this section, we review four topics of the ISM of LAEs.
%
In the first three topics, we introduce three ISM properties of LAEs: gas kinematics and neutral hydrogen column density, dust extinction, and metallicity. All of these properties are explained by the nature of LAEs that have a Ly$\alpha$ escape fraction higher than the other galaxy population, which are detailed in the fourth topic. In the fourth topic, we present that LAEs are sources of ionizing photons with distinguished ISM features, 
i.e.
%
a high ionization state. \\

\begin{figure}[h]
\includegraphics[width=5.06in]{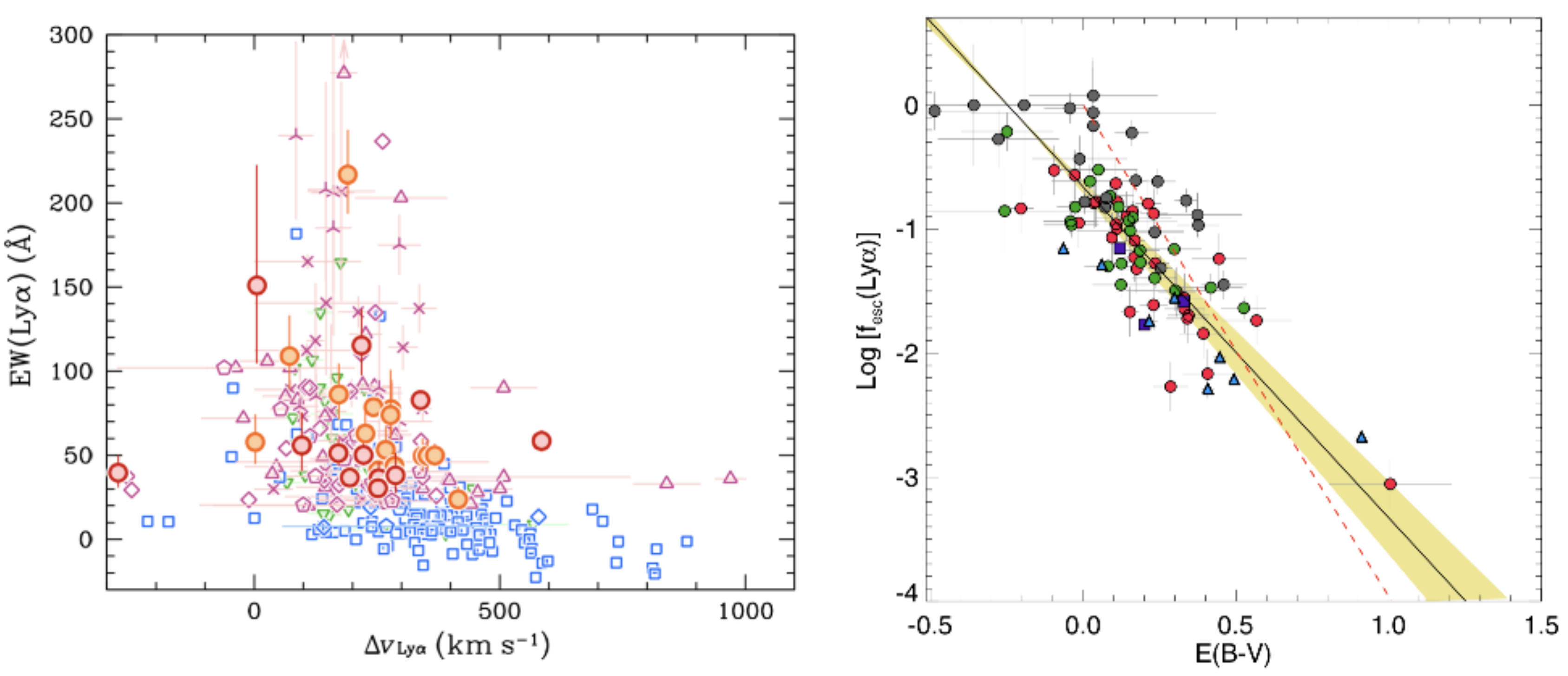}
\caption{Left: Ly$\alpha$ $EW_0$ as a function of Ly$\alpha$ velocity offset $\Delta v_{\rm Ly\alpha}$ \citep{nakajima2018b}. 
The data points are the compilation of the literature data (red, orange, and magenta: LAEs at $z\sim2-3$; blue: LBGs at $z\sim2-3$; green: green pea galaxies at $z\sim0.3$).
Adapted from \citet{nakajima2018b} with permission. Right:  Ly$\alpha$ escape fraction
as a function of color excess, $E(B-V)$ \citep{atek2014}. The data points represent 
measurements of SFGs at $z\sim0-0.3$ taken from the literature
(see also Section \ref{sec:Lya_emitter_obs}).
%
The differences of the data-point colors indicate differences of the references.
%
The black line and the yellow shade present the best-fit linear function and its $1\sigma$ fitting uncertainty, respectively. The red dashed line denotes 
the MW extinction law for a foreground-screen case that ignores Ly$\alpha$ resonant scattering.
%
Note that a similar relation is also found for LAEs at $z=2-4$ \citep{blanc2011}.
%
Adapted from \citet{atek2014} with permission.}
\label{fig_dvlya_ewlya_ebv_fesc_lya}
\end{figure}

\noindent {\it Gas Kinematics and Neutral Hydrogen Column Density} --- Analyses of emission and absorption lines suggest that LAEs show a strong gas outflow and a low neutral hydrogen column density $N_{\rm HI}$. The gas kinematics is evaluated from the velocity offset of the low-ionization interstellar (LIS) UV metal absorption lines $\Delta v_{\rm IS}$ with respect to the systemic redshift $z_{\rm sys}$. Deep NIR spectroscopic observations have found that the LIS UV metal absorption lines are typically blueshifted from $z_{\rm sys}$ by $\sim200$ km s$^{-1}$ for LAEs (\citealt{hashimoto2013, shibuya2014b}; Figure \ref{fig:LAE_Concept}$a$), suggesting that there exists a notable gas outflow with an outflow velocity of $V_{\rm exp} \sim 200$ km s$^{-1}$ in LAEs that is similar to
LBGs, majority of which are SFGs with weak/no Ly$\alpha$ emission
\citep{steidel2010}. Another important tool characterizing the gas kinematics is a Ly$\alpha$ emission line. 
Observations of Ly$\alpha$ emission find various shapes of Ly$\alpha$ profiles, a Ly$\alpha$ emission velocity offset (redshift) from the systemic velocity and double Ly$\alpha$ emission line peaks.
Several studies have measured 
the Ly$\alpha$ velocity offset
%
$\Delta v_{\rm Ly\alpha}$ for LAEs \citep{mclinden2011, guaita2013}, and have identified an anti-correlation between Ly$\alpha$ $EW_0$ and $\Delta v_{\rm Ly\alpha}$ \citep{hashimoto2013, shibuya2014b, erb2014}. As shown in the left panel of Figure \ref{fig_dvlya_ewlya_ebv_fesc_lya}, LAEs have a Ly$\alpha$ velocity offset of $\Delta v_{\rm Ly\alpha}\sim 200$ km s$^{-1}$ (see a typical spectrum in Figure \ref{fig:LAE_Concept}$e$), which is systematically smaller than $\Delta v_{\rm Ly\alpha}\sim 400$ km s$^{-1}$ of LBGs (\citealt{steidel2010}; see a typical spectrum in Figure \ref{fig:LAE_Concept}c). 
%
%
According to the observational results, LAEs have, on average, a Ly$\alpha$ velocity offset comparable with the outflow velocity, $\Delta v_{\rm Ly\alpha} \sim V_{\rm exp} \sim 200$ km s$^{-1}$, while LBGs show $\Delta v_{\rm Ly\alpha} \sim 2 \, V_{\rm exp} \sim 400$ km s$^{-1}$.

If the ES model (Section \ref{sec:Lya_model}) is applicable to LAEs and LBGs, various profiles of the observed Ly$\alpha$ emission lines including the Ly$\alpha$ velocity offsets and double peaks can be beautifully explained. The ES models best-fit to the observational data are presented with red lines in Figure \ref{fig:LAE_Concept}$c,e$. Moreover, the Ly$\alpha$ $EW_0-\Delta v_{\rm Ly\alpha}$ anti-correlation can be interpreted by the difference in $N_{\rm HI}$ between LAEs and LBGs.
%
In the ES model, the relation of $\Delta v_{\rm Ly\alpha} \sim V_{\rm exp}$ is 
obtained,
%
if the backscattered Ly$\alpha$ emission is weak due to the small H{\sc i} column density of the shell, $N_{\rm HI} \lesssim 10^{20}$ cm$^{-2}$.
%
In contrast, $\Delta v_{\rm Ly\alpha} \sim 2 \, V_{\rm exp}$ is reproduced, if the amount of Ly$\alpha$ back scattering is large in the case of a high H{\sc i} column density
in the shell,
%
$N_{\rm HI} \gtrsim 10^{20}$ cm$^{-2}$.
%
%
The interpretation based on the ES model suggests that LAEs have a {\sc Hi} column density
of the shell
%
lower than those of LBGs, which are quantitatively concluded by ES model fitting results \citep{hashimoto2015}. This difference in $N_{\rm HI}$ would produce the Ly$\alpha$ $EW_0-\Delta v_{\rm Ly\alpha}$ anti-correlation.

The low H{\sc i} column density for LAEs is also explained with measurements of the covering fraction, $f_{\rm c}$, of LIS gas estimated from the depth of LIS metal absorption lines. Medium-high resolution spectroscopy find a possible anti-correlation between Ly$\alpha$ $EW_0$ and $f_{\rm c}$ \citep{jones2013, shibuya2014b}.
Because $f_{\rm c}$ positively correlates with the {\sc Hi} gas density on average, the Ly$\alpha$ $EW_0-f_{\rm c}$ anti-correlation supports the idea that a large $EW_0$ (i.e. a large Ly$\alpha$ escape) is given by the low $N_{\rm HI}$ gas clouds.\\ 

\noindent {\it Dust extinction} --- LAEs are known as dust poor galaxies. The stellar and nebular extinction values are estimated to be $E(B-V) \sim 0-0.2$ for majority of LAEs by stellar-population synthesis model fitting and Balmer decrement analyses, respectively \citep{ono2010a, kojima2017}.
%
%
The dust-poor nature of LAEs is confirmed with measurements of the UV-continuum spectral slope $\beta$ that is defined by the power-law approximation of the UV-continuum flux, $f_\lambda \propto \lambda^\beta$. LAEs have typically a blue UV-continuum spectrum of $\beta\sim-2$ that is systematically smaller than those of LBGs in the same UV magnitude range \citep{stark2010}, supporting the fact that LAEs are dust poor. 

The dust extinction law of a galaxy can be evaluated by combining $\beta$ and the $IRX$ ratio, 

\begin{equation}
IRX = \frac{L_{\rm IR}}{L_{\rm UV}}, 
\end{equation}

\noindent where $L_{\rm IR}$ and $L_{\rm UV}$ are the total infrared and UV luminosities, respectively. Recent studies using {\it Spitzer} and {\it Herschel} have revealed that LAEs at $z\sim2$ have typically low $IRX$ values at a given $\beta$ value similar to that of the Small Magellanic Cloud (SMC) 
or even below those of Calzetti's local starbursts \citep{wardlow2014, kusakabe2015}. This low $IRX$ values are consistent with results that LAEs at $z>5$ have faint  far-infrared (FIR) continua at the observed-frame $1$mm band found by Atacama Large Millimeter/sub-millimeter Array (ALMA) observations 
\citep{knudsen2016}. 
These $IRX$ and FIR studies also confirm that LAEs have the low dust extinction. \\

\noindent {\it Metallicity} --- LAEs have a gas-phase metallicity lower than SFGs with a low Ly$\alpha$ EW on the basis of deep spectroscopy and two kinds of metallicity estimates,
strong line and
direct electron temperature $T_{\rm e}$ methods. 
The strong line method using, e.g., H$\alpha$, H$\beta$, [O {\sc iii}]5007,4959, is widely applied to LAEs at $z\sim2-3$. Observational studies have estimated the typical gas-phase metallicity to be $Z\sim0.1-0.5Z_\odot$ for LAEs at $z\sim2-3$ with the N2 and R23 indices \citep{finkelstein2011a, nakajima2012}. These metallicity measurements are comparable to or lower than those of the galaxy mass metallicity relation at $z\sim2-3$ \citep{finkelstein2011a}, and consistent with the SFR-mass metallicity relation \citep{nakajima2012}. 
Similarly low metal abundances, $Z \sim 0.1-0.3 Z_\odot$, are obtained for $z\sim 2-4$ LAEs by the $T_{\rm e}$ methods with the successful detections of faint $T_{\rm e}$-sensitive emission lines,
{\sc [Oiii]}4363 \citep{trainor2016} and O {\sc iii}]1661,1666 \citep{kojima2017}.
%
%
These studies of strong line and direct $T_{\rm e}$ methods indicate that the gas-phase metallicity of typical LAEs falls in a range of $Z\sim0.1-0.5 Z_\odot$ that is similar to or slightly lower than the typical metallicity of LBGs with the same UV-continuum luminosity \citep{steidel2014}.

\begin{figure}[h]
\begin{minipage}{2.6in}
\begin{center}
\includegraphics[width=2.6in]{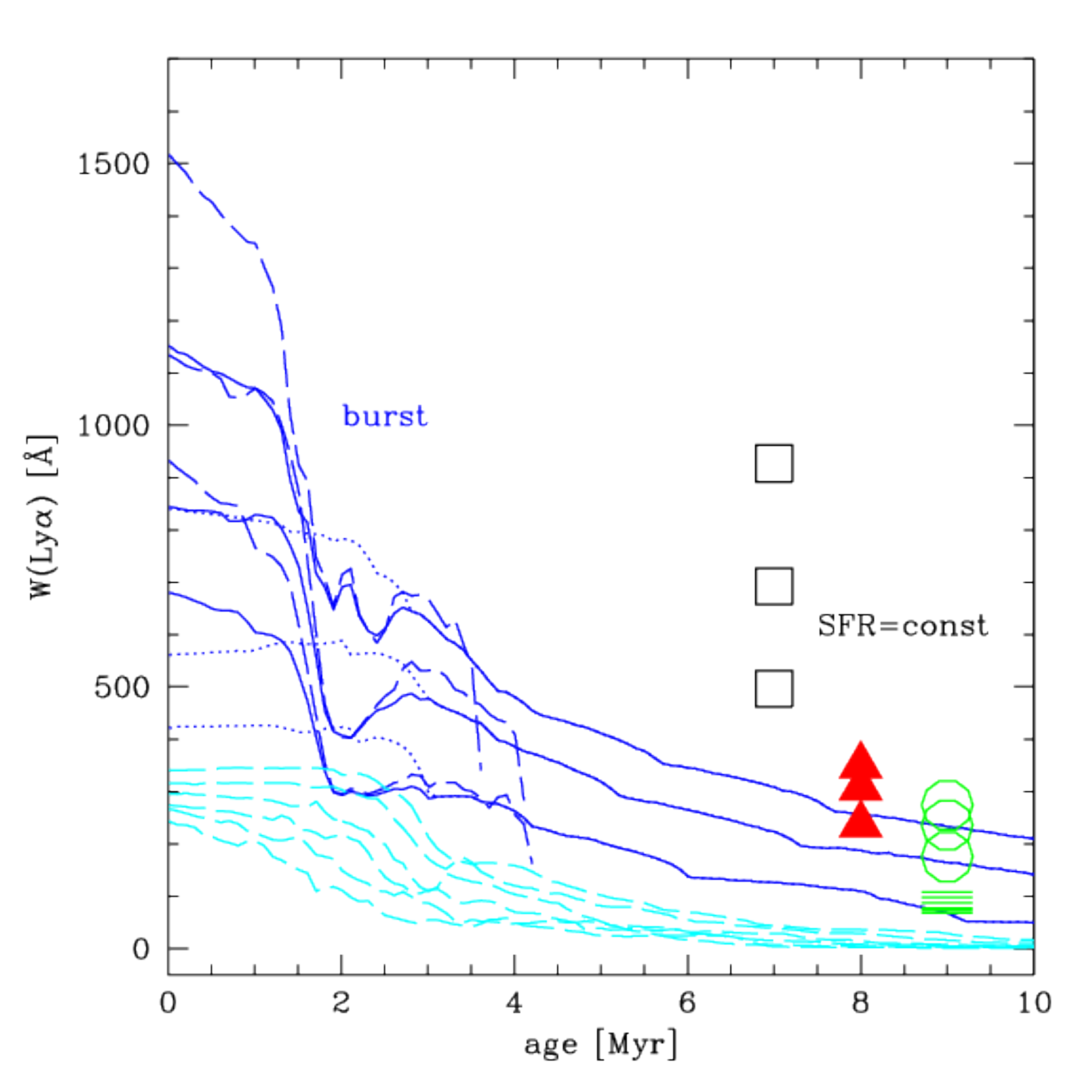}
\end{center}
\end{minipage}
\begin{minipage}{2.6in}
\begin{center}
\includegraphics[width=2.6in]{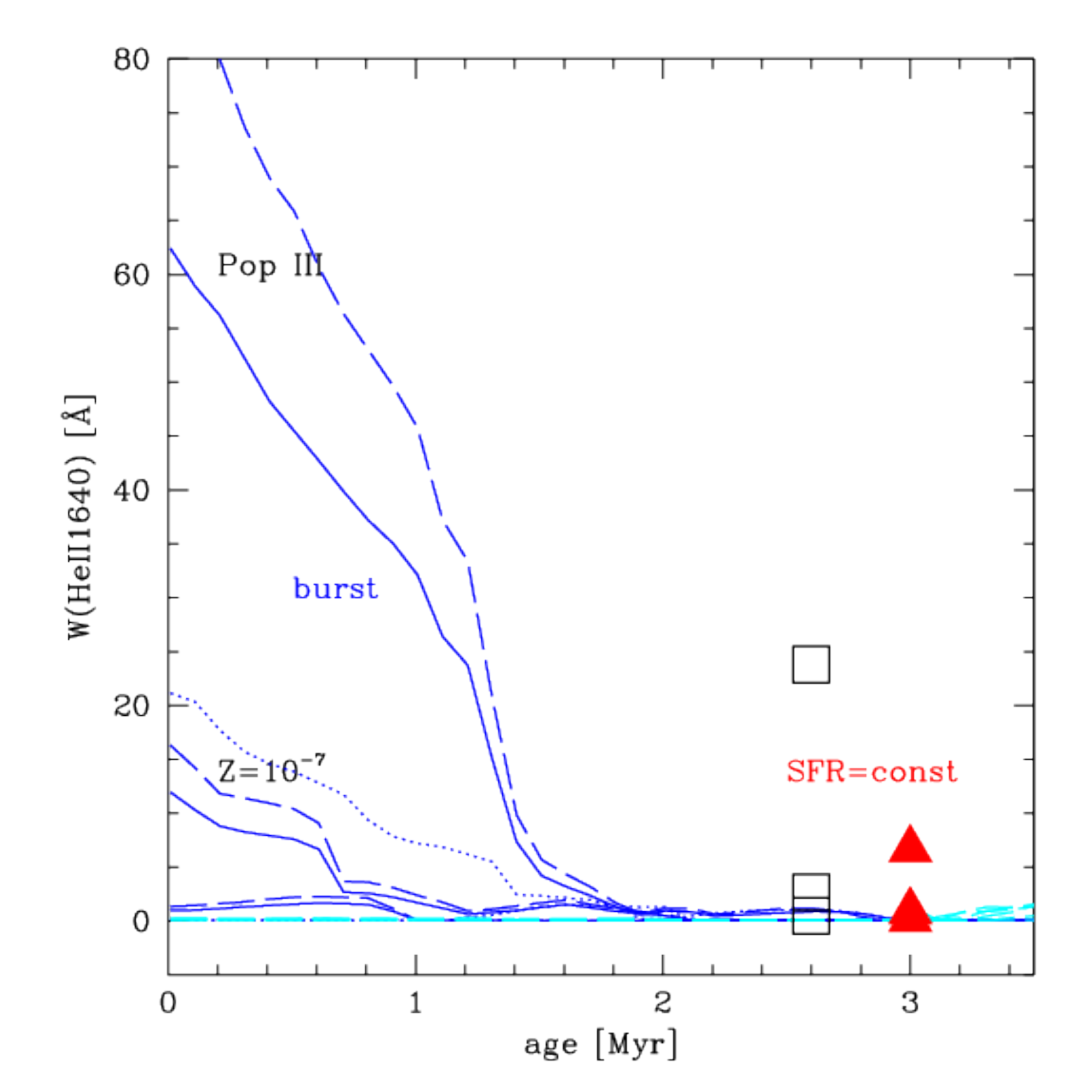}
\end{center}
\end{minipage}
\caption{
Left: Ly$\alpha$ $EW_0$ as a function of stellar age 
for models of stellar evolution and photoionization \citep{schaerer2003}. The three blue dashed lines indicate pop III ($Z=0$) instantaneous starburst models for a Salpeter IMF with mass ranges of $50-500 M_\odot$ (top), $1-500 M_\odot$ (middle), and $1-100 M_\odot$ (bottom). The blue solid (dotted) lines are the same as the blue dashed lines, but for a metallicity of $Z=10^{-7}$ ($10^{-5}$). The cyan dashed lines represent instantaneous starbursts of Salpeter IMF $1-100$ $M_\odot$ models with $Z=0.0004$, $0.001$, $0.004$, $0.008$, $0.020$, and $0.040$ from top to bottom. The three squares denote constant SFR models of the $50-500 M_\odot$ Salpeter IMF with metallicities of $Z=0$, $10^{-7}$, and $10^{-5}$ from top to bottom, where the stellar ages are arbitrary. The red triangles (green circles) are the same as the squares, but for the mass range $1-500 M_\odot$ ($1-100 M_\odot$). The green lines are the same as the green circles but for metallicities of $Z=0.0004$, $0.001$, $0.004$, $0.008$, $0.020$, and $0.040$ from top to bottom. Right: Same as the left panel, but for He{\sc ii}1640 $EW_0$. 
Adapted from \citet{schaerer2003} with permission.
}
\label{fig:LyaEWHeIIEW_age}
\end{figure}

LAEs with a large Ly$\alpha$ $EW_0$ are thought to be candidates of metal-free, i.e. pop III, galaxies. A large Ly$\alpha$ $EW_0$ of $\gtrsim240$\,\AA\, cannot be reproduced by recombination processes in a normal SFG with the solar metallicity and a Salpeter initial mass function 
(IMF; \citealt{malhotra2002}).
%
The left panel of Figure \ref{fig:LyaEWHeIIEW_age} presents theoretical predictions for Ly$\alpha$ $EW_0$ of young galaxies including pop III galaxies.  
%
The large Ly$\alpha$ $EW_0$ can be accomplished by
a young extremely metal poor galaxy with
a top heavy IMF that efficiently produces ionizing photons for strong Ly$\alpha$ via recombination 
\citep{schaerer2003}.
Large-area imaging and deep spectroscopic observations have identified a large number of large-Ly$\alpha$ EW sources \citep{hashimoto2017, shibuya2018a}.
A strong He{\sc ii}1640 emission line is another useful tool to distinguish pop III galaxies from pop II SFGs, because a production of He{\sc ii} line requires metal-poor massive stars emitting high-energy ($>54$ eV) photons that ionize He$^+$.
The right panel of Figure \ref{fig:LyaEWHeIIEW_age} shows He{\sc ii}1640 $EW_0$ as a function of stellar age for models of young galaxies including pop III galaxies, and indicates that a moderately strong He{\sc ii}1640 ($EW_0 \gtrsim 5$\AA) line can be observed in a young extremely metal poor galaxy with a top heavy IMF. Although He{\sc ii}1640 emission can be also found in AGNs and SFGs with a notable amount of Wolf-Rayet stars, such objects, except type 2 AGNs and high-mass X-ray binaries, can be removed by a presence of their broad line ($\gtrsim 1000$\AA) component of He{\sc ii} emission.
There are several reports of detecting such strong and narrow He{\sc ii}1640 emission in large-Ly$\alpha$ EW sources \citep{prescott2009, sobral2015}. However, subsequent spectroscopy and multi-wavelength studies suggest that these objects are not pop III galaxies,
due to the detections of metal lines or the lack of evidence on the strong He{\sc ii}1640 emission (\citealt{prescott2015a, shibuya2018b, bowler2017}; See Section 4.3 of \citealt{ouchi2019} for a summary).
So far, no promising candidates of pop III galaxies have remained.\\

\noindent {\it Ly$\alpha$ Escape Fraction} --- As described above, LAEs have the ISM with $N_{\rm HI}$, $E(B-V)$, and $Z$ lower than those of other types of galaxies with a low Ly$\alpha$ EW at the similar redshift. These characteristics of the ISM 
are plausibly related
%
to the high Ly$\alpha$ escape fraction ($f_{\rm esc}^{\rm Ly\alpha}$; Section \ref{sec:Lya_emitter_obs}) of LAEs. The low $N_{\rm HI}$ and $E(B-V)$ allow Ly$\alpha$ photons to easily escape from the ISM of LAEs due to a small number of the Ly$\alpha$ resonant scattering and absorption. Here, the low $Z$ is necessary for the low $E(B-V)$.

The connection between $f_{\rm esc}^{\rm Ly\alpha}$ and the ISM properties has been understood 
on the individual and statistical bases.
The Ly$\alpha$ escape fraction has been estimated individually for LAEs at $z\sim0-4$ \citep{atek2014, blanc2011}. The right panel of Figure \ref{fig_dvlya_ewlya_ebv_fesc_lya} presents that $f_{\rm esc}^{\rm Ly\alpha}$ anti-correlates with $E(B-V)$, indicating that Ly$\alpha$ photons are heavily absorbed in a dusty galaxy. 
%
%
On the other hand, as discussed (shown) in Section \ref{sec:Lya_emitter_obs} (Figure \ref{fig:LyaLD}), the cosmic average $f_{\rm esc}^{\rm Ly\alpha}$ values increase monotonically from $z\sim0$ to $z\sim6$ by two orders of magnitude \citep{hayes2010, konno2016}. 
%
The strong (two orders of magnitude) increase in
$f_{\rm esc}^{\rm Ly\alpha}$ 
is explained neither by 1) IGM absorption, 2) stellar population, 3) outflow velocity, 4) clumpy ISM, nor 5) dust extinction of a simple screen dust model \citep{konno2016}. 

%
Historically, it was often discussed that 
the large $f_{\rm esc}^{\rm Ly\alpha}$ found in SFGs at high redshift was explained by the scenario
of selective dust attenuation for Ly$\alpha$ and UV-continuum in the clumpy ISM  (a.k.a. Neufeld's effect; \citealt{neufeld1991}). 
%
In the clumpy ISM, Ly$\alpha$ photons are resonantly scattered on the surface of clumpy gas clouds with a negligible dust absorption 
(see Figure 1 of \citealt{neufeld1991}).
By the resonance scattering, the total number of Ly$\alpha$ photons is conserved.
%
On the other hand, 
UV-continuum photons penetrate multiple clumpy gas clouds in the foreground whose centers have dusty molecular gas, being heavily absorbed by the dust.
%
%
%
The difference of Ly$\alpha$ and UV-continuum extinction produces a high Ly$\alpha$ $EW_0$.
%
The validity of the Neufeld's effect had not been tested in realistic ISM conditions, but only in a simple case of the static and very clumpy/dusty ISM since the publication of \citet{neufeld1991}. Recent radiative transfer simulations have conducted extensive calculations with various physical parameters, and found that the Neufeld's effect exists, but emerges only under special physical conditions, a low outflow velocity, very high extinction, and an extremely clumpy gas distribution (most gas is locked up in clumps), many of which do not agree with the observed properties of LAEs \citep{larusen2013, duval2014}. The clumpy ISM scenario is probably not the reason of the high $f_{\rm esc}^{\rm Ly\alpha}$ values at high redshift. 

Although the physical origin of the strong increase in $f_{\rm esc}^{\rm Ly\alpha}$ has not been definitively concluded,
the rise in $f_{\rm esc}^{\rm Ly\alpha}$ is probably caused by another scenario,
%
6) the decrease of H{\sc i} column density towards high-$z$ that reduces the resonant scattering and thus dust attenuation of Ly$\alpha$. 
This scenario can explain 
the increase of $f_{\rm esc}^{\rm Ly\alpha}$ by two orders of magnitude \citep{konno2016}.\\


\begin{figure}[h]
\includegraphics[width=5.0in]{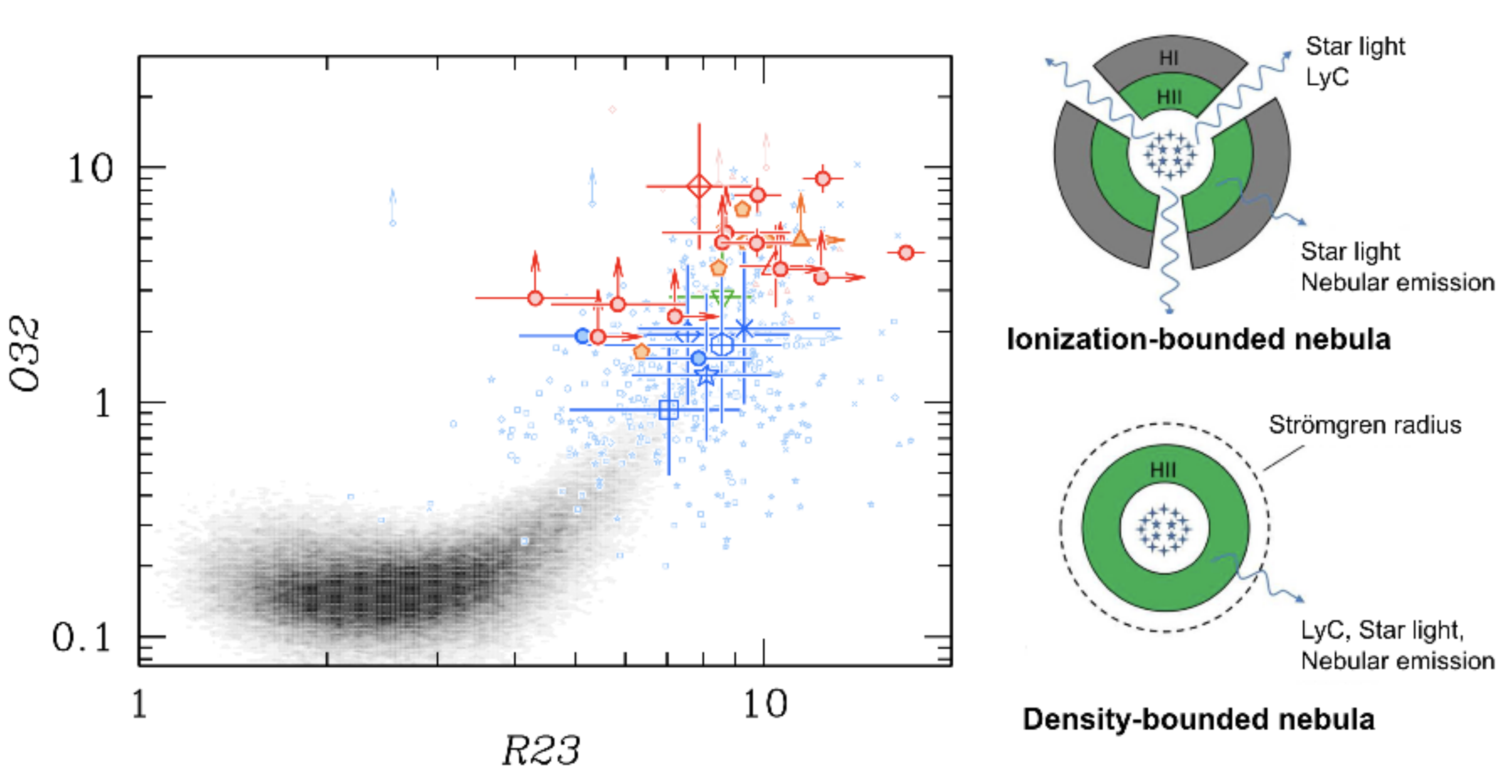}
\caption{Left: Diagram of $O32$ vs. $R23$ indices for high-$z$ and local galaxies \citep{nakajima2014,nakajima2016}. The red and blue symbols represent LAEs and LBGs at $z\sim2-4$, respectively. The orange data points indicate Lyman-continuum leakers. The green triangle denotes a green pea galaxy. The gray shade shows SDSS local galaxies. 
The large open symbols and the cross indicate average measurements.
This figure is reproduced and provided by K. Nakajima.
Right:  Conceptual figure for ionization-bounded (top) and density-bounded (bottom) nebulae \citep{zackrisson2013}. Adapted from \citet{zackrisson2013} with permission.}
\label{fig_ionization}
\end{figure}

\noindent {\it Ionization State} --- Recent optical and NIR spectroscopy find that LAEs have ionizing states significantly different from those of other galaxy populations. The left panel of Figure \ref{fig_ionization} shows that the $O32$ ratio
\footnote{
In this review, we use the definition, $O32\equiv f({\rm [OIII]5007})/f({\rm [OII]3727})$, including Figure \ref{fig_ionization}. Note that some studies use $O32=f({\rm [OIII]5007,4959})/f({\rm [OII]3727})$, and that the the ratio of $f({\rm [OIII]5007})$ to $f({\rm [OIII]4959})$ is 3, a constant value determined by atomic physics \citep{storey2000}.
}
, the line flux ratio of 
[O{\sc iii}]5007
%
to [O{\sc ii}]3727, of $z\sim2-3$ LAEs is $\sim10-100$ times higher than those of local galaxies and even higher than those of LBGs at the similar redshifts \citep{nakajima2014}. Based on comparisons with photoionization models in the $O32$ vs. $R23$
\footnote{
$R23\equiv \{f({\rm [OIII]5007,4959})+f({\rm [OII]3727})\}/f({\rm H}\beta)$.
} 
diagram, the high $O32$ ratios of LAEs indicate high ionization parameters, $q_{\rm ion} = 1-9\times10^8$ cm s$^{-1}$. Detections of high ionization lines of 
{\sc Ciii}]1907,1909 
and {\sc Civ}1548 also indicate that the ionization state of LAEs is very high \citep{stark2014}. On the other hand, recent ALMA studies have suggested that the [{\sc Cii}]158$\mu$m emission of LAEs are systematically fainter than the local SFR-$L_{\rm [CII]}$ relation (\citealt{ouchi2013, harikane2018}, cf. \citealt{carniani2018a}). Although a faint [{\sc Cii}]158$\mu$m luminosity for a given SFR could be explained by a presence of AGN or the collisional de-excitation of C$^+$, LAEs show neither AGN activity nor gas density high enough for the collisional de-excitation.

The two properties of the high $q_{\rm ion}$ and weak [{\sc Cii}]158$\mu$m emission may be consistently explained by the ISM of density-bounded nebula \citep{nakajima2014}. The right panel of Figure \ref{fig_ionization} shows a conceptual diagram of density-bounded and ionization-bounded nebulae. 
In the density-bounded (ionization-bounded) nebula, the size of ionized regions are determined by the amount of gas around ionizing sources (the number of ionizing photons).
%
%
In contrast with the ionization-bounded nebula, the density-bounded nebula has a small outer shell of ionized gas emitting low-ionization lines including [{\sc Oii}]3727, while the density-bounded nebula has a well-developed inner core of highly ionized gas producing high-ionization lines (e.g., {\sc Ciii}]1907,1909, [{\sc Oiii}]5007,4959, and [{\sc Oiii}]88$\mu$m). In this case, [{\sc Cii}]158$\mu$m-emitting photo-dissociation region (PDR) does not exist. 
More realistically, a non-zero $\sim 10$\% covering fraction of PDR can explain both small {\sc [Cii]}158$\mu$m/SFR and large {\sc [Oiii]}88$\mu$m/SFR ratios found in high-$z$ SFGs, a majority of which are LAEs \citep{harikane2020}.
%
Moreover, the density-bounded nebula would help ionizing photons escape from the ISM of LAEs, contributing to the cosmic reionization (\citealt{nakajima2014, jaskot2014}; see Section \ref{sec:reionization}). \\

%
%

There are many open questions about the ISM of LAEs. Although the density-bounded nebula is an interesting scenario,
no direct evidence of density-bounded nebulae has been obtained. 
Another problem of the ISM is that some LAEs have an extremely low $IRX$ value for a given UV slope $\beta$ even below the SMC's dust extinction curve \citep{capak2015}.
It is unclear how such a low $IRX$ value is reproduced, because realistic dust sizes cannot make the flat $IRX$-$\beta$ relation. 
Besides these detailed physical questions, one should push high Ly$\alpha$ EW LAE searches
\footnote{
See the text of the metallicity topic in this Section.
}
for a promising candidate of pop III galaxy whose identification is key for understanding the first stage of galaxy formation. 
%

%
%

\section{Circum-Galactic Medium (CGM) and the Large-Scale Structures Traced by Diffuse Ly$\alpha$}
\label{sec:CGM}

\begin{figure}[h]
\includegraphics[width=4.5in]{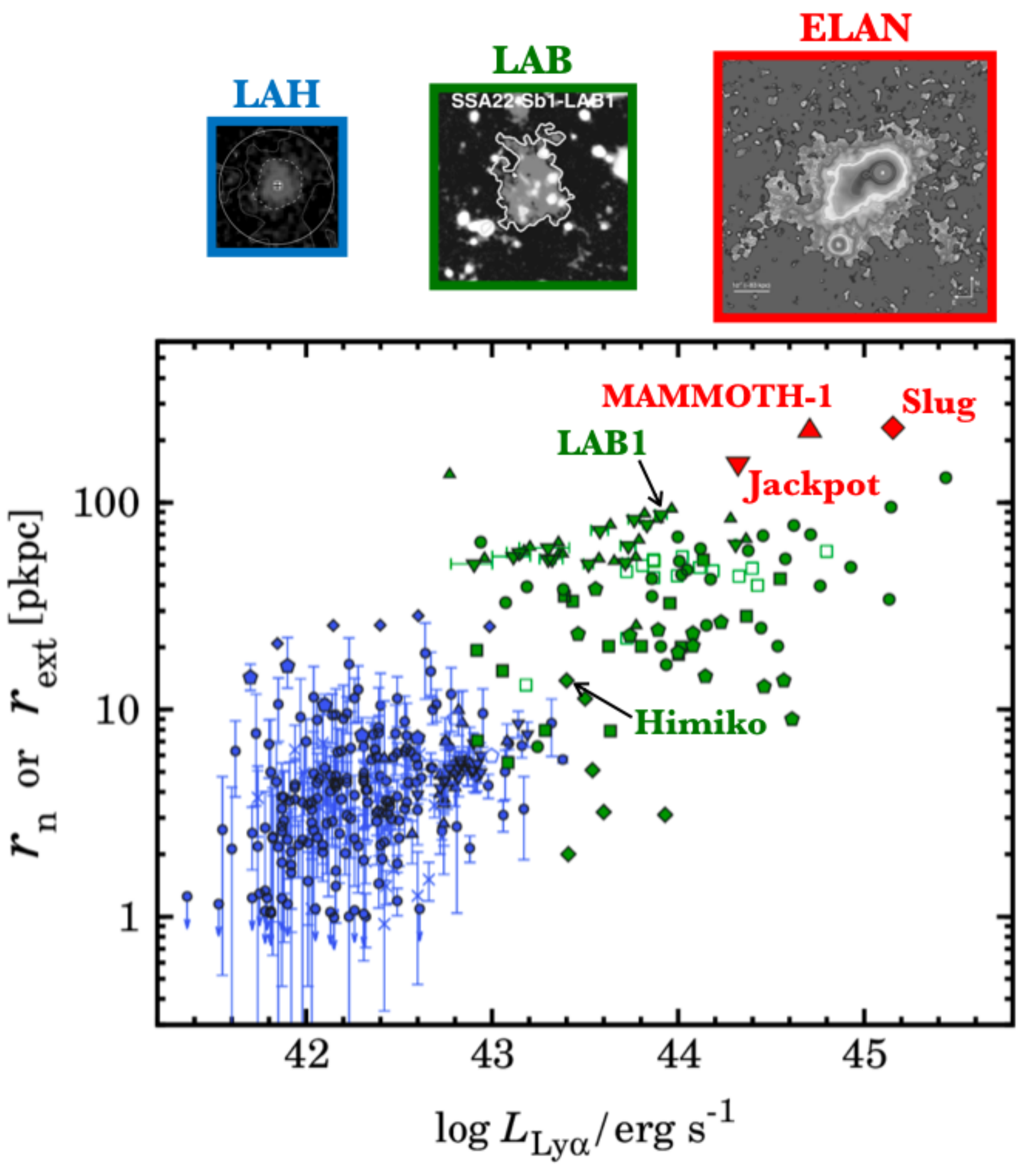}
\caption{Ly$\alpha$ scale length $r_{\rm n}$ or extent $r_{\rm ext}$ of extended Ly$\alpha$ emission as a function of Ly$\alpha$ luminosity. The blue, green, and red symbols denote LAHs, LABs, and ELANe, respectively, taken from the literature 
\citep{leclercq2017, wisotzki2016, steidel2011, momose2014, momose2016, xue2017, matsuda2011, zhang2019, borisova2016, cantalupo2014, hennawi2015, cai2016}, 
while the definitions of $r_{\rm ext}$ and Ly$\alpha$ luminosity depend on various studies in the literature. 
Here, the $r_{\rm ext}$ values are converted from the end-to-end Ly$\alpha$ extent \citep{borisova2016} multiplied by a factor of $0.5$ to mitigate the difference in measurement techniques. The top three images show 
sample
%
snapshots of an LAH \citep{leclercq2017}, LAB \citep{matsuda2011}, and ELAN \citep{cantalupo2014} from left to right that are
adapted with permission.
}
\label{fig_llya_rn}
\end{figure}

This section presents the diffuse and spatially extended Ly$\alpha$ emission around SFGs from a small scale to a large scale that are known as Ly$\alpha$ halos (LAHs), Ly$\alpha$ blobs (LABs), enormous Ly$\alpha$ nebulae (ELANe), and the large-scale Ly$\alpha$ emission in the IGM. The extended Ly$\alpha$ emission has a large diversity in size, ranging from $\sim1$ pkpc to $\gtrsim1,000$ pkpc,
which are summarized in Figure \ref{fig_llya_rn}. Note that LAHs, LABs, and ELANe are not clearly defined by physical quantities, but roughly classified by the Ly$\alpha$ luminosity and size.
Although the physical origins of the extended Ly$\alpha$ emission are poorly understood, there are six scenarios for the extended Ly$\alpha$ emission: 1) resonant scattering of Ly$\alpha$ emission from central star-forming regions and/or AGNs, 2) photoionization/recombination in unresolved dwarf satellite galaxies, 3) outflowing gas, 4) infalling gas, 5) fluorescence, and 6) galaxy mergers. In the following paragraphs, we review properties and possible physical origins of the extended Ly$\alpha$ emission.\\

\begin{figure}[h]
\includegraphics[width=4.5in]{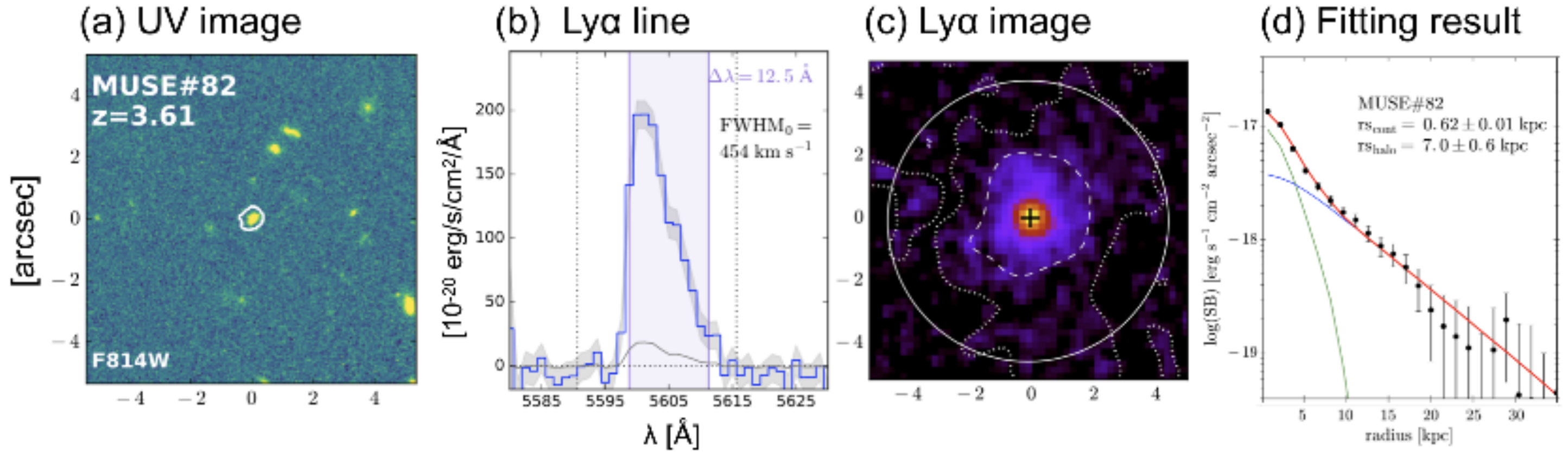} 
\includegraphics[width=4.0in]{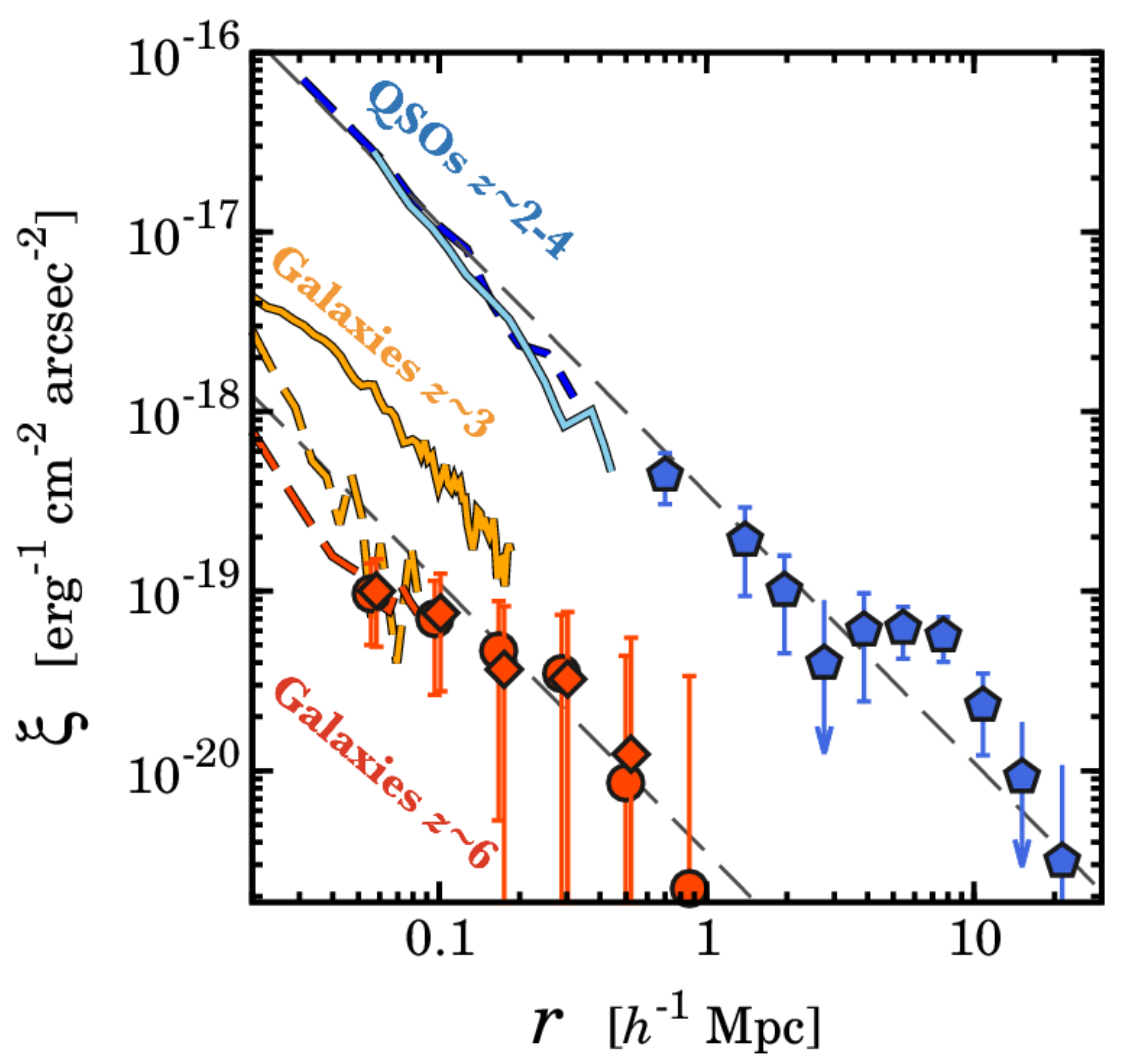}
\caption{Top panels (a)-(d): Images, spectrum, and radial SB profiles of a galaxy with an LAH at $z=3.61$ that are taken from \citet{leclercq2017} with permission. (a) UV-continuum image. (b) Ly$\alpha$ spectrum. (c) Ly$\alpha$ image. (d) Radial Ly$\alpha$ SB profile. In the panel (d), the black circles with the error bars present the observed Ly$\alpha$ SB profile, while the (red) green and blue curves indicate the best-fit Ly$\alpha$ SB profiles of (total) core and LAH components, respectively. Bottom: Cross-correlation functions of Ly$\alpha$ emission with SDSS BOSS quasars at $z\sim2-4$ \citep[blue pentagons; ][]{croft2018} and with LAEs \citep[red diamonds: $z=5.7$; red circles: $z=6.6$; ][]{kakuma2019}. The curves represent radial Ly$\alpha$ SB profiles of an ELAN \citep[cyan solid line; ][]{cantalupo2014}, radio-quiet quasars \citep[blue dashed line; ][]{borisova2016}, and galaxies at $z\sim3$ (orange solid line: \citealt{steidel2011}; orange dashed line: \citealt{leclercq2017}) and at $z\sim6$ \citep[red dashed line: ][]{leclercq2017}. The two black dashed lines are power law functions of $\xi \propto r^{-1.5}$
that are shown for references.}
\label{fig_lya_halo_intensity_map}
\end{figure}

\noindent --- {\it Ly$\alpha$ halos (LAHs)} are the $1-10$ pkpc-scale extended Ly$\alpha$ emission found around SFGs at $z\sim2-7$ (top panels of Figure \ref{fig_lya_halo_intensity_map}). Although the Ly$\alpha$ luminosity and size 
of LAHs depend on physical properties of hosting SFGs,
%
most of high-$z$ SFGs ubiquitously have LAHs. The diffuse Ly$\alpha$ emission of LAHs have been identified by image stacking analyses on the statistical basis \citep{hayashino2004, steidel2011, matsuda2012} and deep spectroscopic observations on the individual basis \citep{wisotzki2016, leclercq2017}. The Ly$\alpha$ radial SB profile is approximated with an 
exponential function,

\begin{equation}
S(r) = C_{\rm n} \exp{(-r / r_{\rm n})}, 
\end{equation}

\noindent where $S(r)$, $r$, and $C_{\rm n}$ are the Ly$\alpha$ SB, the radius from the galaxy center, and the normalization factor, respectively 
\footnote{
Although the exponential profile is an empirical model, a Ly$\alpha$ radial SB profile
similar to the exponential profile can be derived on the basis of a physical picture of Ly$\alpha$ resonant scattering in a gaseous halo with a simple power-law distribution of {\sc Hi} gas, $f_c\propto r^{-\gamma}$, where $\gamma$ is a power-law index \citep{steidel2011}.
}.
The parameter of $r_{\rm n}$ is the Ly$\alpha$ halo scale length, quantifying the extent of LAHs. Several studies find that $r_{\rm n}$ correlates with physical properties of LAEs, providing useful hints for physical origins of LAHs. 
In SFGs,
the Ly$\alpha$ halo scale length 
decreases with increasing the Ly$\alpha$ luminosity of the central part of the SFGs at $r<8$ pkpc
%
($L_{\rm cent,Ly\alpha}$;  \citealt{momose2016}), 
indicating that a large Ly$\alpha$ halo is found in $L_{\rm cent,Ly\alpha}$-faint galaxies.
(In other words, Ly$\alpha$ halos are small in LAEs whose central $<8$ pkpc part of Ly$\alpha$ luminosity is bright.)
%
The $L_{\rm cent,Ly\alpha}$-faint galaxies may be explained by strong Ly$\alpha$ resonant scattering (scenario 1) in the thick {\sc Hi} clouds near the galaxy center, which eventually produce a largely extended Ly$\alpha$ halo. 
On the other hand, radiative transfer and hydrodynamical simulations suggest that the Ly$\alpha$ resonant scattering effect (scenario 1) cannot reproduce the largely-extended Ly$\alpha$ SB profiles of LAHs, while the cooling radiation (scenario 4) may be needed to explain Ly$\alpha$ emission at the outer part of LAHs \citep{lake2015}. In contrast with the Ly$\alpha$ emission, UV-continuum emission is not spatially extended in observational data \citep{momose2014, leclercq2017}. This compact UV-continuum emission would rule out a significant contribution from unresolved dwarf satellite galaxies (scenario 2). Recently, stacked ALMA data reveal that SFGs at $z\sim5-7$ have a 10 pkpc-scale extended carbon [{\sc Cii}]$158\mu$m emission whose SB profile shape is very similar to the one of the Ly$\alpha$ emission \citep{fujimoto2019}. The existence of the extended emission of [{\sc Cii}] similar to Ly$\alpha$ might suggest that LAHs are originated from the neutral gas expelled by outflows that are carbon enriched (scenario 3). 
Deep observations identify that the $r_{\rm n}$ value is almost constant over the redshift range of $z\sim 2-6$, 
perhaps suggesting no significant redshift evolution of physical properties of LAHs 
at $z<6$
%
\citep{momose2014, leclercq2017}. There is a hint of an increase of $r_{\rm n}$ towards the reionization epoch of $z=6.6$, but the increase is found only at the $1\sigma$ level \citep{momose2014}.\\
%


\noindent ---  {\it Ly$\alpha$ blobs (LABs)} are spatially extended Ly$\alpha$ nebulae  with a physical scale of $\sim10-100$ pkpc and a Ly$\alpha$ luminosity of $\sim10^{43}$ erg s$^{-1}$ \citep{steidel2000, matsuda2004, shibuya2018a}. Such extended Ly$\alpha$ emission has been found around various types of sources: e.g., SFGs \citep{steidel2000, ouchi2009}, radio-loud galaxies \citep{mccarthy1987, vanojik1997}, and radio-quiet QSOs \citep{borisova2016}.  

Physical properties of LABs have been investigated by multi-wavelength observational and theoretical studies. Deep X-ray observations find that a majority of LABs show no clear AGN activity, while $\sim20$\% of LABs host an AGN \citep{basuzych2004, geach2009}. Similarly, $\sim30$\% of LABs are detected at radio wavelengths \citep{ao2017}. These X-ray and radio results would indicate that AGNs contribute to the extended Ly$\alpha$ emission in some LABs via Ly$\alpha$ resonant scattering and/or fluoscence (scenarios 1 and 5). A tangential polarization signal up to $20$\% is detected in the Ly$\alpha$ emission around an LAB \citep{hayes2011}, which might suggest that LABs are produced by Ly$\alpha$ resonant scattering (scenario 1; \citealt{dijkstra2008}; cf. \citealt{trebitsch2016}). According to theoretical studies \citep{fardal2001, dijkstra2009, goerdt2010}, the extended Ly$\alpha$ emission of LABs can be reproduced by the cooling radiation (scenario 4). However, LABs have a moderately large Ly$\alpha$ velocity offset from the systemic velocity, several hundreds km s$^{-1}$ \citep{yang2014}, 
that may be evidence for strong outflow (scenario 3). HST observations reveal that some high-$z$ LABs have multiple stellar components \citep{ouchi2013, sobral2015}, which 
suggests
that starbursts caused by galaxy mergers (scenario 6) create the extended Ly$\alpha$ emission. The galaxy merger scenario might be supported by the fact that LABs reside in galaxy overdense regions \citep{matsuda2004, yang2010, kikuta2019}.\\

 
\noindent --- {\it Enormous Ly$\alpha$ nebulae (ELANe)} are extended Ly$\alpha$ emission harboring QSOs \citep{cantalupo2014, hennawi2015, cai2016}. The Ly$\alpha$ emission of ELANe extends to a scale of several hundreds pkpc ($>400$ pkpc for some ELANe) that is larger than a virial radius of a dark-matter halo hosting the QSOs. The existences of energetic QSOs suggest that ELANe are fluorescently illuminated (scenario 5). However, very high density clumps
are needed to produce the observed high Ly$\alpha$ SB of $\sim10^{-18}-10^{-16}$ erg s$^{-1}$ cm$^{-2}$ arcsec$^{-2}$ \citep{cantalupo2014}. Alternatively, the Ly$\alpha$ resonant scattering may be also important 
in ELANe (scenario 1). Distinguishing the two scenarios of 1) and 5), one should observe non-resonant hydrogen recombination lines such as H$\alpha$
in ELANe 
as demonstrated by
\citet{leibler2018}. 

NB imaging observations and the Keck/Palomar Cosmic Web Imagers (KCWI/PCWI) spectroscopy find that ELANe tend to reside in galaxy overdense regions \citep{cai2017a, cai2018}. The environments of ELANe indicate that ELANe would exist in progenitors of massive galaxy clusters.\\

\noindent --- {\it The large-scale Ly$\alpha$ emission} extending to $\sim1-15$ comoving Mpc (cMpc) is identified by the cross-correlation Ly$\alpha$ intensity mapping technique (bottom panel of Figure \ref{fig_lya_halo_intensity_map}). The cross-correlation Ly$\alpha$ intensity mapping technique is to obtain the spatial cross correlation between objects with known redshifts (e.g. galaxies and QSOs) and faint Ly$\alpha$ emission below a detection limit of imaging/spectroscopic data, and to detect the faint Ly$\alpha$ emission, where the spatial cross correlation analysis systematically removes signals unrelated to the objects. 
\citet{croft2018} report the detection of a very faint Ly$\alpha$ emission around QSOs at a large scale of $\sim1-15$ cMpc by Ly$\alpha$ intensity mapping of cross correlation between SDSS BOSS QSOs at $z\sim2-3.5$ and Ly$\alpha$ emission in a large number of SDSS spectra (with no Ly$\alpha$ detections on individual basis).
%
%
The radial SB profile of these large-scale Ly$\alpha$ emission is smoothly connected to the one of ELANe at $r\sim1$ cMpc (bottom panel of Figure \ref{fig_lya_halo_intensity_map}). The connection with ELANe indicates that the large-scale Ly$\alpha$ emission may be produced by the fluorescent radiation from QSOs (scenario 5). In addition to QSOs, \citet{kakuma2019} conduct Ly$\alpha$ intensity mapping of cross correlation between LAEs at $z\sim 6$ and Ly$\alpha$ emission in NB image pixels, and identify extremely faint Ly$\alpha$ emission around LAEs at a spatial scale beyond a virial radius of the LAE hosting dark-matter halo ($\sim 0.15$ cMpc) up to $\sim 1$ cMpc.
This cross-correlation signals of LAEs are consistent with the extrapolation of Ly$\alpha$ radial SB profiles of LAHs (\citealt{leclercq2017}; bottom panel of Figure \ref{fig_lya_halo_intensity_map}). Comparisons with numerical simulations suggest that the large-scale Ly$\alpha$ emission is not fully explained by the combination of resonant scattering (scenario 1) and unresolved dwarf satellite galaxies (scenario 2), but contributed by Ly$\alpha$ emission created by other mechanisms that possibly includes the cold accretion.\\

%
%

Due to the diverse physical properties, it is still under debate how the extended Ly$\alpha$ emission is produced in LAHs, LABs, and ELANe. 
It is also unclear whether the possible increase in $r_{\rm n}$ to the epoch of reionization truly exists
(see the topic of LAHs),
%
which is thought to be an indicator of the IGM neutral hydrogen fraction increase \citep{jeesondaniel2012}.

%
%

\begin{figure}[h]
\includegraphics[width=5.0in]{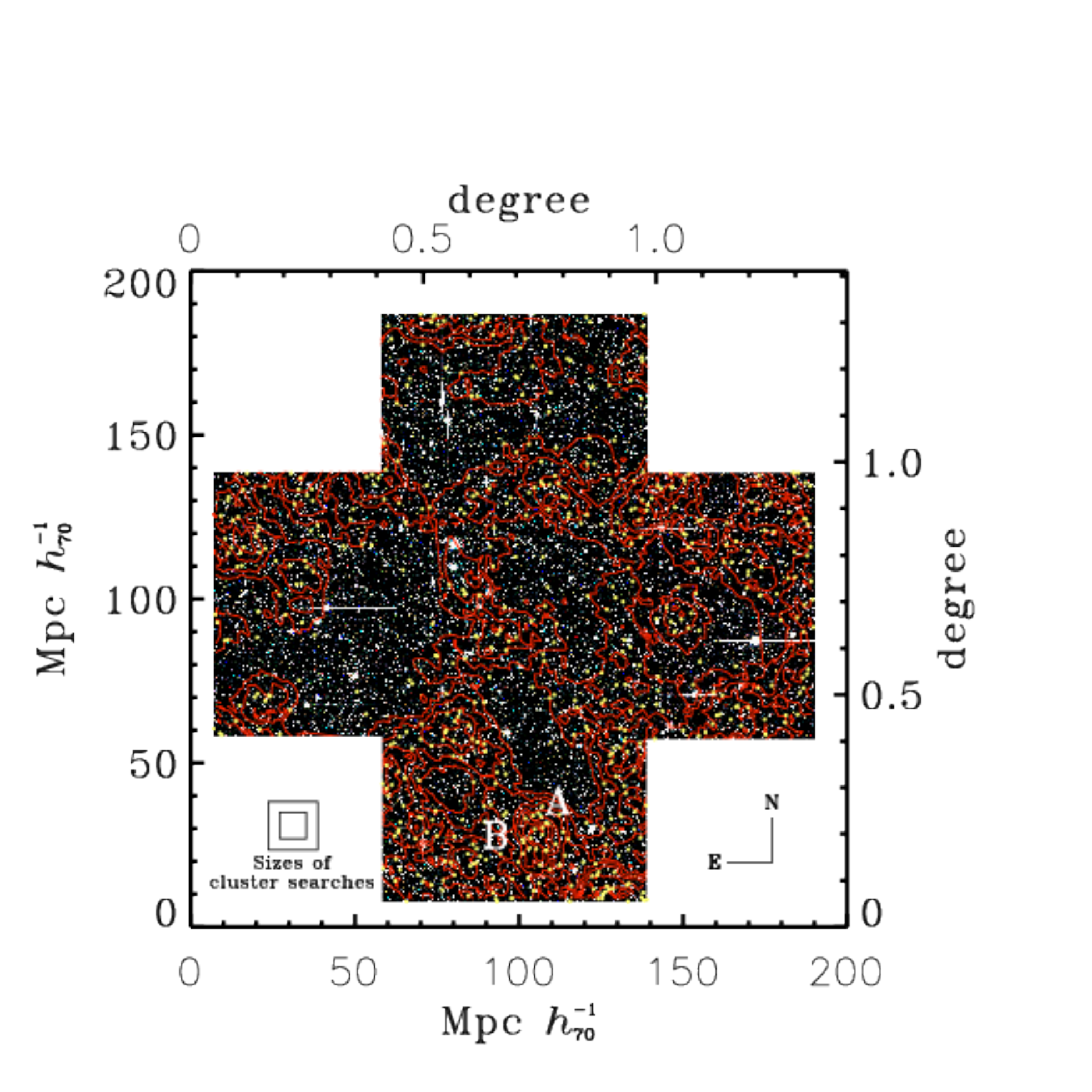}
\caption{
Map of LAEs at $z=5.7\pm 0.05$ over the scale of $1.5$ deg$^2$ that corresponds to $\sim 200$ cMpc \citep{ouchi2005a}. The yellow 
dots
%
represent positions of the LAEs on the survey area of a black cross-hair region that is indicated with the color composite sky image. The red contours denote the surface density of the LAEs. The positions of the LAE overdensity regions are marked with 
the characters
, A and B.
Adapted from \citet{ouchi2005a} with permission.
}
\label{fig:LSS}
\end{figure}

\section{Clustering of LAEs}
\label{sec:LAE_clustering}


Distant cosmic structures including primeval large-scale structures and progenitors of galaxy clusters (a.k.a. protoclusters) are investigated with LAEs. LAEs are advantageous to map out the spatial distribution of high-$z$ galaxies efficiently by 
NB imaging and spectroscopic surveys \citep{ouchi2005a,yamada2012a}. 
%
Figure \ref{fig:LSS}
is the map of LAEs at $z\sim 6$ over the scale of $1.5$ deg$^2$ corresponding to $\sim 200$ cMpc,
showing primeval galaxy clusters, filaments, and voids.
%

The systematic large-scale structure survey is being conducted by HETDEX (Section \ref{sec:Lya_emitter_obs}). HETDEX is a cosmology survey investigating cosmic structures at $z\sim 2-3$, measuring the baryon acoustic oscillation on the scale of $\sim 100$ cMpc for understanding the equation state of dark energy.

\begin{table}[b]
\tabcolsep7.5pt
\caption{List of Protoclusters}
\label{tab:protocluster}
\begin{center}
\scalebox{0.63}{
\begin{tabular}{@{}l|c|c|c|c|c|c@{}}
\hline
Name & $z$ & $N_{\rm spec}$ & $\delta$ & Sample & $M_{\rm total}$ & Ref. \\
(1) & (2) & (3) & (4) & (5) & (6) & (7) \\
\hline
z66OD & 6.59 & 12  & $14.3\pm 2.1$ & LAE & $5.4\times 10^{14}$ & H19 \\
HSC-z7PCC26 & 6.54 & 14 & $6.8^{+6.1}_{-3.7}$ & LAE & $8.4\times 10^{14}$ & C17,19,Hi19 \\
SDF & 6.01 & 10 & $16\pm 7$ & LBG & $(2-4)\times 10^{14}$ & To14 \\
z57OD & 5.69 & 44 & $11.5\pm 1.6$ & LAE & $4.8\times 10^{14}$ & O05,J18,H19 \\
SPT2349-56 & 4.31 & 14 & $>1000$ & SMG & $1.16\times 10^{13}$ & M18 \\
TNJ1338-1942 & 4.11 & 37 & $3.7^{+1.0}_{-0.8}$ & LAE/LBG & $(6-9)\times 10^{14}$ & V07,M04,Z05,Ov08 \\
DRC-protocluster & 4.00 & 10 & $\sim 5.5-11.0$ & SMG & $3.2-4.4\times 10^{13}$ & O18 \\
PC217.96+32.3 & 3.79 & 65 & $14\pm 7$ & LAE & $(0.6-1.3)\times 10^{15}$ & Lee14,D16,S19 \\
D4GD01 & 3.67 & 11 & $...$ & LBG & $...$ & To16 \\
ClJ0227-0421 & 3.29 & 19 & $10.5\pm 2.8$ & Spec & $(1.9-3.3)\times 10^{14}$ & Lem14 \\
TNJ2009-3040 & 3.16 & $>11$ & $0.7^{+0.8}_{-0.6}$ & LAE & $...$ & V07 \\
MRC0316-257 & 3.13 & 31 & $2.3^{+0.5}_{-0.4}$ & LAE & $(3-5)\times 10^{14}$ & V07 \\
SSA22FLD & 3.09 & $>15$ & $3.6^{+1.4}_{-1.2}$ & LBG/LAE/SMG & $(1.0-1.4)\times 10^{15}$ & S00,M05,Y12,U18 \\
\hline
\end{tabular}
}
\end{center}
\begin{tabnote}
Note: 
Spectroscopically-confirmed protoclusters having $>10$ galaxies with spectroscopic redshifts. The protoclusters are listed in order of redshift. This table is the abstract of Table 3 of \citet{harikane2019}.
(1) Protocluster name. 
(2) Redshift.
(3) Number of member galaxies confirmed by spectroscopy. 
(4) Galaxy overdensity.
Note that the length scales of the overdensity measurements differ.
%
(5) Types of member galaxies so far identified. 
(6) Total (expected) mass of the overdensity at the redshift ($z=0$) in units of solar masses.  
(7) Reference: 
(
C17,19:
\citealt{chanchaiworawit2017,chanchaiworawit2019}, 
D16:
\citealt{dey2016},
Hi19:
\citealt{higuchi2019},
H19:
\citealt{harikane2019},
J18:
\citealt{jiang2018}
\citealt{lee2014},
Lee et al. 2014, 
Lem14:
\citealt{lemaux2014},
M04:
\citealt{miley2004},
M05:
\citealt{matsuda2005},
M18:
\citealt{miller2018},
O05:
\citealt{ouchi2005},
Ov08:
\citealt{overzier2008},
O18:
\citealt{oteo2018},
S00:
\citealt{steidel2000},
S19:
\citealt{shi2019},
To14,16:
\citealt{toshikawa2014,toshikawa2016}
U18:
\citealt{umehata2018}
V07:
\citealt{venemans2007},
Y12:
\citealt{yamada2012}
Z05:
\citealt{zirm2005}
)

\end{tabnote}
\end{table}

\begin{figure}[h]
\begin{minipage}{2.5in}
\begin{center}
\includegraphics[width=2.5in]{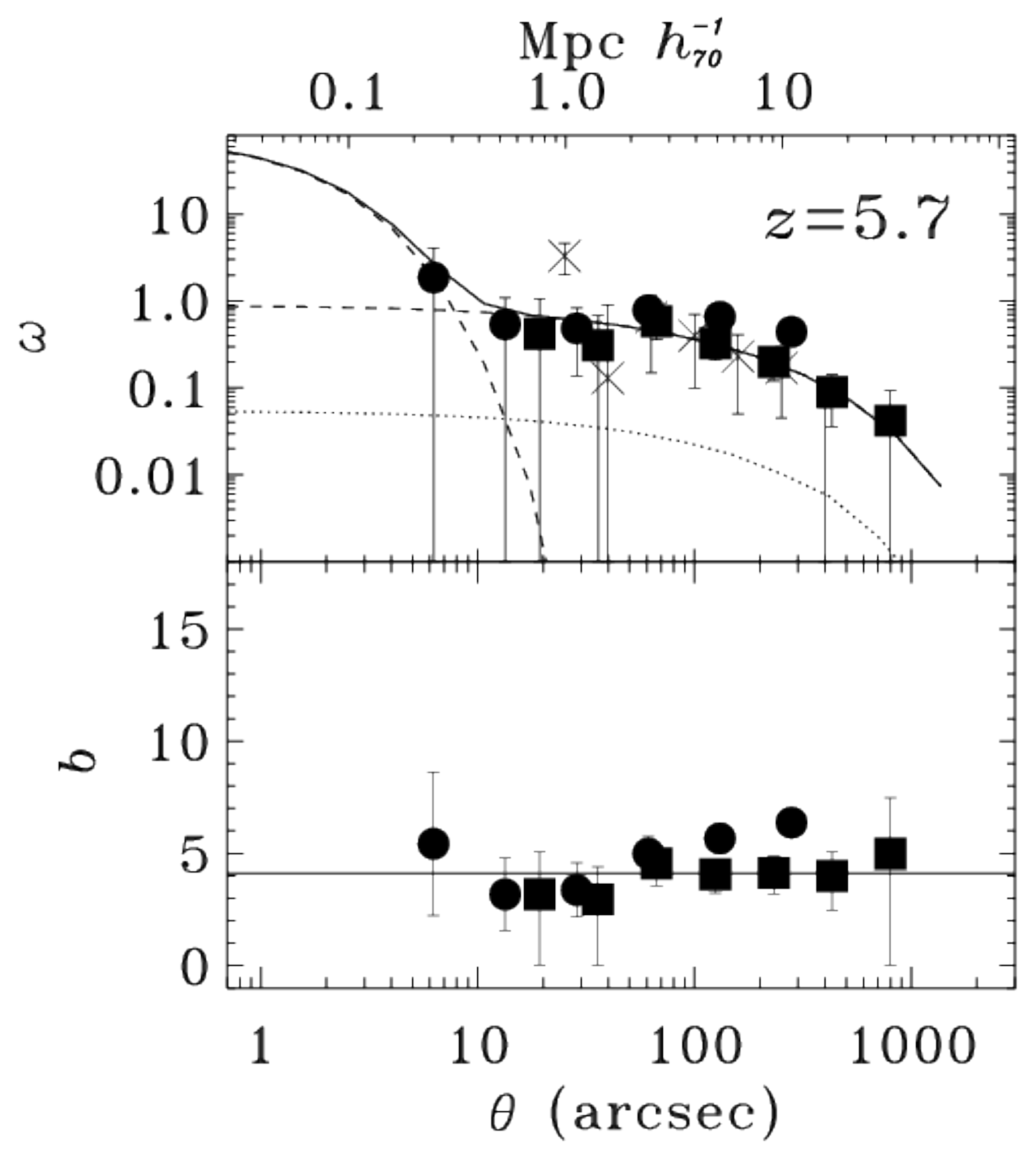}
\end{center}
\end{minipage}
\begin{minipage}{2.5in}
\begin{center}
\includegraphics[width=2.5in]{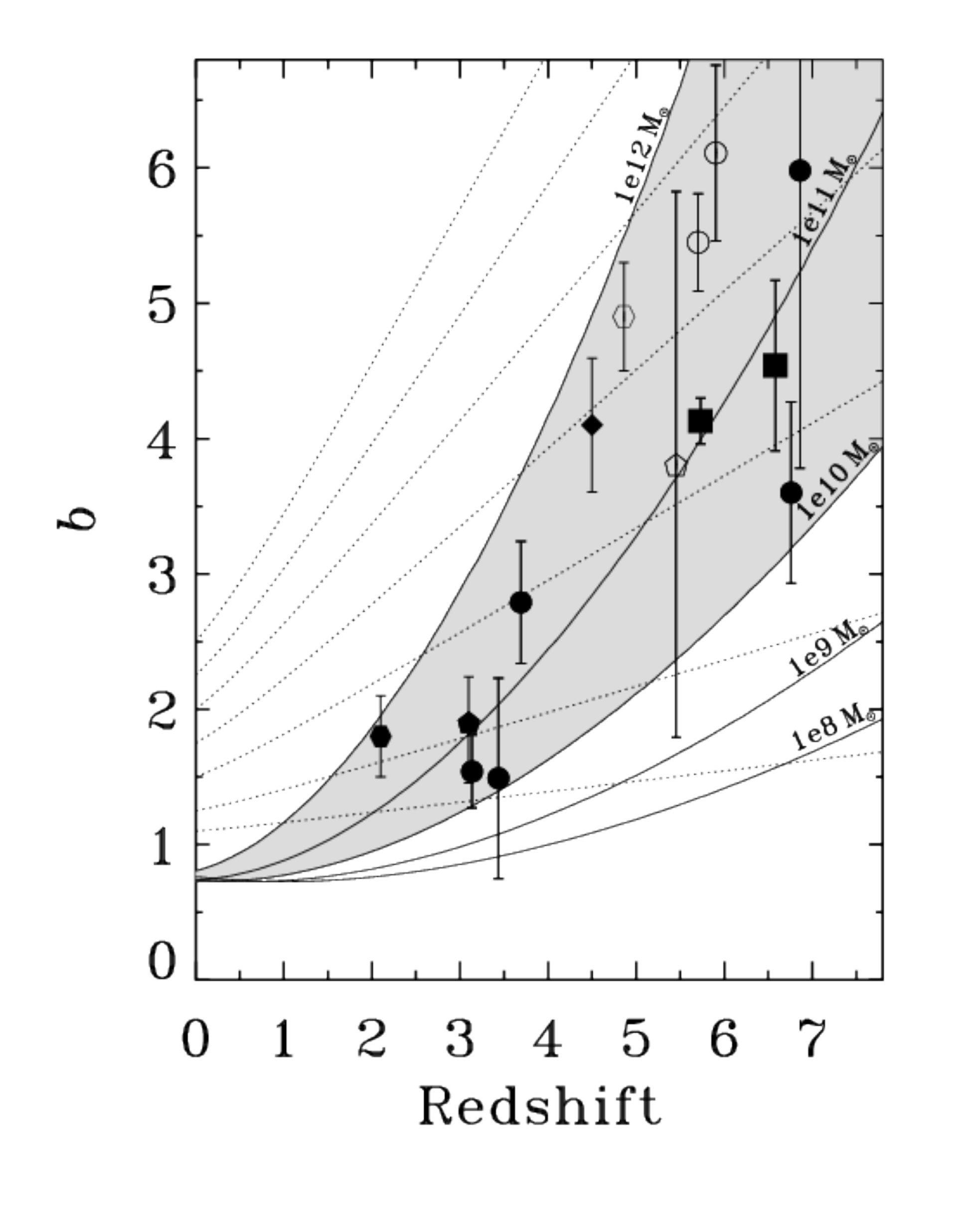}
\end{center}
\end{minipage}
\caption{
%
Left:
%
%
ACF and bias of $\gtrsim L^*$ LAEs at $z=5.7$ \citep{ouchi2018} that are shown in the top and bottom panels, respectively. The data points indicate the ACFs (top) and bias (bottom) of the LAEs.
In the top panel, the solid line presents the best-fit HOD model. The 1-halo and 2-halo terms of the best-fit HOD model are shown with the dashed lines that overlap with the solid line at the small and large scales, respectively. The dotted line denotes the underlying dark matter predicted by the linear theory.
In the bottom panel, the horizontal line represents the average bias of the LAEs.
The top axis denotes the projected distance in comoving megaparsecs. 
Right: 
Bias of $\gtrsim L^*$ LAEs as a function of redshift \citep{ouchi2018}. The data points represent the bias of the LAEs at $z=2-7$. The solid lines indicate bias of dark-matter halos with a halo mass of $10^{8}$, $10^{9}$, $10^{10}$, $10^{11}$, and $10^{12} M_\odot$ in the case of one-to-one correspondence between galaxies and dark-matter halos. The gray region shows the dark-matter halo mass range of $10^{10}-10^{12} M_\odot$ where the bias of LAEs at $z=2-7$ fall, while the number-weighted average mass is $10^{10}-10^{11} M_\odot$. The dotted lines are evolutionary tracks of bias in the case of the galaxy-conserving model. 
}
\label{fig:acorr_biasevol}
\end{figure}

In the scale of galaxy clusters, LAEs are tracers of the galaxy distribution, which is the concept same as the one of the HETDEX cosmology survey.
The overdensity $\delta$ is defined as
\begin{equation}
\delta = \frac{n - \bar{n}}{\bar{n}},
\label{eq:delta}
\end{equation}
where $n$ ($\bar{n}$) is the (average) number density of LAEs in a volume or an area.
Because LAEs are low-mass galaxies (Section \ref{sec:stellar_population}),
LAEs show overdensity regions whose $\delta$ values are smaller than those of LBG and submillimeter-galaxy (SMG) overdensity regions. However, overdensity regions of mass distribution are pinpointed with an LAE sample, if the LAE sample is large enough to provide $\delta$ measurements with a high statistical accuracy distinguishing the overdensity regions from the field. 

Theoretical models suggest that LAE overdensities with large $\delta$ values are progenitors of clusters, 
i.e. protoclusters 
\citep{chiang2013}. Protoclusters of LAE overdensities  are identified at $z\sim 2-7$ by blank-field surveys such as large-area 
NB observations
\citep{ouchi2005a,yamada2012a,harikane2019}. Table \ref{tab:protocluster} 
summarizes the spectroscopically-confirmed protoclusters identified by the LAE blank-field surveys, and compares protoclusters identified with the other types of galaxies such as LBGs and SMGs. 
%


Clustering properties of LAEs are quantified with correlation functions in two or three-dimensional space. The top left panel of Figure \ref{fig:acorr_biasevol} presents the two-dimensional space correlation functions $\omega (\theta)$, i.e. the angular correlation functions (ACFs), of $\gtrsim L^*$ LAEs at $z=5.7$ and $6.6$ \citep{ouchi2018}. Clearly, the ACFs of the LAEs are stronger than those of underlying dark matter $\omega_{\rm DM} (\theta)$ predicted by the linear theory of the $\Lambda$CDM model (top left panel of Figure \ref{fig:acorr_biasevol}). In the $\Lambda$CDM framework, the excess of $\omega (\theta)$ above $\omega_{\rm DM} (\theta)$ is evaluated by the bias $b$ that is defined as 
\begin{equation}
b^2 \equiv \frac{\omega (\theta)}{\omega_{\rm DM} (\theta)}
\label{eq:bias}
\end{equation}
%
The definition of the bias is also given by $b^2 \equiv \xi (r) /\xi_{\rm DM} (r)$, where $\xi (r)$ and $\xi_{\rm DM} (r)$ are the same as $\omega (\theta)$ and $\omega_{\rm DM} (\theta)$, respectively, but for the three-dimensional space (spatial) correlation functions at the scale of $r$.
The bottom left panel of Figure \ref{fig:acorr_biasevol} shows bias of the LAEs as a function of angular distance. The bias of the LAEs evolves by redshift, and increases towards high-$z$ (right panel of Figure \ref{fig:acorr_biasevol}).
The average bias of $L^*$ LAEs is $b\sim 1-2$ at $z\sim 2-3$ and $b\sim 4-5$ at $z\sim 6-7$ \citep{ouchi2018}. This increase of bias towards high $z$ is explained by the physical picture that a galaxy forms only at a rare peak of density fluctuations, where dark-matter halos are made,
at the early stage of the cosmic structure formation \citep{bardeen1986}.
Based on the average bias evolution of Figure \ref{fig:acorr_biasevol}, the low bias values indicate that LAEs at $z\sim 2-3$ may be progenitors of today's Milky-Way like galaxies \citep{gawiser2007}, while the high bias values suggest that LAEs at $z\sim 5-7$ are progenitors of present-day massive elliptical galaxies \citep{ouchi2010}.
%

There are large scatters in the bias measurements of LAEs obtained to date (see the right panel of Figure \ref{fig:acorr_biasevol}). The scatters would be larger than the statistical uncertainties. It is suggested that the large scatters are made by the sample variance, a.k.a. cosmic variance, originated by the small survey volumes \citep{kusakabe2018}. Although the survey volumes of LAEs are generally small, $10^6$ cMpc$^3$ or less,
on-going 
Subaru HSC 
and HETDEX observations are providing the measurements of bias with negligibly small cosmic variance effects. 


The correlation function is modeled by the power law or the halo occupation distribution (HOD) model. The power law is the empirical relation that has been used since the early measurements of local galaxy correlation functions were obtained \citep{totsuji1969}. The HOD model is the parameterized model providing the relation between observed galaxies and hosting dark-matter halos of the $\Lambda$CDM structure formation  \citep{cooray2002}. 
The parameters of the HOD model define the occupation of galaxies in a dark-matter halo as a function of mass, including the dark-matter halo mass limit for hosting a galaxy and the power-law slope (and the scatter) of the occupation number depending on dark-matter halo mass. Once the occupation of galaxies in a dark-matter halo is determined, a correlation function and abundance of galaxies can be predicted with the HOD model. The correlation function consists of two components, 1-halo and 2-halo terms in small ($\lesssim 1$ comoving Mpc; cMpc) and large ($\gtrsim 1$ cMpc) scales, respectively (top left panel of Figure \ref{fig:acorr_biasevol}). The 1-halo term signal is originated from clustering of galaxies within one dark-matter halo, while the 2-halo term signal is made by clustering of galaxies hosted by different dark-matter halos. The correlation function and the abundance of galaxies, thus predicted, are compared with those of observational results, which determine the HOD model best-fit parameters. The HOD model with the best-fit parameters reveals properties of dark-matter halos hosting LAEs, and indicates that the average mass of dark-matter halos $\left < M_{\rm h} \right >$ are moderately small, $\left < M_{\rm h} \right >=10^{10}-10^{11} M_\odot$ for the LAEs (top left panel of Figure \ref{fig:acorr_biasevol}; \citealt{ouchi2018}).
%
Because this is the average value of the dark-matter halo masses, there should exist more LAEs with masses higher and lower than the average mass in overdensity (i.e. proto-cluster) and underdensity regions, respectively.
Once the mass of the hosting dark-matter halos is constrained by HOD modeling, the number density of the hosting dark-matter halos $n_{\rm DMH}$ can be estimated with the $\Lambda$CDM model. The $n_{\rm DMH}$ value can be compared with the number density of the observed LAEs $n_{\rm Ly\alpha}$. The Ly$\alpha$ duty cycle of LAEs, $DC_{\rm Ly\alpha}$ is defined as the ratio of $n_{\rm Ly\alpha}$ to $n_{\rm DMH}$,
\begin{equation}
DC_{\rm Ly\alpha} = \frac{n_{\rm Ly\alpha}}{n_{\rm DMH}}.
\label{eq:dc_lya}
\end{equation}
$DC_{\rm Ly\alpha}$ is the fraction of the Ly$\alpha$ emitting galaxies to the dark-matter halos for a given mass. In the physical picture of LAEs, the Ly$\alpha$ duty cycle is determined by two effects, the intermittent star-formation activity and time-dependent Ly$\alpha$ escape.  Observational results suggest that $DC_{\rm Ly\alpha}$ is about $1$\% \citep{gawiser2007,ouchi2010} that is comparable with the predictions of the numerical simulations, $DC_{\rm Ly\alpha} \sim 1-10$\% \citep{nagamine2010}. 
As detailed in this section, clustering measurements of LAEs are not only useful to probe the cosmic structures, but also to understand the physical properties of LAEs, hosting dark-matter halos and Ly$\alpha$ duty cycles.

\begin{figure}[h]
\includegraphics[width=6.2in]{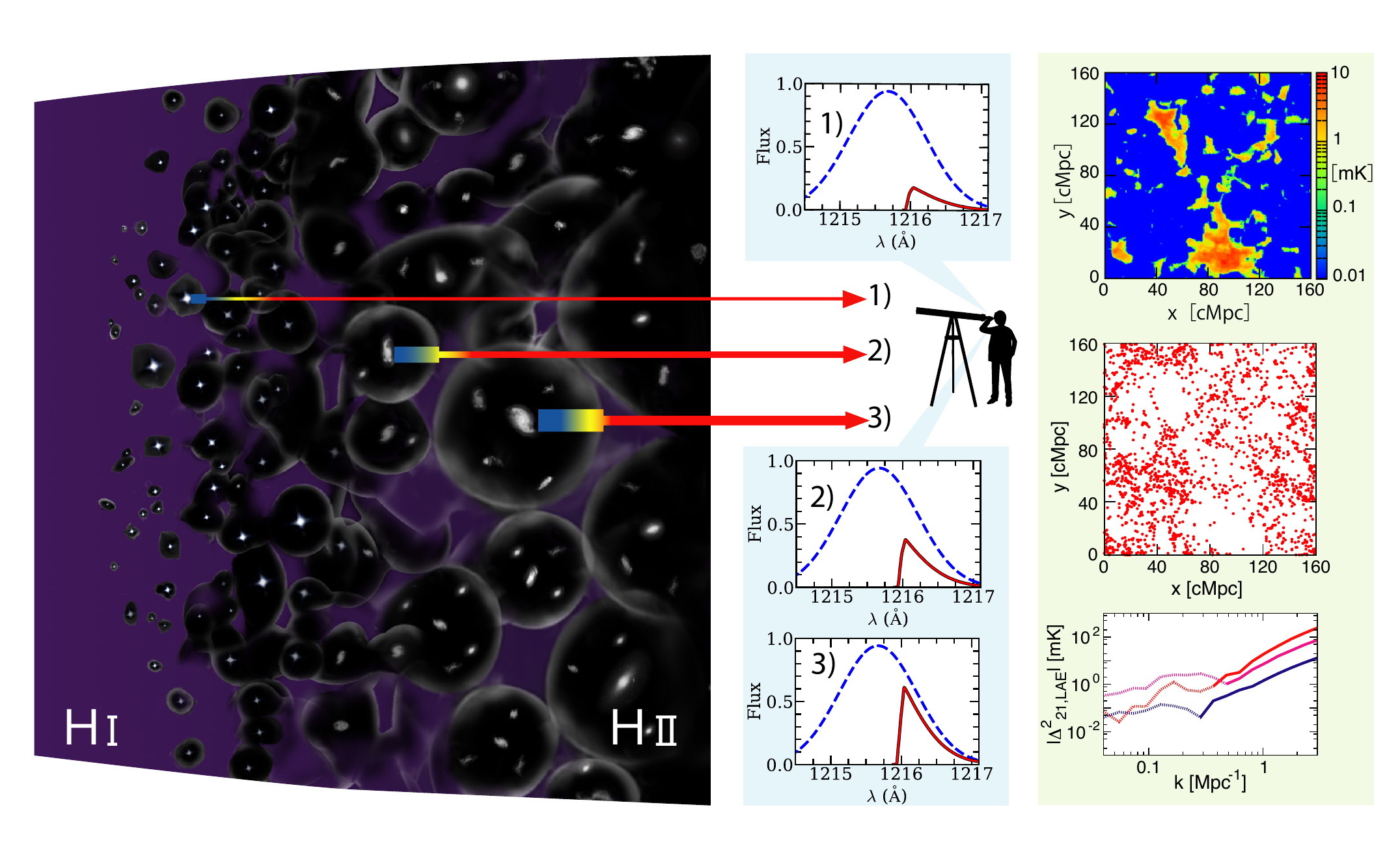}
\caption{
%
%
Left: Conceptual diagram of observations for LAEs at the EoR in the expanding universe. The purple and black colors indicate the neutral and ionized IGM, respectively. LAEs in the ionized IGM (i.e. ionized bubbles) emit Ly$\alpha$ photons that are scattered by the neutral hydrogen IGM. The Ly$\alpha$ photons emitted in the large ionized bubbles can escape, being dimmed by a small amount of scattering, due to the Ly$\alpha$ velocity redshifted from the IGM velocity of the Hubble flow.
In the diagram, wavelengths of Ly$\alpha$ emission are expressed with the spectral bands of the blue to red colors, and the widths of the spectral bands indicate Ly$\alpha$ emission intensities.
%
Center: Model Ly$\alpha$ spectra \citep{dijkstra2007}. The Ly$\alpha$ lines produced by galaxies are shown with the blue dashed curves. The observed Ly$\alpha$ lines are shown with the red curves. These galaxies reside at ionized bubbles whose radii are small (0 pMpc, i.e. fully neutral IGM; top), medium (2 pMpc; middle), and large (10 pMpc; bottom),
where pMpc stands for physical Mpc.
%
Right: 21cm brightness temperature map at $z=6.6$ predicted by numerical simulations (top; \citealt{kubota2018}). The blue regions of 0 mK correspond to the fully ionized regions. The middle panel shows positions of the model LAEs (red circles) in the cosmic volume same as the one of the top panel. The bottom panel presents LAE-21cm cross-power spectra calculated on the basis of the models of the top and the middle panels
for three conditions of neutral hydrogen fractions, 1.7\% (blue line), 31\% (purple line), and 60\% (red line). The solid and dotted lines indicate the positive and negative cross-power spectra, respectively. For display purposes, the signs of the negative cross-power spectra (dotted lines) are changed to the positive signs.
%
The data of center and right panels are adapted from \citet{dijkstra2007} and \citet{kubota2018}, respectively, with permission.
}
\label{fig:Lya_reionization}
\end{figure}

\section{Cosmic Reionization and Ly$\alpha$}
\label{sec:reionization}



Deep observations for galaxies have reached the EoR at $z\gtrsim 6$, when the neutral hydrogen of the IGM is ionized (\citealt{fan2006}; left panel of Figure \ref{fig:Lya_reionization}).


Ly$\alpha$ photons from a galaxy 
are 
scattered by the partly neutral hydrogen at the EoR. 
Ly$\alpha$ lines are redshifted from the systemic velocity of a galaxy by $\sim 100-200$ km s$^{-1}$ on average (Section \ref{sec:ISM}), avoiding the extremely strong resonant scattering 
at the rest-frame $1216$\AA\ in the partly neutral IGM. However, the redshifted Ly$\alpha$ is still scattered by the neutral hydrogen of the IGM via the Ly$\alpha$ damping wing (DW). Ly$\alpha$ DW is a Ly$\alpha$ profile of the long wavelength tail given by the natural broadening (originated by the quantum effect).

There is another important mechanism of Ly$\alpha$ scattering in the neutral IGM at the EoR that is illustrated in Figure \ref{fig:Lya_reionization}. Theoretical models of reionization predict that UV radiation of a galaxy make an ionized bubble around the galaxy in the neutral IGM, and that the galaxy resides within the ionized bubble (left panel of Figure \ref{fig:Lya_reionization}). In this theoretical picture, Ly$\alpha$ photons can escape from the partly neutral IGM via the ionized bubble. 
At the back of the bubble on the border between the ionized and neutral hydrogen IGM,
%
the neutral hydrogen gas is redshifted from the galaxy by the Hubble flow, which help Ly$\alpha$ photons of the galaxy escape from the partly neutral IGM to the observer
(center column panels
%
of Figure \ref{fig:Lya_reionization}).

Constraining the amount of the Ly$\alpha$ scattering in the IGM by observations, various studies have estimated the neutral hydrogen fraction $x_{\rm HI}$ or the ionized fraction $Q_{\rm HII}$,
\begin{eqnarray}
x_{\rm HI} & = & \frac{n_{\rm HI}}{(n_{\rm HI} + n_{\rm HII})}\\
Q_{\rm HII} & = & 1-x_{\rm HI},
\label{eq:xHI}
\end{eqnarray}
where $n_{\rm HI}$ and $n_{\rm HII}$ are the densities of neutral and ionized hydrogen, respectively. A volume-averaged $x_{\rm HI}$ (or $Q_{\rm HI}$) is derived by Ly$\alpha$ studies.

Recent observational studies have identified signatures of Ly$\alpha$ photons scattered by Ly$\alpha$ DW of the neutral hydrogen gas at the EoR. Bright continuum objects, Gamma-ray bursts (GRBs) and QSOs, at the EoR are good probes of Ly$\alpha$ DW absorption with an intrinsic continuum spectrum modeled with a power law and an average spectrum of low-$z$ QSOs, respectively \citep{totani2016,banados2018}. However, the small numbers of GRBs and quasars can probe the neutral hydrogen of the IGM on the small number of sight lines. For example, a theoretical model suggests that a single-object estimate of GRB gives a systematic bias at the moderately high level of $\Delta x_{\rm HI}\sim 0.3$, due to the patchy distribution of the neutral hydrogen at the EoR \citep{mcquinn2008}. Moreover, there is a well-known systematics of the ionization state of the IGM around a QSO (i.e. proximity effect \citealt{Bajtlik1988}). Here, observations of LAEs can complement the one with the bright continuum objects, GRBs and QSOs. Although there are a variety of Ly$\alpha$ emission spectral shapes in LAEs, one can investigate a large number of LAEs distributed from high to low density regions. The large statistics of LAEs allows us to 
understand the evolution of
%
the average neutral hydrogen fraction with negligible bias raised by the diversity of Ly$\alpha$ emission spectral shapes \citep{weinberger2019}.

\begin{figure}[h]
\includegraphics[width=6.2in]{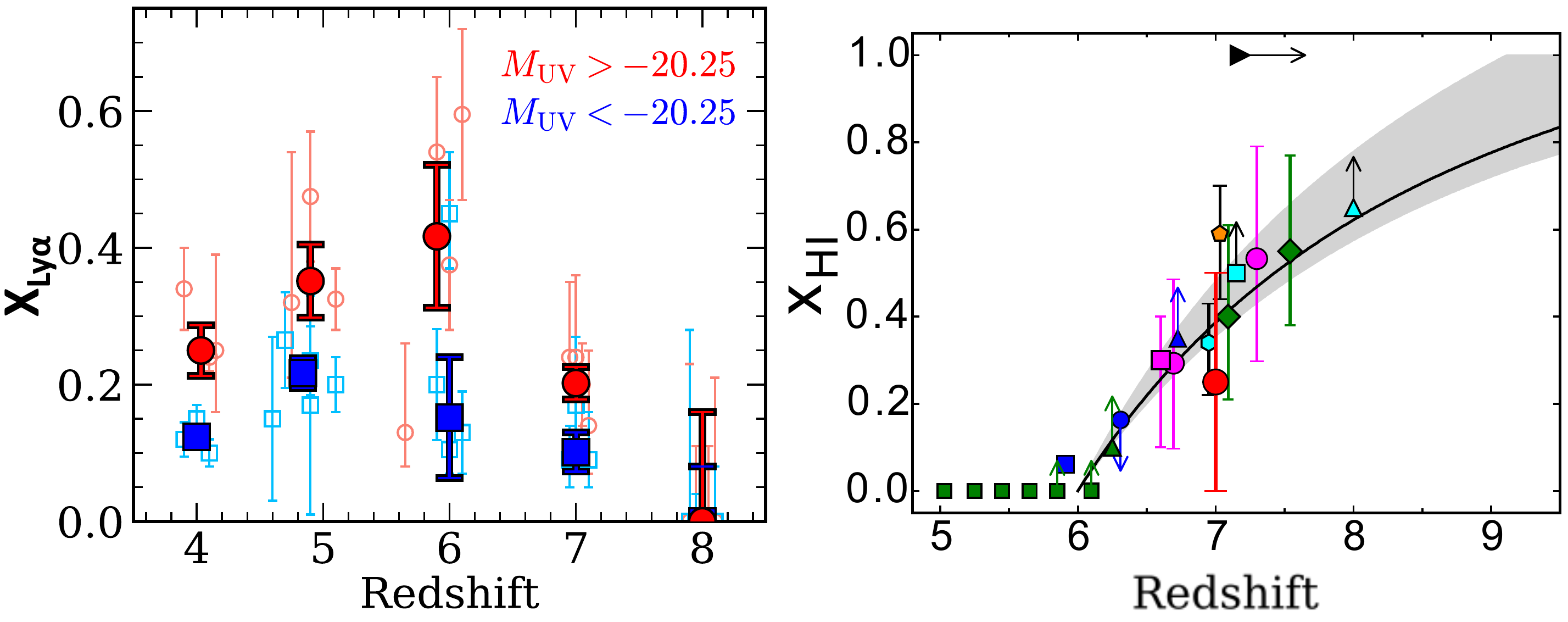}
\caption{
Left: Fraction of Ly$\alpha$ emitting galaxies with Ly$\alpha$ $EW_0\ge 25$\AA, $X_{\rm Ly\alpha}$, as a function of redshift. The red filled circles and the blue filled squares indicate the average $X_{\rm Ly\alpha}$ values for galaxies with the faint ($M_{\rm UV}>-20.25$) and bright ($M_{\rm UV}<-20.25$) UV-continuum magnitudes, respectively. These average $X_{\rm Ly\alpha}$ values are estimated 
with the compilation data of the red open circles ($M_{\rm UV}>-20.25$) and the blue open squares ($M_{\rm UV}<-20.25$) 
that are so far obtained by various observational studies \citep{arrabalharo2018,caruana2018,cassata2015,curtislake2012,debarros2017,kusakabe2020,mallery2012,mason2019,ono2012,pentericci2018,schenker2014,stark2011,tilvi2014,treu2013}.
%
Right: Neutral hydrogen fraction as a function of redshift \citep{itoh2018}. The red and magenta circles represent $x_{\rm HI}$ estimates given by the Ly$\alpha$ LD evolution obtained with LF 
measurements. The cyan circle, square, and triangle are the constraints of $x_{\rm HI}$ given by the Ly$\alpha$ emitting galaxy fraction measurements. The orange pentagon shows the $x_{\rm HI}$ estimate calculated with the Ly$\alpha$ EW distribution of dropout galaxies. The magenta square denotes the $x_{\rm HI}$ value estimated with the LAE clustering. The blue triangle, square, and circle indicate $x_{\rm HI}$ values given by GRB Ly$\alpha$ damping wing (DW) measurements. The green diamonds and triangle show $x_{\rm HI}$ constraints obtained by the QSO Ly$\alpha$ DW observations, while the green squares represent those of QSO Gunn-Peterson optical depths. See Figure 14 of \citet{itoh2018} for the references for the measurements. The black triangle presents the $1 \sigma$ lower limit obtained with the CMB Thomson scattering optical depth measurement. The black curve and the gray shade denote the best-estimate $x_{\rm HI}$ and the error given with eqs. (\ref{eq:ionization_equation}) and (\ref{eq:n_ion_galaxy}) and galaxy $\rho_{\rm UV}$ measurements \citep{ishigaki2018}.
The panel b is adapted from \citet{itoh2018} with permission.
}
\label{fig:Lyafrac_xHI}
\end{figure}

There are two popular techniques that study the average amount of Ly$\alpha$ damping wing absorption in the IGM with LAEs. One technique is the Ly$\alpha$ 
LF 
via the comparison with the UV-continuum 
LF \citep{malhotra2004,kashikawa2011,itoh2018}. Ly$\alpha$ and UV-continuum LDs are derived with the 
LFs. 
Note that the redshift evolution of the Ly$\alpha$ LD is determined by galaxy evolution and Ly$\alpha$ opacity of the IGM. Because the UV-continuum LD is a tracer of 
SFR 
and the dust extinction in the ISM, the Ly$\alpha$ LD decrease faster than the UV-continuum LD evolution suggests an increase of Ly$\alpha$ opacity of the IGM,
%
which is shown in the top panel of Figure \ref{fig:LyaLD}
(see Figure 11 of \citealt{itoh2018} for statistically significant results).
%
Ly$\alpha$ LD rapidly drops from $z\sim 6.5$ to $7.5$, while the evolution of the UV-continuum LD is milder than the one of the Ly$\alpha$ LD. This evolutional difference allows us to estimate the Ly$\alpha$ opacity via the comparisons with analytical and numerical models of cosmic reionization \citep{furlanetto2006,mcquinn2007}. Another technique is to investigate a fraction of Ly$\alpha$ emitting galaxies to all of the UV continuum-selected galaxies \citep{stark2011,pentericci2011,ono2012}. The left panel of Figure \ref{fig:Lyafrac_xHI} 
and Table \ref{tab:lya_emitting_galaxy_fraction}
summarize
the measurements of the Ly$\alpha$ emitting galaxy fractions for galaxies with the two different limits of the UV-continuum magnitudes. 
%
In Figure \ref{fig:Lyafrac_xHI} and Table \ref{tab:lya_emitting_galaxy_fraction}, we show the average Ly$\alpha$ emitting galaxy fractions with standard errors that are estimated 
from the compilation of the data so far obtained by various observational studies \citep{arrabalharo2018,caruana2018,cassata2015,curtislake2012,debarros2017,kusakabe2020,mallery2012,mason2019,ono2012,pentericci2018,schenker2014,stark2011,tilvi2014,treu2013}.
%
The Ly$\alpha$ emitting galaxy fraction is increasing from $z\sim 4$ to $6$, while this fraction clearly decreases from $z\sim 6$ to $8$, indicating that the Ly$\alpha$ opacity is increasing from $z\sim 6$ to $8$. The decrease of the Ly$\alpha$ emitting galaxy fraction is modeled, providing estimates of the cosmic average $x_{\rm HI}$. The $x_{\rm HI}$ estimates, thus obtained, are summarized in the right panel of Figure \ref{fig:Lyafrac_xHI} indicating the cosmic reionization history. The neutral hydrogen gas fraction increases from $z\sim 6$ towards high-$z$, and the mid-point of cosmic reionization, $x_{\rm HI}=50$\%, is estimated to be $z\sim 7.5$.



\begin{table}[h]
\tabcolsep7.5pt
\caption{
Fraction of Ly$\alpha$ Emitting Galaxies with $EW_0\ge 25$\AA
}
\label{tab:lya_emitting_galaxy_fraction}
\begin{center}
\begin{tabular}{@{}l|c|c|c@{}}
\hline
Sample & Magnitude & Redshift & $X_{\rm Ly\alpha}$ \\
\hline
Bright & $M_{\rm UV}<-20.5$ & 4.0 & $0.12\pm0.02$\\
       &                    & 4.8 & $0.22\pm0.02$\\
       &                    & 6.0 & $0.15\pm0.09$\\
       &                    & 7.0 & $0.10\pm0.03$\\
       &                    & 8.0 & $<0.08$\\
Faint & $M_{\rm UV}>-20.5$ & 4.0 & $0.25\pm0.04$\\
      &                    & 4.9 & $0.35\pm0.05$\\
      &                    & 5.9 & $0.42\pm0.10$\\
      &                    & 7.0 & $0.20\pm0.02$\\
      &                    & 8.0 & $<0.16$\\
\hline
\end{tabular}
\end{center}
\end{table}

Ly$\alpha$ emission is the strongest line found in the optical wavelength for majority of objects at $z\sim 2-7$. Although this characteristics of Ly$\alpha$ emission is very advantageous for studies of high-$z$ objects, this advantage may not be true for sources at the middle or early EoR. This is because a chance of a detection of Ly$\alpha$ emitting galaxy decreases towards high-$z$ (left panel of Figure \ref{fig:Lyafrac_xHI}), due to the fact that Ly$\alpha$ photons from galaxies are scattered by the neutral hydrogen of the IGM at the EoR $z>6$. If LAEs and the IGM are static and uniformly distributed, no Ly$\alpha$ photons can escape from the neutral universe. However, in reality, Ly$\alpha$ photons come out of the highly neutral IGM with the help of the Ly$\alpha$ velocity shifts (from the neutral IGM) given by the peculiar motions of galaxies as well as the ionized bubbles around clustered galaxies. Theoretical models indicate that about 10\% of Ly$\alpha$ fluxes can escape from the highly neutral IGM by the help of these physical effects \citep{gnedin2004}. 
%
%
There is an open question whether a redshift limit of a Ly$\alpha$ detection exists and what redshift is the limit, if it exists. A future accurate measurement of the Ly$\alpha$ observability at the early EoR ($z\sim 10$) will allow us to understand the physical properties of the LAE dynamics and the ionized bubbles.
%
%



Theoretical studies predict that LAEs are key sources for identifying signals of the {\sc Hi} 21cm emission originated from neutral hydrogen at the EoR \citep{lidz2009,sobacchi2015,hutter2018}. Although there are several high-sensitiviy radio telescopes targeting the EoR 21cm emission (e.g. LOw Frequency ARray, LOFAR; \citealt{jelic2014}), no signals of the EoR 21cm emission have been detected so far, mainly due to the bright foreground emission of the Earth's ionosphere.
%
%
Here, LAEs can be used as signposts for the EoR 21cm emission detections. The large sample of LAEs with known positions and redshifts allows us to conduct cross-correlation analysis with EoR 21cm data \citep{lidz2009}, and to identify a signal of the EoR 21cm emission, removing the problematic foreground emission. The bottom right panel of Figure \ref{fig:Lya_reionization} presents the LAE-21cm cross-power spectra predicted by numerical simulations \citep{kubota2018}.
These cross-power spectra are calculated with the simulation data of
the 21cm brightness temperature (LAE distribution) map shown in
the top (middle) right panel of Figure \ref{fig:Lya_reionization}.
The cross-power spectrum consists of two components at small and large scales. The large-scale component is the negative signal (i.e. anti-correlation) representing ionizing sources (LAEs) that make ionized bubbles (i.e. no neutral hydrogen around LAEs), while the small-scale component is the positive signal indicating that LAEs form in an overdensity of neutral hydrogen. There exists a transition between the negative and positive signals, due to the ionized bubbles around LAEs. A detection of the negative and positive signals will confirm an existence of ionized bubbles at the EoR. In summary, the LAE-21cm cross correlation analysis will help the detection of the EoR 21cm signal, and characterize the cosmic reionization process via the ionized bubble topology.



LAEs are young 
SFGs 
that have abundant massive stars producing ionizing photons contributing to cosmic reionization. 
For faint SFGs with $M_{\rm UV}> -20.25$, the number fraction of LAEs reaches $\sim 50$\% of the faint SFGs
%
at the redshift of the EoR $z\sim 6$ 
\footnote{
Although the fraction of LAEs decreases from $z\sim 6$ towards high-$z$, this decrease is just made by a low observability given by the external effect of the neutral IGM that scatters Ly$\alpha$ photons.
}
(left panel of Figure \ref{fig:Lyafrac_xHI}),
indicating that LAEs are dominant population of 
SFGs 
that supply ionizing photons for cosmic reionization.
In fact, the UV-continuum luminosity function of LAEs is comparable to the one of LBGs at $z\sim 6$, while those of LAEs fall below those of LBGs by an order of magnitude at $z\sim 3-4$ (See Figure 22 of \citealt{ouchi2008}).

With the ionizing photon production rate $\dot{n}_{\rm ion}$ that is defined by the number of ionizing photons per volume and time, the budget of ionizing photons in cosmic reionization is modeled in the one zone model of the ionization equation \citep{madau1999,robertson2015}.
Here the derivative of ionized fraction (eq. \ref{eq:xHI}) with respect to time $\dot{Q}_{\rm HII}$ is written as
%
\begin{equation}
\dot{Q}_{\rm HII} = \frac{ \dot{n}_{\rm ion} }{ \left< n_{\rm H} \right> } - \frac{ Q_{\rm HII} }{ t_{\rm rec} },
\label{eq:ionization_equation}
\end{equation}
where $\left< n_{\rm H} \right>$ and $t_{\rm rec}$ are the average hydrogen number density and the recombination time, respectively, defined by
\begin{eqnarray}
\label{eq:hydrogen_number_density}
\left< n_{\rm H} \right>  & = & \frac{X_{\rm p} \Omega_{\rm b} \rho_{\rm c}}{m_{\rm H}}\\
\label{eq:recombination_time1}
t_{\rm rec} & = & \frac{1}{C_{\rm HII} \alpha_{\rm B}(T) (1+Y_{\rm p}/4X_{\rm p}) \left< n_{\rm H} \right> (1+z)^3}.
\end{eqnarray}
The parameters, $m_{\rm H}$, $\rho_{\rm c}$, and $X_{\rm p}$ ($Y_{\rm p}$) are the mass of the hydrogen atom, the critical density, and the primordial mass fraction of hydrogen (helium), respectively. In eq. (\ref{eq:recombination_time1}), 
the parameters of $C_{\rm HII}$ and $\alpha_{\rm B} (T)$ represent the clumping factor and the case B hydrogen recombination coefficient for the IGM temperature $T$ at a mean density, respectively. All of these parameters are determined by physics and cosmology, except for $C_{\rm HII}$ and $\dot{n}_{\rm ion}$. The parameter of $\dot{n}_{\rm ion}$ is closely related to galaxy formation, and estimated by

\begin{eqnarray}
\dot{n}_{\rm ion} & = & \int^{+\infty}_{0}
f_{\rm esc}^{\rm ion} (L_{\rm UV})\ \xi_{\rm ion} (L_{\rm UV})\ \phi(L_{\rm UV})\ L_{\rm UV}\ dL_{\rm UV}\\
 & = & f_{\rm esc}^{\rm ion}\ \xi_{\rm ion}\ \rho_{\rm UV} \ \ \ \ \ \ \ \ \ \ \ \ \ {\rm [for \ no\ L_{\rm UV}\ dependences]},
\label{eq:n_ion_galaxy}
\end{eqnarray}
%
%
%
%
where 
$f_{\rm esc}^{\rm ion}$ and $\xi_{\rm ion}$ are the ionizing photon escape fraction
\footnote{
Note that this parameter of $f_{\rm esc}^{\rm ion}$ is different from the escape fraction of Ly$\alpha$ photons (eq. \ref{eq:f_esc}).
}
and the ionizing photon production efficiency, respectively. 
The values of $L_{\rm UV}$, $\phi(L_{\rm UV})$, $\rho_{\rm UV}$ are the UV-continuum luminosity, the UV-continuum 
LF, 
and the UV-continuum LD, respectively (Section \ref{sec:Lya_emitter_obs}). 
%
The equation  (\ref{eq:n_ion_galaxy}) assumes no luminosity dependence in the parameters of $f_{\rm esc}^{\rm ion}$ and $\xi_{\rm ion}$.

There are three major parameters for $\dot{n}_{\rm ion}$; i) $\rho_{\rm UV}$, ii) $\xi_{\rm ion}$, and iii) $f_{\rm esc}^{\rm ion}$.
%
%
i) The LAEs' contribution to $\rho_{\rm UV}$ is large at high redshift. As discussed in Section \ref{sec:Lya_emitter_obs}, the fraction of LAEs to all galaxies in $\rho_{\rm UV}$ increases from the local universe towards the EoR $z\sim 6$. This increase indicates that LAEs' contribution to cosmic SFRDs increases \citep{ouchi2008,ciardullo2012}.
The fraction of the Ly$\alpha$ emitting galaxies reaches nearly 50\% at $z\sim 6$ \citep{stark2011}. Because LAEs are abundant and major galaxies at the EoR, ionizing photon emission properties of LAEs are critical to understand sources of cosmic reionization.
ii) For the photon production, it is recently claimed that the LAEs' $\xi_{\rm ion}$ is higher than other high-$z$ galaxy populations. The typical ionizing photon production efficiency of LAEs is 
$\log(\xi_{\rm ion}/{\rm [Hz\ erg^{-1}]}\sim 25.5$, significantly higher than those of LBGs  at a given UV magnitude \citep{harikane2018}.
iii) It is suggested that the average ionizing photon escape fraction is also high for LAEs.
Deep spectroscopy and 
NB imaging observations reveal that the average $f_{\rm esc}^{\rm ion}$ value is $\sim 20$\% for LAEs at $z\sim 3$ (\citealt{shapley2006}; see also \citealt{vanzella2016,debarros2016}).
In other words, a positive correlation between $f_{\rm esc}^{\rm ion}$ and Ly$\alpha$ $EW_0$ is suggested \citep{iwata2009,nestor2013}
\footnote{
Because Ly$\alpha$ emission is produced by recombination in hydrogen gas of the ISM that absorbs ionizing photons, strong Ly$\alpha$ emission should not be found in galaxies emitting ionizing photons.
However, even under this physical picture, model calculations suggest that there is a positive correlation between $f_{\rm esc}^{\rm ion}$ and Ly$\alpha$ 
$EW_0$
in the regime of the moderately low Ly$\alpha$ 
$EW_0$
values \citep{nakajima2014}.
}.
Deep 
NIR spectroscopy studies find that the line flux ratios of {\sc [Oiii]}5007 to {\sc [Oii]}3727, $O_{32}$, of LAEs at $z\sim 2$ are large, $O_{32}\sim 10$ \citep{nakajima2013}. Because such galaxies with large $O_{32}$ values show a high ionizing photon escape fraction, $f_{\rm esc}^{\rm ion} \sim 5-50$\% \citep{izotov2016,vanzella2016}, LAEs are thought to be major emitters of ionizing photons at high redshift. For the physical origin of ionizing photon emission of LAEs, various scenarios including density-bounded nebula are suggested  \citep{nakajima2014}. However, the physical reason of escaping ionizing photons is still under debate.

\section{Open Questions and Future Prospects}
\label{sec:open_question}
In observational studies of Ly$\alpha$ emission, there are five major questions
\footnote{
In addition to these five major questions, various observational measurements should be refined and revisited. One example is the determination of Ly$\alpha$ 
LFs, 
where the flux contributions of extended Ly$\alpha$ halos are not included in many of previous studies 
\citep{herenz2019}.
}
that are listed below.

\begin{enumerate}

    \item What are dominant sources of Ly$\alpha$ photons in LAEs including ELANe, LABs, and LAHs? Are the majority of Ly$\alpha$ photons produced in {\sc Hii} regions of star formation and/or AGN? Are there any major contributions of Ly$\alpha$ photons from gravitational cooling? (Section \ref{sec:physical_picture})
    
    \item How faint 
    LAEs 
    exist? What is the dark-matter low-mass limit of 
    SFGs that emit Ly$\alpha$? What is the faint-end slope $\alpha$ of the Ly$\alpha$ 
    LF? (Section \ref{sec:Lya_emitter_obs})
 
    \item
    %
    What is the major physical origin of the high Ly$\alpha$ emissivity of LAEs such characterized by the large Ly$\alpha$ $EW_0$ and the high $f_{\rm esc}^{\rm Ly\alpha}$? What makes LAEs different from other SFGs with weak/no Ly$\alpha$ emission? (Sections \ref{sec:Lya_emitter_obs}-\ref{sec:LAE_clustering})
 
    \item What are the ionization state of the ISM and CGM of LAEs? Is the density-bounded picture possible in the ISM? Do regular LAEs (that do not include ELANe with an active AGN) have the CGM gas mainly consisting of neutral hydrogen? What makes the observational trend of the high fraction of Lyman-continuum leaking for LAEs that is important for sources of reionization? (Sections \ref{sec:ISM}-\ref{sec:CGM} and \ref{sec:reionization})

    \item Do pop III LAEs exist, as originally predicted by \citet{partridge1967}? If pop III LAEs exist, are they observable at the post EoR (late epoch) and/or the EoR (early epoch) that may or may not have a mechanism allowing the Ly$\alpha$ escape from the partly neutral universe? Does the Ly$\alpha$ escape from the partly neutral universe depend on reionization topology? (Sections \ref{sec:ISM} and \ref{sec:reionization})
  
\end{enumerate} 

There are a number of on-going and next generation instrument projects that are critical for resolving these five questions. Such instruments include those for statistical studies, Subaru/HSC, Prime-Focus Spectrograph (PFS), and HETDEX, high sensitivity observations, VLT/MUSE, and 
Keck/KCWI (\citealt{morrissey2018}), 
and submm/mm wavelength observations,
ALMA. Ly$\alpha$ emission studies will be boosted with the next generation large telescopes, 
\textit{James Webb Space Telescope} (JWST) 
and extremely large telescopes (ELTs) by the great sensitivities. Moreover, the {\sc Hi} 21cm observations with Hydrogen Epoch of Reionization Array (HERA) and Square Kilometre Array (SKA) will complement the Ly$\alpha$ emission studies at the EoR (Section \ref{sec:reionization}). Because the future imaging and spectroscopic data such taken with HSC, PFS, and HETDEX are too large to be analyzed by visual inspections for quality assessments, one will need new analysis technique of artificial intelligence that includes machine learning.



 
In the past two decades, observations have targeted 
SFGs 
emitting Ly$\alpha$ that are easily detected. Recent studies change the observational targets to diffuse Ly$\alpha$ emission around galaxies and fillaments of the large-scale structures that are new observational frontiers.
%
%
%
Although Ly$\alpha$ emission traces both {\sc Hii} and {\sc Hi} gas with a great sensitivity, the origins of the Ly$\alpha$ emission cannot be clearly distinguished between, e.g. recombination and resonant scattering.
%
The future observing programs should combine the Ly$\alpha$ emission results with the complementary non-resonant hydrogen emission (H$\alpha$ and H$\beta$) and {\sc Hi} absorption observations \citep{lee2014},
%
%
exploiting the high sensitivity instruments and the next generation large telescopes. Including the 21cm observations at the EoR, future observations will reveal the {\sc Hi} and {\sc Hii} gas distribution in the universe in the scales from star-forming regions to the large-scale structures.
%
%

\section{Summary and Final Thoughts}
\label{sec:summary}

Since the first discoveries of strong Ly$\alpha$ emitting high-$z$ galaxies in the late 1990s, Ly$\alpha$ emission has been one of the major probes for the high redshift universe. 
Defining LAEs as galaxies with Ly$\alpha$ $EW_0\gtrsim 20$\AA, more than 1,000 (20,000) LAEs are spectroscopically (photometrically) identified to date.
%
%
%
Observations suggest that LAEs emerge at high redshift, and the fraction of LAEs to the other galaxies increases (for a given UV-continuum luminosity) towards high redshift up to the EoR (Section \ref{sec:Lya_emitter_obs}). LAEs are galaxies of a major population at high redshift. This emergence of LAEs are understood by the increase of Ly$\alpha$ escape fraction for high-$z$ 
SFGs. 

At high redshift ($z\gtrsim 2$), a majority of LAEs with a $L^*$ Ly$\alpha$ luminosity are low-mass ($M_*\sim 10^8-10^9 M_\odot$ and $M_{\rm h}\sim 10^{10}-10^{11} M_\odot$) galaxies with a 
%
star-forming activity ($1-10\ M_\odot$ yr$^{-1}$; Figure \ref{fig:SEDfit}). Such typical LAEs are high-$z$ analogs of dwarf galaxies having a young dust-poor stellar population with a sub-solar metallicity in the ISM gas, showing a compact disky morphology 
($n_{\rm s}\sim 1$) 
with an effective radius of $\sim 1$ pkpc (Sections \ref{sec:morphology}-\ref{sec:ISM}).  These physical properties of LAEs can be explained by the nature of the faint continuum 
(sub $L_{\rm UV}^*$;
%
Section \ref{sec:Lya_emitter_obs}). In fact, LAEs follow the size-mass relation and the SFR-mass metallicity relation.
%
LAEs are probably 
SFGs similar to those of dropout galaxies or LBGs with weak or no Ly$\alpha$ emission. 
In this case, the difference of LAEs from the other galaxies is characterized by the Ly$\alpha$ emission duty cycle of $\sim 1$\% (Section \ref{sec:LAE_clustering}) caused by the combination of intermittent star formation and time-dependent Ly$\alpha$ escape. 
%


Ly$\alpha$ emission is not only observed at the ionized hydrogen clouds by recombination, but also at the neutral hydrogen clouds by resonant scattering.
The resonance nature of Ly$\alpha$ allows us to probe diffuse gas in the ISM/CGM of a galaxy as well as the IGM. For the ISM, the skewed profiles of galaxy Ly$\alpha$ emission suggest the existence of 
neutral hydrogen gas outflowing from Ly$\alpha$ sources (such as star-forming regions) that is explained by the
ES model (Sections \ref{sec:physical_picture} and \ref{sec:ISM}). 
For the CGM (Section \ref{sec:CGM}), diffuse Ly$\alpha$ emission around an 
SFG, 
i.e. LAH, is ubiquitas. Although it is not understood whether the LAH is made by the recombination process of the ionized clouds or the resonant scattering of the neutral clouds, the CGM includes metal at least up to $\sim 10$ pkpc
such identified by the extended carbon emission with a radial profile similar to those of LAHs,
%
indicative that the CGM is metal-enriched by outflow.
%
The LAH extends beyond the virial radius of a galaxy dark-matter halo up to $1$ cMpc or more, perhaps originating from filaments in the large-scale structures.
%
%


Strong Ly$\alpha$ emission is easily detected by observations, being used as a signpost of high-$z$ objects. 
Reports of Ly$\alpha$-emission detection have been made for galaxies up to $z\sim 9$, to date. The redshift frontier of galaxy observations has entered into the EoR, where Ly$\alpha$ is scattered by the partly neutral hydrogen IGM. 
Large statistics of galaxy Ly$\alpha$ emission provides the average amount of Ly$\alpha$ scattered by the neutral IGM at the EoR, having revealed the cosmic reionization history, so far, up to the middle point of reionization $x_{\rm HI}\sim 50$\% at $z\sim 7-8$ (Section \ref{sec:reionization}).
The Ly$\alpha$ escape from the partly neutral hydrogen IGM is not only a probe of cosmic reionization, but also a signal of the galaxies and ionized-bubble relation (determined by galaxy clustering, star-formation duty cycle, and galaxy peculiar motions).
One critical question about Ly$\alpha$ sources is how early epoch, e.g. $z\sim 10$ or $z\sim 15$, Ly$\alpha$ emission can escape (Section \ref{sec:open_question}).

{\it Throughout this review}, we have devoted some pages to explain the observational differences of LAEs from the other SFGs emitting weak/no Ly$\alpha$, discussing possible physical origins of the high Ly$\alpha$ emissivity of LAEs such characterized by the large Ly$\alpha$ $EW_0$ and the high $f_{\rm esc}^{\rm Ly\alpha}$. Although the major physical origin of the Ly$\alpha$ emissivity difference is an open question as raised in the item 3 of Section \ref{sec:open_question}, we have a physical picture that a column density of {\sc Hi} gas towards a Ly$\alpha$ source (such as star-forming regions) is a key physical quantity that controls the Ly$\alpha$ emissivity. For LAEs, the Ly$\alpha$ offset velocities are comparable with outflow velocities ($\Delta v_{\rm Ly\alpha} \sim V_{\rm exp} \sim 200$ km s$^{-1}$), and this kinematic property of LAEs can be explained by a small {\sc Hi} column density producing no significant backscattered Ly$\alpha$ photons (Section \ref{sec:ISM}). The high $O32$ ratios and the existence of the strong high ionization lines suggest that the ISM of LAEs are highly ionized, and this photoionization property of LAEs indicates a small fraction of {\sc Hi} gas in LAEs (Section \ref{sec:ISM}). There is an anti-correlation between Ly$\alpha$ $EW_0$ and the Ly$\alpha$ spatial offset $\delta_{\rm Ly\alpha}$ on the sky, and this Ly$\alpha$ morphological property can be explained by the long mean-free path (less resonant scattering) of Ly$\alpha$ photons for LAEs with a small H{\sc i} column density towards the observer (Section \ref{sec:morphology}). This explanation for the Ly$\alpha$ $EW_0$-$\delta_{\rm Ly\alpha}$ anti-correlation is consistent with the fact that the Ly$\alpha$ halos are small for LAEs whose central part ($r<8$ pkpc) of Ly$\alpha$ luminosity is bright (Section \ref{sec:CGM}).
%
%
%
%
%
In this way, all of the independent observational properties of kinematics, photoionization, and morphology consistently support the physical picture that a column density of {\sc Hi} gas is a key physical quantity for the Ly$\alpha$ emissivity, and that {\it LAEs are different from the other SFGs emitting weak/no Ly$\alpha$ due to the small H{\sc i} column density of LAEs}.
In this physical picture, we would witness SFGs at the galaxy evolutionary phase of the highly-ionized {\sc Hi} poor state as LAEs, when the SFGs are free from dusty and gaseous starbursts such triggered by gas rich mergers. Because the cosmic average of $f_{\rm esc}^{\rm Ly\alpha}$ (i.e. average Ly$\alpha$ emissivity) increases from $z=0$ to $6$ by two orders of magnitudes (Section \ref{sec:ISM}), in galaxy formation a number of early young galaxies should experience the highly-ionized {\sc Hi} poor (LAE) phase.
%
%
%
%
%

{\it As reviewed in this article}, 
observations of Ly $\alpha$ emission have driven various studies of the high-redshift universe for more than two decades. Ly$\alpha$ observations have been pioneering the redshift frontier and investigating low-mass high-$z$ galaxies. Moreover, recently, Ly$\alpha$ observations have attracted attention for studying cosmic structures traced by diffuse neutral and ionized hydrogen gas of the CGM and the IGM,
serving as a new observational role.
Because the studies of the CGM and the IGM have been (will be) conducted by hydrogen and metal absorption observations (radio low-frequency 21cm observations), forthcoming studies combining these absorption (radio) observations with Ly$\alpha$ emission from neutral and ionized hydrogen will open up a new territory of study, cosmic structures and its connection to galaxy formation that are uncovered to date.



\section*{ACKNOWLEDGMENTS}

We thank 
Hakim Atek,
Sebastiano Cantalupo,
Mark Dijikstra,
Alex Hagen,
Takuya Hashimoto, 
Ryohei Itoh, 
Akira Konno,
Kirsten Kraiberg Knudsen, 
Kenji Kubota,
Floriane Leclercq,
Yuichi Matsuda,
Kimihiko Nakajima, 
Paola Santini, 
Daniel Schaerer,
Dan Stark,
Yuma Sugahara, 
Anne Verhamme,
Hidenobu Yajima, 
and
Erik Zackrisson
for sharing their data and figures 
and giving us helpful comments. 
This publication is based upon work
supported by World Premier International Research
Center Initiative (WPI Initiative), MEXT, Japan, and
KAKENHI (15H02064, 15K17602, 17H01110, 17H01114, and 19K14752) 
Grant-in-Aid for Scientific Research (A), 
Young Scientists (B)
and 
Early-Career Scientists 
through Japan Society for the Promotion of Science.

\bibliographystyle{ar-style2}
\bibliography{ref_ono,ref_ouchi,ref_shibuya}

\end{document}